\documentclass[twocolumn,aps,pra,longbibliography,superscriptaddress,nofootinbib,10pt]{revtex4-1}
\usepackage[latin1]{inputenc}
\usepackage{graphicx,dcolumn,bm}
\usepackage{multirow}
\usepackage{array}
\usepackage{arydshln}
\usepackage{color}
\usepackage[colorlinks,linkcolor=blue,citecolor=blue,hyperindex]{hyperref}
\usepackage{amsfonts}
\usepackage{amssymb}
\usepackage{amsmath}
\usepackage{latexsym}
\usepackage{simplewick}
\usepackage{epstopdf}
\usepackage{multirow}
\usepackage{float}
\usepackage[table]{xcolor}
\usepackage{multibib}
\usepackage{mathrsfs}
\usepackage[sans]{dsfont}
\usepackage[mathscr]{eucal}
\usepackage{epsfig}
\usepackage{palatino}
\usepackage{physics}

\definecolor{Blue}{rgb}{0.0,0.0,1}
\definecolor{Red}{rgb}{1,0.0,0.0}
\definecolor{Green}{rgb}{0,0.5,0.0}

\usepackage{tikz}
\usetikzlibrary{decorations.pathmorphing}
\usetikzlibrary{arrows}
\usetikzlibrary{intersections,shapes.arrows}
\usetikzlibrary{calc}
\usetikzlibrary{quotes,angles}
\usepackage{nicefrac}
\usepackage{pgfplots}
\usepgfplotslibrary{fillbetween}
\pgfplotsset{compat=1.13,colormap={violetnew}{rgb=(0.293416, 0.0574044, 0.529412) rgb=(0.394818,0.233715,0.671945) rgb =(0.49622,0.410025,0.814477) rgb=(0.588672,0.567494,0.910066) rgb=(0.663226,0.687282,0.911765) rgb=(0.73778,0.807069,0.913465) rgb=(0.807267,0.861883,0.894034) rgb=(0.874222,0.884211,0.864039) rgb=(0.941176, 0.906538, 0.834043)}}
\usepgfplotslibrary{groupplots} 
\usepgfplotslibrary[groupplots] 
\usetikzlibrary{pgfplots.groupplots} 
\usetikzlibrary[pgfplots.groupplots] 
\usepgfplotslibrary{statistics}
\usepackage{pgfplotstable}
\tikzset{jumpdot/.style={mark=*,solid},excl/.append style={jumpdot,fill=white},incl/.append style={jumpdot,fill=black}}

\begin{document}

\title{Bounding generalized relative entropies: Nonasymptotic quantum speed limits}
\author{Diego Paiva Pires}
\affiliation{International Institute of Physics and Departamento de F\'{i}sica Te\'{o}rica e Experimental, Universidade Federal do Rio Grande do Norte, Natal, RN, 59078-970, Brazil}
\author{Kavan Modi}
\affiliation{School of Physics \& Astronomy, Monash University, Clayton, Victoria 3800, Australia}
\author{Lucas Chibebe C\'{e}leri}
\affiliation{Department of Physical Chemistry, University of the Basque Country UPV/EHU, Apartado 644, E-48080 Bilbao, Spain}
\affiliation{Institute of Physics, Federal University of Goi\'{a}s, 74.690-900 Goi\^{a}nia, Goi\'{a}s, Brazil}

\begin{abstract}
Information theory has become an increasingly important research field to better understand quantum mechanics. Noteworthy, it covers both foundational and applied perspectives, also offering a common technical language to study a variety of research areas. Remarkably, one of the key information-theoretic quantities is given by the relative entropy, which quantifies how difficult is to tell apart two probability distributions, or even two quantum states. Such a quantity rests at the core of fields like metrology, quantum thermodynamics, quantum communication and quantum information. Given this broadness of applications, it is desirable to understand how this quantity changes under a quantum process. By considering a general unitary channel, we establish a bound on the generalized relative entropies (R\'{e}nyi and Tsallis) between the output and the input of the channel. As an application of our bounds, we derive a family of quantum speed limits based on relative entropies. Possible connections between this family with thermodynamics, quantum coherence, asymmetry and single-shot information theory are briefly discussed. 
\end{abstract}


\maketitle


\section{Introduction}
\label{sec:introd000_xxx_0001}

Since its formulation decades ago by Shannon~\cite{Shannon1948}, information theory has played a major role in both applied and fundamental science, ranging from neuroscience~\cite{Dimitrov2011} to quantum gravity~\cite{Kempf2018,Konishi_2020}, and along the way has impacted thermodynamics~\cite{Goold2016}, finance~\cite{Zhou2013}, and evolutionary biology~\cite{Seoane2018}. A central e\-le\-ment in this theory is the Shannon entropy, which measure how much information is contained in a probability distribution. Shannon entropy also plays a role on the speed of evolution of classical stochastic processes~\cite{PhysRevLett.121.070601}. However, when the basic assumptions of the theory do not hold, e.g., extensivity or very large data sets (nonasymptotic regime), other information measures appear as gene\-ra\-li\-za\-tions of the Shannon entropy. Indeed, such a family of information-theoretic measures include the paradigmatic cases of Tsallis~\cite{Tsallis1988} and R\'{e}nyi~\cite{Renyi1961} entropies.

Each of these developments is based on the idea that a physical process could be understood as an information processing protocol. In such tasks, dis\-tin\-gui\-shing classical probability distributions or quantum states plays a fundamental role. The relative entropy (RE)~\cite{umegaki1962}, also called divergence~\cite{6832827ErvenHarremos}, stand as a remarkable information-theoretic distin\-guishability metric, thus exhibiting distinct operational meanings in se\-ve\-ral fields. For ins\-tance, relative entropy quantifies the dissipated work in a driven evolution~\cite{Kawai2007}, the amount of entanglement~\cite{Vedral1997} and quantum coherence in a given state~\cite{Bu2017,0253-6102-67-6-631}. Moreover, it unveils the role of entropy production in thermal relaxation processes~\cite{PhysRevLett.123.110603,arXiv:2008.10764,PhysRevResearch.2.023377}, and also the asymmetry of a state or process~\cite{Marvian2014}. 

R\'{e}nyi relative entropy (RRE) determine an entire family of second laws of thermodynamics in the quantum regime~\cite{Brandao3275}, which also applies to black hole physics~\cite{Bernamonti2018}, and cut-off rate in the hypothesis testing theory~\cite{Csiszar1995}, and quantum Gaussian states~\cite{doi:10.1063/1.5007167}, to name only a few. Furthermore, RRE is linked to an entropic energy-time uncertainty relation for time-independent systems~\cite{PhysRevLett.122.100401}, also being related to the concept of multiple quantum coherences~\cite{PhysRevA.102.012429}.

Tsallis entropy is mainly considered in the field of nonextensive statistical mechanics~\cite{Tsallis2004}. However, important applications of this theory also appear in several other areas~\cite{doi:10.1063_1.1805729}. Interestingly, it has been shown that Tsallis relative entropies (TRE) define a {\it bona fide} quantum coherence quantifier~\cite{PhysRevA.93.032136}. Furthermore, TRE satisfies a class of bounds derived from Pinsker and Fannes-type inequalities~\cite{ARastegin_MathPhysAnalGeom_16_213}.

Here we consider the fundamental pro\-blem of boun\-ding the chan\-ge in the generalized relative entropies under an arbitrary unitary process. Specifically, we derive an upper bound on both asymmetric and symmetric versions of RRE and TRE, between the initial and final states. As an application of this result, we show that this upper bound implies an entirely family of quantum speed limits (QSLs). 

The importance of our results is twofold. First, it establishes a bound on entropic quantifiers that are employed in distinct fields, from quantum communication to bio\-logy~\cite{RevModPhys.74.197,Baez_2016}. In general, since the computation of relative entropies are usually difficult, our main result can directly be applied in all of these fields by providing bounds on central quantities. Second, our family of QSLs provide nonasymptotic bounds on the time evolution of quantum systems in the sense of the so-called single-shot information theory~\cite{arXiv:0512258}. Furthermore, due to the broadness application of RRE, it provides a bridge among the speed of quantum evolution, thermodynamics~\cite{PhysRevE.99.050101} and quantum resources, e.g., entanglement, coherence and asymmetry~\cite{PhysRevA.93.052331}. Importantly, since our results apply to TRE, it also provides a nonextensive version of the QSL, which can found several applications, both on fundamental and practical aspects~\cite{Abe2001}.

The paper is organized as follows. In Sec.~\ref{sec:sec0001} we briefly review the main properties of R\'{e}nyi and Tsallis relative entropies, which in turn can be recasted in terms of a generalized entropy. In Sec.~\ref{sec:sec0002}, we introduce the physical setting and present an upper bound on those generalized entropies. Next, in Sec.~\ref{sec:sec0003} we discuss an entire family of QSLs derived from the referred upper bound on R\'{e}nyi and Tsallis relative entropies. In Sec.~\ref{eq:singlequbit000000001} we illus\-tra\-te our findings via the prototypical case of the single-qubit state, thus presenting analytical results for the family of QSLs, and also discussing the tightness of the main bound on generalized entropies. Finally, in Sec.~\ref{sec:conclusions000xxx001} we close the paper discussing these results and comment on possible applications.


\section{Generalized relative entropies}
\label{sec:sec0001}

Let us start by de\-fi\-ning our physical system, which is described by a finite-dimensional Hilbert space $\mathcal{H}$, with $d = \text{dim}(\mathcal{H})$. In ge\-ne\-ral, the state of the system will be given by a density matrix $\rho \in \Omega$, where $\Omega = \{\rho \in \mathcal{H} \mid {\rho^{\dagger}} = \rho,~\rho\geq 0,~\text{Tr}(\rho) = 1\}$ defines the convex space of density operators. In this setting, given two states $\rho$, $\omega\in\Omega$, the R\'{e}nyi (RRE) and Tsallis (TRE) relative entropies are defined, respectively, as~\cite{10.1063.1.4838856}
\begin{equation}
\label{eq:renyi001}
\text{R}_{\alpha}(\rho\|\omega) = \frac{1}{\alpha - 1}\ln\left[g_{\alpha}(\rho,\omega) \right] ~,
\end{equation}
and
\begin{equation}
\label{eq:rtsallis001}
\text{H}_{\alpha}(\rho\|\omega) = \frac{1}{1 - \alpha}\left[ 1 - g_{\alpha}(\rho,\omega)\right] ~,
\end{equation}
where ${g_{\alpha}}(\rho,\omega) = \text{Tr}\left(\rho^{\alpha}\omega^{1 - \alpha}\right)$ is the $\alpha$-relative purity, also called Petz-R\'{e}nyi relative quasi-entropy~\cite{doi:10.1063/1.5007167}, with the parameter $\alpha \in (0,1)\cup(1,+\infty)$ labeling the fa\-mi\-ly of quantum relative entropies~\cite{DBLPjournals_qic_Audenaert01}. Equation~\eqref{eq:renyi001} is also called Petz-R\'{e}nyi relative entropy~\cite{PETZ198657}, standing as the first quantum extension of the classical RRE. Indeed, due to the noncommutativity of quantum states, the nonuniqueness of quantum information-theoretic quantifiers has triggered the search for a plethora of quantum entropies, e.g., sandwiched R\'{e}nyi relative entropy~\cite{10.1063.1.4838856, doi:10.1063/1.4838835, doi:10.1063/1.4838855, Datta_2014}, and $\alpha$-$z$-relative R\'{e}nyi entropy~\cite{10.1063.1.4906367}.

Importantly, relative purity satisfies the property ${g_{\alpha}}(\omega,\rho) = {g_{1 - \alpha}}(\rho,\omega)$, i.e., it is skew symmetric with respect to $\alpha$. In particular, when $\rho = \omega$ we have $g_{\alpha}(\rho,\rho) = 1$ for all $\alpha$, and thus one gets $\text{R}_{\alpha}(\rho\|\rho) = \text{H}_{\alpha}(\rho\|\rho) = 0$. Noteworthy, for $\alpha = 1/2$ one recovers the so-called quantum affinity, which is related to Hellinger angle~\cite{PhysRevA.69.032106}. In turn, Hellinger angle is associated to Wigner-Yanase skew information me\-tric and cha\-rac\-te\-rizes the length of the geodesic path connecting states $\rho$, $\omega\in\Omega$~\cite{Gibilisco_Isola_10.1063_1.1598279, Jencova_10.1063_1.1689000}.

In the following we summarize the main properties of RRE and TRE. A more complete presentation can be found in Ref.~\cite{Amigo2018}. Starting with RRE, the limit $\alpha \rightarrow 1$ recovers the well-known quantum relative entropy $\text{R}_1(\rho\|\omega) = \text{S}(\rho\|\omega) := \text{Tr}(\rho\ln\rho - \rho\ln\omega)$. For $\alpha = 0$, RRE reduces to the min-relative entropy $\text{R}_0(\rho\|\omega) = -\ln \text{Tr}(\Pi_{\rho}\, \omega)$, with $\Pi_{\rho}$ being the projector onto the support of the state $\rho$~\cite{4957651_Datta}. Noteworthy, for $0 \leq \alpha \leq 2$, RRE satisfies the data-processing ine\-qua\-li\-ty, i.e., $\text{R}_{\alpha}(\Lambda(\rho)\|\Lambda(\omega)) \leq \text{R}_{\alpha}(\rho\|\omega)$, thus being monotonic under any completely positive and trace pre\-ser\-ving map $\Lambda(\bullet)$~\cite{5730573}. This is a fundamental inequality not only within information theory, but also for physics (see, for instance, Ref.~\cite{Sagawa2012} where the second law of thermodynamics is obtained from such inequality). Moving to Tsallis relative entropy, it has been shown that, for $0 \leq \alpha < 1$, TRE is (i) nonnegative, i.e., $\text{H}_{\alpha} (\rho\|\omega) \geq 0$ for all $\rho,\omega \in \Omega$, with the equality holding if and only if $\rho = \omega$; (ii) jointly convex; (iii) nonaditive; and (iv) contractive under completely positive and trace preserving maps~\cite{ABE2003336, PhysRevA.68.032302, doi:10.1063_1.1805729}. Importantly, TRE also recovers the standard quantum relative entropy in the limit $\alpha \rightarrow 1$, i.e., $\text{H}_1(\rho\|\omega) = \text{S}(\rho\|\omega)$.

We shall stress that RRE and TRE are asymmetric with respect to states $\rho,\sigma\in\Omega$. However, a {\it bona fide} distance measure within the information geometry theory is usually symmetric. For instance, the so-called quantum Jensen-Shannon divergence, i.e., the square-root of symmetrized quantum relative entropy, was proved to be a metric on the space of density matrices~\cite{arXiv:1910.10447}. For the case at hand, the aforementioned entropies can be symmetrized as
\begin{equation}
\label{eq:idea0000xxxxx000x1}
\mathcal{O}_{\alpha}(\rho:\omega) := \mathcal{O}_{\alpha}(\rho\|\omega) + \mathcal{O}_{\alpha}(\omega\|\rho) ~,
\end{equation}
where index $\mathcal{O} \equiv \{ \text{R} , \text{H} \}$ labels RRE and TRE, respectively. We are now ready to present our main result.


\section{Bounds on generalized relative entropies}
\label{sec:sec0002}

The dynamics of our system is governed by a time-dependent Hamiltonian $H_t \in \mathcal{B}(\mathcal{H})$, with $\mathcal{B}(\mathcal{H})$ being the set of bounded operators acting on $\mathcal{H}$. In general, the Hamiltonian $H_t$ is not self-commuting at different times, i.e., $[H_s,H_t] \neq 0$ for $s \neq t$. The initial state $\rho_0 \in \Omega$ undergoes the unitary evolution $\rho_{t} = U_t \, \rho_{0} U^{\dagger}_t$, for $t \in [0,\tau]$, where ${U_t} = \mathcal{T} {e^{- i {\int_0^t} \, d s\, {H_s}}}$ is the time-ordered unitary evolution operator satisfying the equation $-i(d{U_t}/dt) = {H_t}{U_t}$. From now on, we will work in natural units, $\hbar = k_{B} = 1$.

Based on this physical setting, our goal is to provide a class of nontrivial upper bounds for RRE and TRE. In Appendix~\ref{sec:B1A} we have proved that, for $\alpha\in(0,1)$, RRE and TRE satisfy the ine\-qua\-lity
\begin{equation}
\label{eq:idea0000xxxxx010x1}
\mathcal{O}_\alpha({\rho_{\tau}}\|{\rho_0}) \leq \frac{\tau\, {\langle\!\langle {\mathcal{G}^{\mathcal{O}}_{\alpha}}(t) \rangle\!\rangle_{\tau}} }{|1 - \alpha|} ~,
\end{equation}
where $\langle\!\langle \bullet \rangle\!\rangle_{\tau} = \tau^{-1} \, {\int_0^{\tau}} \bullet \, d t$ stands for the time average, and
\begin{equation}
\label{eq:idea0000xxxxx010x10002}
{\mathcal{G}^{\mathcal{O}}_{\alpha}}(t) :=  {\Phi_{\alpha}^{\mathcal{O}}} \, {{\|{\rho_0^{1 - \alpha}}\|}_2} \, {{\| [ {H_t}, {\rho_0^{\alpha}} ]  \|}_2} ~,
\end{equation}
with $\| A \|_2 = \sqrt{\text{Tr}\, (A^{\dagger}A)}$ being the Schatten 2-norm. Here $\Phi_{\alpha}^{\mathcal{O}}$ is an auxiliary function which reads 
\begin{equation}
\label{eq:idea0000xxxxx010x100033333}
{\Phi_{\alpha}^{\mathcal{O}}} = 
\begin{cases}
{|1 + (1 - \alpha) \ln({{\lambda_{\text{min}}}({\rho_0})}) |^{-1}} ~,& \mbox{for ${\mathcal{O}} \equiv {\text{R}}$} \\
1 ~,& \mbox{for ${\mathcal{O}} \equiv {\text{H}}$} ~,
\end{cases}
\end{equation}
where ${\lambda_{\text{min}}}({\rho_0})$ sets the smallest eigenvalue of the input state $\rho_0$. Noteworthy, Eq.~\eqref{eq:idea0000xxxxx010x1} is the first main result of this article. Remarkably, the bound mostly depends on $\rho_0$ and $H_t$. Naturally, a similar bound can be obtained for the case in which the arrangement of states ${\rho_0}$ and ${\rho_{\tau}}$ in Eq.~\eqref{eq:idea0000xxxxx010x1} is swapped, which is given by (see Appendix~\ref{sec:B1A})
\begin{equation}
\label{eq:nonsym000222020202}
{\mathcal{O}_{\alpha}}({\rho_0} \| {\rho_{\tau}}) \leq \frac{\tau\, {\langle\!\langle {\mathcal{G}^{\mathcal{O}}_{1 - \alpha}}(t) \rangle\!\rangle_{\tau}} }{|1 - \alpha|}  ~.
\end{equation}
Furthermore, the corres\-pon\-ding ine\-qua\-lity for symmetrized forms of RRE and TRE is then obtained, roughly speaking, by com\-bi\-ning the two nonsymmetric upper bounds, and reads (see details in Appendix~\ref{sec:B1B})
\begin{equation}
\label{eq:sym000222020202}
{\mathcal{O}_{\alpha}}({\rho_{\tau}} : {\rho_0}) \leq \frac{\tau\, {\langle\!\langle {\mathcal{G}^{\mathcal{O}}_{\alpha}}(t) + {\mathcal{G}^{\mathcal{O}}_{1 - \alpha}}(t) \rangle\!\rangle_{\tau}} }{|1 - \alpha|} ~.
\end{equation}
Remarkably, such bounds do not depend on the time-ordered evolution operator $U_t$, neither on the evolved state of the system. Thus, for states $\rho_0$ and $\rho_{\tau}$ of a given closed quantum system, our results provide a route to estimate both RRE and TRE entropies which will depend mostly on the spectral properties of the initial state $\rho_0$ and the driving Hamiltonian.

Importantly, Eq.~\eqref{eq:idea0000xxxxx010x1} can be recasted in terms of the Fr\"{o}benius norm of the initial state of the system, while being a function of the time average of the Schatten 2-norm of the Hamiltonian. To see this, we first point out that for two arbitrary complex matrices $X$ and $Y$, it has been proved the Schatten 2-norm fulfills the ine\-qua\-lity ${\|{[X,Y]}\|_2} \leq \sqrt{2}\,  {\|{X}\|_2}{\|{Y}\|_2}$~\cite{BOTTCHER20081864,AUDENAERT20101126,FONG20111193}. Hence, from Eq.~\eqref{eq:idea0000xxxxx010x10002} one readily obtains the upper bound ${\mathcal{G}^{\mathcal{O}}_{\alpha}}(t) \leq  \sqrt{2}\, {{\|{\rho_0^{1 - \alpha}}\|}_2} \, {\|{\rho_0^{\alpha}}\|_2} \, {\|{H_t}\|_2}$. Therefore, the bound in Eq.~\eqref{eq:idea0000xxxxx010x1} can be recasted as
\begin{equation}
\label{eq:L0000000010}
\mathcal{O}_\alpha({\rho_{\tau}}\|{\rho_0}) \leq \frac{\sqrt{2}\, \tau \, {\Phi_{\alpha}^{\mathcal{O}}} \, {{\|{\rho_0^{1 - \alpha}}\|}_2}\, {\|{\rho_0^{\alpha}}\|_2} \,  {\langle\!\langle\,  {\|{H_t}\|_2} \rangle\!\rangle_{\tau}}}{|1 - \alpha|}   ~.
\end{equation}
In particular, note that ${\langle\!\langle\,  {\|{H_t}\|_2} \rangle\!\rangle_{\tau}} = {\| H \|_2}$ for the case in which the Hamiltonian is time independent, i.e., $H_t \equiv H$. Overall, the bound in Eq.~\eqref{eq:L0000000010} does requires minimal information about the system, e.g., its initial state $\rho_0$ and the e\-ner\-gy levels of the Hamiltonian $H_t$. Indeed, the latter comes from the fact that, given the spectral decomposition ${H_t} = {\sum_{j = 1}^d}\, {\epsilon_j}(t)|{\phi_j}(t)\rangle\langle{\phi_j}(t)|$, with $d = \dim\mathcal{H}$ the dimension of the Hilbert space, where the eigenvalues $\{  {\epsilon_j}(t) \}_{j = 1,\ldots,d}$ and eigenstates $\{ |{\phi_j}(t)\rangle \}_{j = 1,\ldots,d}$ are time dependent, and thus one obtains ${\|{H_t}\|_2^2} = {{\sum_{j = 1}^d}\, {{\epsilon_j}(t)^2}}$.

Finally, regardless of the simplicity and usefulness of the bound in Eq.~\eqref{eq:L0000000010}, we shall point out the original bound in Eq.~\eqref{eq:idea0000xxxxx010x1} might stand as the general and tighter one. Before discussing the physical significance of this bound, we will make use of it to obtain a family of QSLs.


\section{Quantum speed limits}
\label{sec:sec0003}

The quantum speed limit (QSL) signals the minimum time of evolution between two quantum states undergoing an arbitrary dynamics. Indeed, Mandelstam and Tamm addressed this question around 75 years ago for closed quantum systems, thus showing the QSL time for orthogonal states is given by ${\tau_{\text{QSL}}} = \hbar \pi/(2\Delta{E})$, where ${(\Delta{E})^2} = \langle{\psi_0}|{H^2}|{\psi_0}\rangle - {\langle{\psi_0}|{H}|{\psi_0}\rangle^2}$ stands for the variance of the Hamiltonian with respect to the initial state $|{\psi_0}\rangle$ of the system~\cite{1945_JPhysURSS_9_249}. Later on, Margolus and Levitin derived the QSL time ${\tau_{\text{QSL}}} = \hbar \pi/(2{\langle{\psi_0}|{H}|{\psi_0}\rangle})$ for the same physical se\-tting, i.e., closed quantum systems evolving between orthogonal states~\cite{1992_PhysicaD_120_188}. Importantly, both lower bounds can be combined accordingly onto a tighter one as $\tau \geq \text{max}\{ \hbar \pi/(2\Delta{E}), \hbar \pi/(2{\langle{\psi_0}|{H}|{\psi_0}\rangle}) \}$~\cite{PhysRevLett.103.160502}. In the last decade, several bounds were introduced in the literature covering the QSL time for different physical settings, e.g., addressing pure and mixed states, for closed and open quantum systems~\cite{2013_PhysRevLett_110_050402,PhysRevLett.110.050403,2013_PhysRevLett_111_010402,Deffner_2017,PhysRevLett.120.070401,PhysRevLett.120.070402,PhysRevResearch.2.023299,PhysRevA.102.042606,PhysRevResearch.2.032020,arXiv:2006.14523,arXiv:2007.15019,arXiv:2009.02231,arXiv:2011.05232}.

Here we will present a family of QSLs by time a\-ve\-ra\-ging the right-hand side of Eq.~\eqref{eq:idea0000xxxxx010x1}, thus followed by a rearrangement of the resulting inequality. Indeed, the time $\tau$ required for an arbitrary unitary evolution driving a closed quantum system from $\rho_{0}$ to ${\rho_{\tau}}$ is lower bounded as 
\begin{equation}
\label{eq:QSL_xxx_0001}
\tau \ge \tau_\alpha^{\mathcal{O}} := \max\{\tau^{\mathcal{O}}_{\alpha}(\rho_\tau \| \rho_0), \tau^{\mathcal{O}}_{\alpha} (\rho_0 \| \rho_\tau), \tau^\mathcal{O}_{\alpha} (\rho_0:\rho_\tau)\} ~, 
\end{equation}
where
\begin{equation}
\label{eq:QSL_asymmetric}
\tau^{\mathcal{O}}_{\alpha}({\rho_{\tau}}\|{\rho_0}) := \frac{| 1 - \alpha | \, \mathcal{O}_{\alpha}({\rho_{\tau}}\|{\rho_0})}{ \langle\!\langle {\mathcal{G}^{\mathcal{O}}_{\alpha}}(t) \rangle\!\rangle_{\tau}} ~,
\end{equation}
and
\begin{equation}
\label{eq:QSL_asymmetric2}
\tau^{\mathcal{O}}_{\alpha}({\rho_0}\|{\rho_{\tau}}) := \frac{| 1 - \alpha | \, \mathcal{O}_{\alpha}({\rho_0}\|{\rho_{\tau}})}{ \langle\!\langle {\mathcal{G}^{\mathcal{O}}_{1 - \alpha}}(t) \rangle\!\rangle_{\tau}} ~,
\end{equation}
while the QSL time due to symmetrized relative entropies reads [see Appendix~\ref{sec:B1B}]
\begin{equation}
\label{eq:QSL_symmetric}
{\tau_{\alpha}^{\mathcal{O}}}(\rho_0:\rho_\tau):= \frac{| 1 - \alpha | \, \mathcal{O}_{\alpha}(\rho_0 : \rho_\tau)}{ \langle\!\langle \, {\mathcal{G}^{\mathcal{O}}_{\alpha}}(t) + {\mathcal{G}^{\mathcal{O}}_{1 - \alpha}}(t) \, \rangle\!\rangle_{\tau}} ~.
\end{equation}

Equation~\eqref{eq:QSL_xxx_0001} stands as our second main result, thus establishing a family of entropic QSLs, i.e., RRE and TRE provide lower bounds on the time of evolution between the initial and final states of the quantum system. Recently, a related family of QSLs, based on the relative entropy, were derived bounding the time it takes to generate or consume a given quantum resource such as entanglement, asymmetry, and athermality~\cite{arXiv:2004.03078}. These bounds, dubbed as \emph{resource speed limits} (RSL) were shown to be tighter than QSLs in several instances. However, as RSLs are constructed using the standard relative entropy ($\alpha \rightarrow 1$), thus being only mea\-ning\-ful in the asymptotic limit. RSLs for single shot scenarios requires working with R\'{e}nyi relative entropies. Here, we have taken the first step in this direction.

We shall stress that the previous discussion is valid for $\alpha \in (0,1)$. In the following, we will discuss the li\-mi\-ting cases $\alpha\rightarrow 1$ and $\alpha\rightarrow 0$, which crucially reduce to the standard relative entropy and the so-called min-relative entropy, respectively. Importantly, these results cannot be simply obtained from the results above. While the case $\alpha \rightarrow 1$ is clearly delicate from the definition of RRE and TRE given in Eqs.~\eqref{eq:renyi001} and~\eqref{eq:rtsallis001}, respectively, the case $\alpha \rightarrow 0$ must be ca\-re\-fully con\-si\-de\-red sin\-ce sim\-ply ta\-king $\alpha=0$ in Eq.~\eqref{eq:idea0000xxxxx010x1} would provide us a trivial bound, independently of the initial state and the dynamics.


\subsection{Limiting case of $\alpha \to 1$}

Let us start by considering the limit $\alpha \rightarrow 1$, in which both RRE and TRE recover the quantum relative entropy, i.e., ${\lim_{\alpha \rightarrow 1}}{\mathcal{O}_{\alpha}}({\varrho}\|{\omega}) = {\text{S}}({\varrho}\|{\omega})$. In this case, it can be proved that the following upper bound applies (see details in Appendix~\ref{app:otocAPD})
\begin{equation}
\label{eq:appc0000033xx1}
{\text{S}}({\rho_{\tau}}\| {\rho_0}) \leq \tau \, {\| \ln{\rho_0} \|_2}\, {\langle\!\langle \, {\| [ {H_t} , {\rho_t} ] \|_2} \rangle\!\rangle_{\tau}} ~.
\end{equation}
In this case, the corresponding QSL family reads 
\begin{equation}
\tau \ge  \tau_1^{\text{RE}} := \max\{\tau_{1}^{\text{RE}}(\rho_\tau \| \rho_0), \tau_{1}^{\text{RE}}(\rho_0 \| \rho_\tau), \tau_{1}^{\text{RE}}(\rho_0 : \rho_\tau) \}
\end{equation}
where
\begin{equation}
\label{eq:QSL_RE0000}
\tau^{\text{RE}}_{1}(A\|B):= \frac{\text{S}(A\| B )}{ {\| \ln{\rho_0} \|_2} \, {\langle\!\langle \, {\| [ {H_t} , {A} ] \|_2} \rangle\!\rangle_{\tau}}} ~,
\end{equation}
and
\begin{equation}
\label{eq:QSL_RE0000v00002}
 \tau^{\text{RE}}_{1}(\rho_0 : \rho_\tau) := \frac{{\text{S}}({\rho_{\tau}}\| {\rho_0}) + {\text{S}}({\rho_0}\| {\rho_{\tau}})}{ {\| \ln{\rho_0} \|_2}\, {\langle\!\langle \, {\| [ {H_t} , {\rho_t} ] \|_2} + {\| [ {H_t} , {\rho_0} ] \|_2} \rangle\!\rangle_{\tau}}} ~.
\end{equation}

In particular, when the Hamiltonian is time-independent, i.e., ${H_t} \equiv H$, one obtains ${\| [ {H} , {\rho_t} ] \|_2} = {\| [ {H} , {\rho_0} ] \|_2} = 2 \sqrt{\mathcal{I}_L(\rho_0, H)}$, where we have used the fact that Scha\-tten 2-norm is unitarily invariant. Here ${\mathcal{I}_L}(\rho_0, H) = (1/4)\, {\| [ H , {\rho_0} ] \|_2^2} = - (1/4) \, \text{Tr}([{\rho_0},H]^2)$ define a time-independent quantum coherence quantifier which sets a lower bound on Wigner-Yanase skew information~\cite{PhysRevLett.113.170401, Wigner1963}. Now, since ${\mathcal{I}_L}(\rho_0,H) \leq (\Delta H)^2$, where $(\Delta H)^2 = \text{Tr}({\rho_0} {H^2}) - [\text{Tr}({\rho_0}H)]^2$ is the squared deviation of the Hamiltonian, thus Eq.~\eqref{eq:QSL_RE0000} implies the lower bounds ${\tau_{1}^{\text{RE}}}({\rho_0}\|{\rho_{\tau}}) \geq  {\text{S}({\rho_0}\| {\rho_{\tau}} )} / ({ 2 \, \Delta{H} \, \| \ln{\rho_0} \|_2})$, and ${\tau_{1}^{\text{RE}}}({\rho_{\tau}}\|{\rho_{0}}) \geq  {\text{S}(\rho_{\tau}\| {\rho_0} )} / ({ 2 \, \Delta{H} \, \| \ln{\rho_0} \|_2})$.


\subsection{Limiting case of $\alpha \to 0$}

Considering now the case $\alpha \rightarrow 0$, in Appendix~\ref{app:otocminEntropy} we have proved that R\'{e}nyi min-relative entropy is upper bounded as
\begin{equation}
\label{eq:min000xxx000x1201}
|{\text{R}_0}({\rho_{\tau}}\|{\rho_0})| \leq \tau \, {\langle\!\langle {\mathcal{Q}_0^t}({\rho_0},{\Pi_{\rho_0}}) \rangle\!\rangle_{\tau}} ~,
\end{equation}
with 
\begin{equation}
\label{eq:min000xxx000x1201aaaaa}
{\mathcal{Q}_0^t}(A,B) := \frac{ {\|A\|_2} \, {\| [{U_t^{\dagger}}{H_t}{U_t},B] \|_2} }{ |\text{Tr}(A{U_t}B{U_t^{\dagger}}) |} ~.
\end{equation}
Here ${\Pi_{\rho_0}}$ is the projector onto the support of the initial state $\rho_0$. From Eq.~\eqref{eq:min000xxx000x1201}, we can derive the QSL time as 
\begin{equation}
\tau \ge {\tau_0^{\text{R}}} := \max\{\tau_{0}^{\text{R}}(\rho_\tau \| \rho_0), {\tau_{0}^{\text{R}}}(\rho_0 \| \rho_\tau), \tau_{0}^{\text{R}}({\rho_0} : {\rho_{\tau}}) \} ~, 
\end{equation}
where 
\begin{equation}
\label{eq:min000xxx000x1200002a}
{\tau_{0}^{\text{R}}}({\rho_{\tau}} \| {\rho_0}) := \frac{|{\text{R}_0}({\rho_{\tau}} \| {\rho_0})|}{\langle\!\langle {\mathcal{Q}_0^t}({\rho_0},{\Pi_{\rho_0}}) \rangle\!\rangle_{\tau}} ~,
\end{equation}
and
\begin{equation}
\label{eq:min000xxx000x1200002b}
{\tau_{0}^{\text{R}}}({\rho_0} \| {\rho_{\tau}}) := \frac{|{\text{R}_0}({\rho_0} \| {\rho_{\tau}})|}{\langle\!\langle {\mathcal{Q}_0^t}({\Pi_{\rho_0}}, {\rho_0}) \rangle\!\rangle_{\tau}} ~,
\end{equation}
while the QSL related to the symmetric min-entropy is given by
\begin{equation}
\label{eq:min000xxx000x1200002c}
\tau_{0}^{\text{R}}({\rho_0} : {\rho_{\tau}}) := \frac{|{\text{R}_0}({\rho_{\tau}} \| {\rho_0}) + {\text{R}_0}({\rho_0} \| {\rho_{\tau}})|}{\langle\!\langle {\mathcal{Q}_0^t}({\rho_0},{\Pi_{\rho_0}}) + {\mathcal{Q}_0^t}({\Pi_{\rho_0}} , {\rho_0}) \rangle\!\rangle_{\tau}} ~.
\end{equation}

Noteworthy, the ``speed'' contribution $\mathcal{Q}_0^t$ is closely related to the QSL derived with respect to Euclidean distance in the Bloch sphere~\cite{PhysRevLett.120.060409, Campaioli2019tightrobust}. Importantly, when the density matrix $\rho_0$ has full-rank, i.e., $\text{dim}({\rho_0}) = \text{supp}({\rho_0})$, thus R\'{e}nyi min-relative entropy vanishes and implies that ${\tau_{\text{R}}^{(0)}} = 0$. To see this, let ${\rho_0} = {\sum_{\ell = 1}^d}\, {p_{\ell}} |{\psi_{\ell}}\rangle\langle{\psi_{\ell}}|$ be the spectral decomposition of the input state, with $d = \text{dim}(\mathcal{H})$. Note the projector $\Pi_{\rho_0}$ onto the support of the full-rank state $\rho_0$ is equal to the identity, ${\Pi_{\rho_0}} = {\sum_{\ell : {p_{\ell}} \neq 0}} \, |{\psi_{\ell}}\rangle\langle{\psi_{\ell}}| =  \mathbb{I}$. Hence, it is straightforward to verify the symmetric min-entropy is identically zero because ${\text{R}_0}({\rho_{\tau}}\|{\rho_0}) = - \ln[\text{Tr}({U_{\tau}^{\dagger}}{\rho_0}{U_{\tau}})] = 0$ and ${\text{R}_0}({\rho_0}\|{\rho_{\tau}}) = - \ln[\text{Tr}({U_{\tau}}{\rho_0}{U_{\tau}^{\dagger}})] = 0$. Furthermore, from Eq.~\eqref{eq:min000xxx000x1201aaaaa} one readily gets ${\mathcal{Q}_0^t}({\rho_0},{\Pi_{\rho_0}}) = 0$, while the functional ${\mathcal{Q}_0^t}({\Pi_{\rho_0}},{\rho_0}) = \sqrt{d}\, {\| [{U_t^{\dagger}}{H_t}{U_t},{\rho_0}] \|_2}$ remains finite for $t \in [0,\tau]$.


\section{Example}
\label{eq:singlequbit000000001}

We now provide an analytical example in order to make clearer the physical implications of our results. Let us consider a single-qubit state, whose Bloch sphere representation is written as $\rho_0 = (1/2)(\mathbb{I} + \vec{r}\cdot\vec{\sigma})$, where $\vec{\sigma} = (\sigma_x,\sigma_y,\sigma_z)$ is the vector of Pauli matrices, $\vec{r} = r\,\hat{r}$ is the Bloch vector, with $\hat{r} = \{\sin\theta\cos\phi,\sin\theta\sin\phi,\cos\theta\}$, $0 < r < 1$, $\theta \in [0,\pi]$ and $\phi \in [0,2\pi]$, while $\mathbb{I}$ is the $2\times 2$ identity matrix. Particularly, for $0 < \alpha < 1$, the operator $\rho_0^{\alpha}$ respective to the initial single-qubit state is written as
\begin{equation}
\label{eq:app000xxx000xxx0001}
{\rho_0^{\alpha}} = \frac{1}{2}\left[ {\xi_{\alpha}^+}\, \mathbb{I} + {\xi_{\alpha}^-}(\hat{r}\cdot\vec{\sigma}) \right] ~,
\end{equation}
where
\begin{equation}
\label{eq:app000xxx000xxx0002}
{\xi_{\alpha}^{\pm}} = {2^{-\alpha}}\left[{(1 + r)^{\alpha}} \pm {(1 - r)^{\alpha}}\right] ~.
\end{equation}
\begin{table}[!t]
\caption{Theoretical-information quantifiers related the single qubit state ${\rho_0} = (1/2)(\mathbb{I} + \vec{r}\cdot\vec{\sigma})$, evolving under the Hamiltonian ${H_t} = \varpi \, \mathbb{I} + {\hat{n}_t}\cdot\vec{\sigma}$. Note that ${\hat{\mu}_t} := ~ {\hat{n}_t} - \sin(2|{\vec{u}_t}|)\, ({\hat{u}_t}\times{\hat{n}_t}) + 2\,{\sin^2}(|{\vec{u}_t}|)\left[  ({\hat{u}_t}\cdot{\hat{n}_t}){\hat{u}_t} - {\hat{n}_t}\right]$, with ${\hat{u}_t} = {\vec{u}_t}/|{\vec{u}_t}|$, and ${\vec{u}_t} = {\int_0^t} \, ds \, {\hat{n}_s}$. If the Hamiltonian is time-independent, i.e., $\hat{n}_t = \hat{n}$, one must apply the changes $\hat{u}_t \rightarrow \hat{n}$, $|\vec{u}_t| \rightarrow t$, and ${\hat{\mu}_t} \rightarrow \hat{n}$ into the listed quantities.}
\begin{center}
\begin{tabular}{cc}
\hline\hline
Quantifier & Analytical value  \\
\hline
${\| [ {H_t}, {\rho_0^{\alpha}} ] \|_2}$ & ${\xi_{\alpha}^-} \sqrt{2\left(1 - {({\hat{n}_t}\cdot\hat{r})^2}\right)}$ \\
${\| [  {H_t} , {\rho_0} ] \|_2}$ & $r \sqrt{2\left(1 - {({\hat{n}_t}\cdot\hat{r})^2}\right)}$ \\
$ {\| [  {H_t} , {\rho_t} ] \|_2}$ & $r \sqrt{2\left(1 - {({\hat{\mu}_t}\cdot\hat{r})^2}\right)}$ \\ 
${\text{S}}({\rho_t}\| {\rho_0})$ & $r \ln\left(\frac{1 + r}{1 - r}\right)\left(1 - {({\hat{u}_t}\cdot\hat{r})^2}\right){\sin^2}( |{\vec{u}_t}| ) $ \\
${\| {\ln{\rho_0}} \|_2}$  & $\sqrt{{\ln^2}\left(\frac{1 - r}{2}\right) + {\ln^2}\left(\frac{1 + r}{2}\right)}$ \\
${{\|{\rho_0^{\alpha}}\|}_2}$ & $\sqrt{{\xi_{2\alpha}^+}} $ \\
\hline\hline
\end{tabular}
\label{tab:table001}
\end{center}
\end{table}

The dynamics of the system is governed by the time-dependent Hamiltonian $H_t = \varpi\, \mathbb{I} + \hat{n}_t\cdot\vec{\sigma}$, where $\hat{n}_t = \{ n_t^x,n_t^y,n_t^z\}$ is a time-dependent unit vector, $|\hat{n}_t| = 1$, and $\varpi\in\mathbb{R}$. In this case, the time ordered evolution operator becomes ${U_t} =  {e^{-it\varpi}}\left[\cos(|{\vec{u}_t}|)\, \mathbb{I} - i \sin(|{\vec{u}_t}|)\, ({\hat{u}_t}\cdot\vec{\sigma})\right]$, where ${\hat{u}_t} = {\vec{u}_t}/|{\vec{u}_t}|$ is a unit vector, with ${\vec{u}_t} := {\int_0^t} \, ds\, {\hat{n}_s}$. In particular, if ${H_t} \equiv H$ is time independent, i.e., ${\hat{n}_t} = \hat{n}$ is a constant unit vector, we directly obtain ${\hat{u}_t} = \hat{n}$. Next, by performing a lengthy but straightforward calculation, one may verify the evolved single-qubit state becomes
\begin{equation}
\label{eq:app000xxx000xxx0005}
{\rho_t^{\alpha}} = {U_t}\, {\rho_0^{\alpha}} {U_t^{\dagger}} = \frac{1}{2}\left[{\xi_{\alpha}^+}\, \mathbb{I} + {\xi_{\alpha}^-} ({\hat{\nu}_t}\cdot\vec{\sigma}) \right] ~,
\end{equation}
with the unit vector ${\hat{\nu}_t} := \hat{r}   +  \sin(2 |{\vec{u}_t}| ) \, (\hat{u}_{t}\times\hat{r}) + 2\,\sin^2(|{\vec{u}_t}| )\left[  ({\hat{u}_t}\cdot\hat{r}){\hat{u}_t} - \hat{r}\right] $. Hence, by considering the range $0 < \alpha < 1$, from Eqs.~\eqref{eq:app000xxx000xxx0001} and~\eqref{eq:app000xxx000xxx0005} we thus have that $\alpha$-relative purity reads
\begin{equation}
\label{eq:exampleQSL_xxx_005cc}
{g_{\alpha}}({\rho_t},{\rho_0}) = 1 - {\xi_{\alpha}^-}{\xi_{1 - \alpha}^-} \left(1 - {({\hat{u}_t}\cdot\hat{r})^2}\right) {\sin^2}(|{\vec{u}_t}| ) ~.
\end{equation}

Interestingly, since relative purity is skew symmetric over the index $\alpha$, Eq.~\eqref{eq:exampleQSL_xxx_005cc} implies that ${g_{\alpha}}({\rho_t},{\rho_0}) = {g_{1 - \alpha}}({\rho_t},{\rho_0}) = {g_{\alpha}}({\rho_0},{\rho_t})$. In turn, both RRE and TRE will satisfy the constraint ${\mathcal{O}_{\alpha}}({\rho_t}\|{\rho_0}) = {\mathcal{O}_{\alpha}}({\rho_0}\|{\rho_t})$ for single-qubit states. Furthermore, from Eq.~\eqref{eq:exampleQSL_xxx_005cc}, note that $\alpha$-relative purity is equal to $1$ for $|{\vec{u}_t}| = n\pi$, with $n\in\mathbb{Z}$ and $t\in[0,\tau]$. Furthermore, ${g_{\alpha}}({\rho_t},{\rho_0}) = 1$ if vectors ${\hat{u}_t}$ and $\hat{r}$ are parallel. Conversely, $\alpha$-relative purity becomes ${g_{\alpha}}({\rho_t},{\rho_0}) = 1 - {\xi_{\alpha}^-}{\xi_{1 - \alpha}^-} {\sin^2}(|{\vec{u}_t}|)$ if vectors ${\hat{u}_t}$ and $\hat{r}$ are orthogonal. In Table~\ref{tab:table001} we summarize the quantities required to evaluate the QSL bounds ${\tau_{\alpha}^{\mathcal{O}}}$, $\tau^{\text{RE}}_{1}$, and $\tau^{\text{R}}_{0}$, for the case of a single-qubit state.

\begin{figure}[t]
\begin{center}
\includegraphics[scale=0.8]{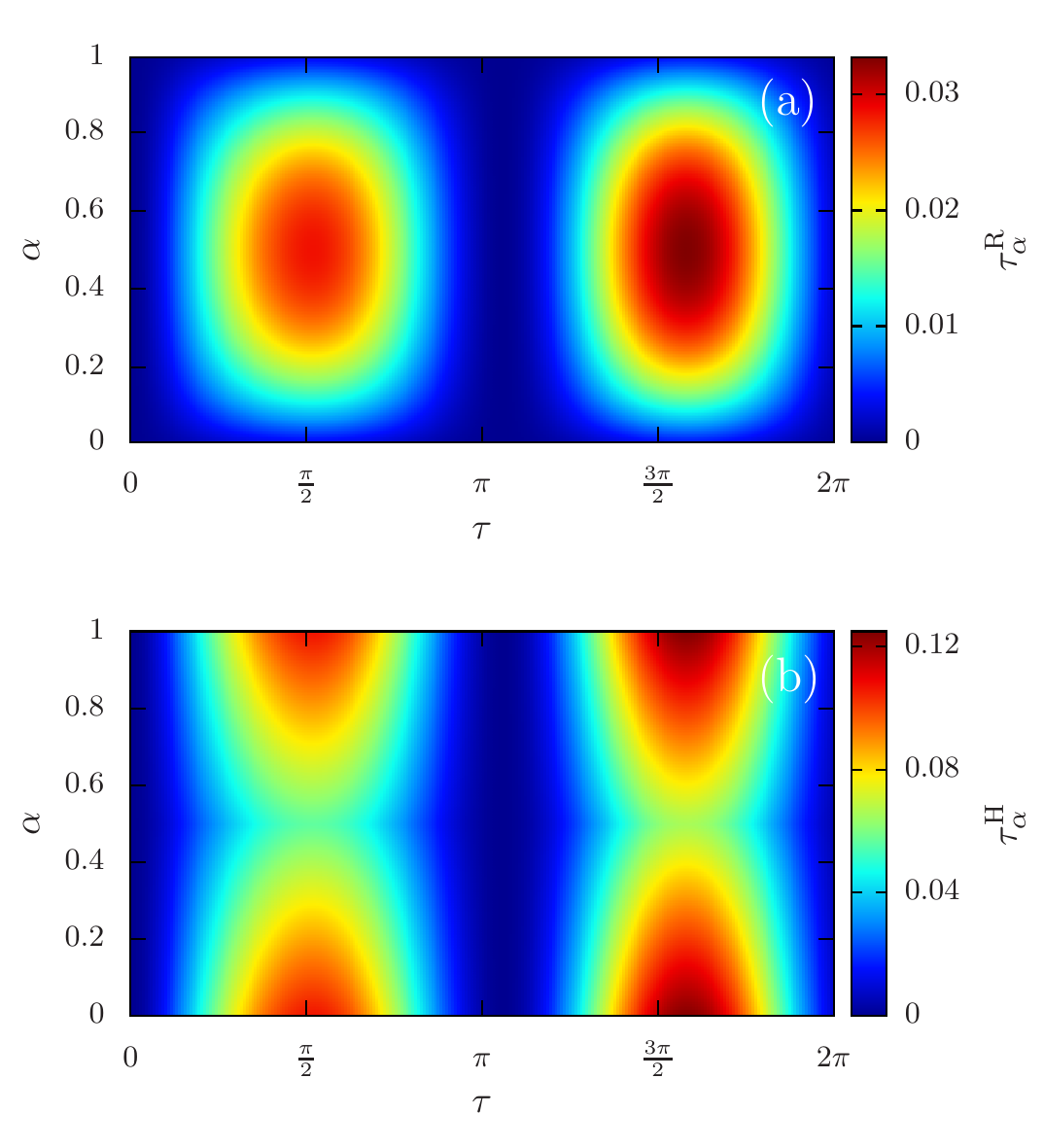}
\caption{(Color online) Density plot of QSL (a) ${\tau_{\alpha}^{\text{R}}}$, and (b) ${\tau_{\alpha}^{\text{H}}}$, as a function of time $\tau$ and $\alpha$, for the time-dependent Hamiltonian ${H_t} = {\hat{n}_t}\cdot\vec{\sigma}$, with $\hat{n}_t = \gamma^{-1}\{ \Delta,0,vt \}$, and $\gamma := \sqrt{{\Delta^2} + {(vt)^2}}$. Here the initial state is defined by $\{r,\theta,\phi\} = \{1/4,\pi/4,\pi/4\}$, and the ratio $\Delta/v = 0.5$.}
\label{fig:figure000xxx00xxxx000}
\end{center}
\end{figure}

For simplicity, let us focus on a Hamiltonian with $\varpi = 0$ and $\hat{n}_t = \gamma^{-1}\{ \Delta,0,vt \}$, with $\gamma := \sqrt{{\Delta^2} + {(vt)^2}}$, where $v$ stands as a ``level velocity'' of the energies of the system, and $\Delta$ is the level splitting~\cite{PhysRevA.96.020301}. Figure~\ref{fig:figure000xxx00xxxx000} shows the QSL $\tau_{\alpha}^{\text{R}}$ and $\tau_{\alpha}^{\text{H}}$ as function of time $\tau$ and the parameter $\alpha$, for the initial single-qubit state with $\{r,\theta,\phi\} = \{1/4,\pi/4,\pi/4\}$, also setting the ratio $\Delta/v = 0.5$. In Appendix~\ref{app:examplesAPD} we provide a complementary numerical study, thus exploiting in details the qualitative behavior depicted in Fig.~\ref{fig:figure000xxx00xxxx000} for RRE and TRE.

\begin{figure*}[t]
\begin{center}
\includegraphics[scale=0.625]{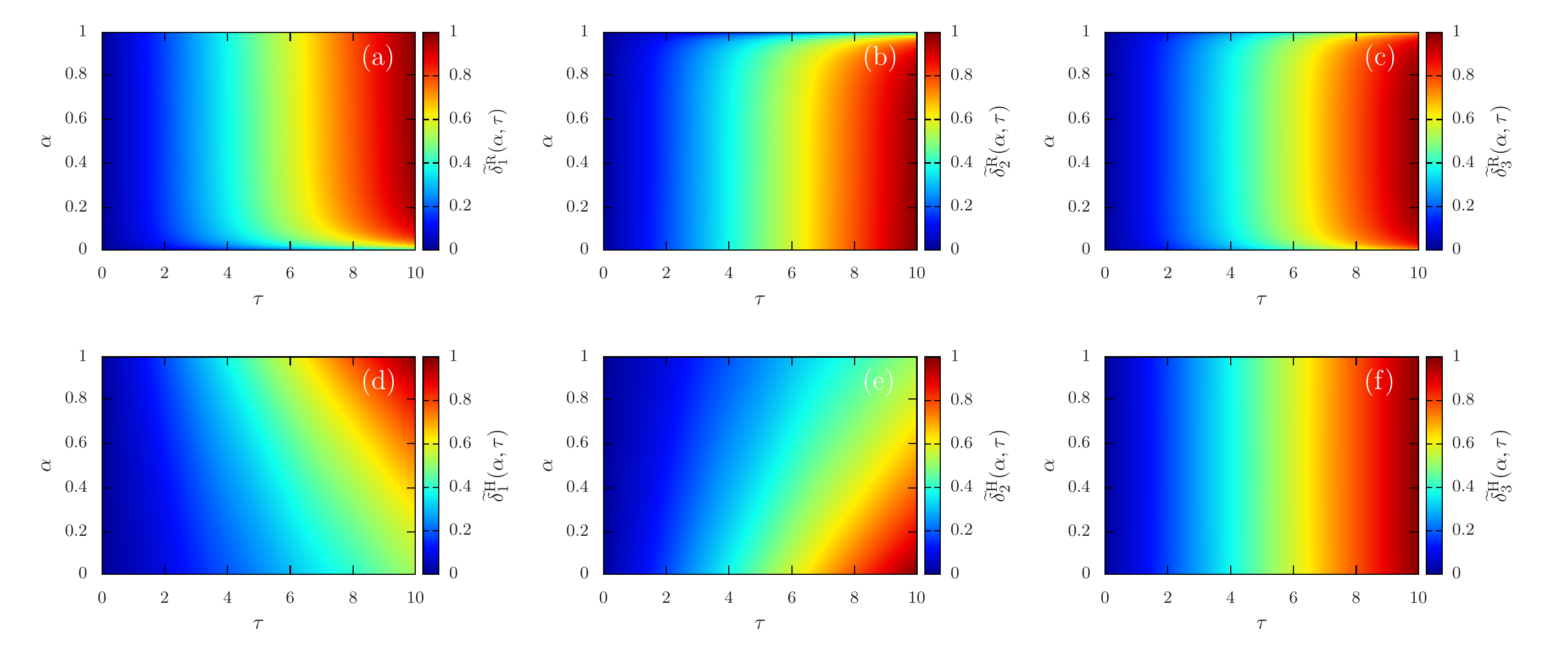}
\caption{(Color online) Density plot of the normalized figure of merit ${\widetilde{\delta}^{\mathcal{O}}_l}(\alpha,\tau) = {{\delta}^{\mathcal{O}}_l}(\alpha,\tau)/\text{max}\left[{{\delta}^{\mathcal{O}}_l}(\alpha,\tau)\right]$, with $l = \{1,2,3\}$ [see Eqs.~\eqref{eq:idea0000xxxxx010x1aaaaaaaa01},~\eqref{eq:idea0000xxxxx010x1aaaaaaaa02}, and~\eqref{eq:idea0000xxxxx010x1aaaaaaaa03}], for R\'{e}nyi relative entropy (a)~${\widetilde{\delta}^{\text{R}}_1}(\alpha,\tau)$, (b)~${\widetilde{\delta}^{\text{R}}_2}(\alpha,\tau)$, and (c)~${\widetilde{\delta}^{\text{R}}_3}(\alpha,\tau)$; and for Tsallis relative entropy (d)~${\widetilde{\delta}^{\text{H}}_1}(\alpha,\tau)$, (e)~${\widetilde{\delta}^{\text{H}}_2}(\alpha,\tau)$, and (f)~${\widetilde{\delta}^{\text{H}}_3}(\alpha,\tau)$. Here we set the initial single-qubit state with $\{r,\theta,\phi\} = \{1/4,\pi/4,\pi/4\}$, and also the ratio $\Delta/v = 0.5$.}
\label{fig:figure000xxx00xxxx0002222}
\end{center}
\end{figure*}
Finally, we will investigate the tightness of our bounds for nonsymmetric and symmetric relative entropies. Overall, the tightness of the bounds is related with the tightness of the QSL. To see this, we introduce the following figures of merit for nonsymmetric RRE and TRE
\begin{equation}
\label{eq:idea0000xxxxx010x1aaaaaaaa01}
{\delta^{\mathcal{O}}_1}(\alpha,\tau) := {\tau\,  {\langle\!\langle \, {\mathcal{G}^{\mathcal{O}}_{\alpha}}(t) \rangle\!\rangle_{\tau}} } - {|1 - \alpha|}{\mathcal{O}_\alpha}({\rho_{\tau}}\|{\rho_0}) ~,
\end{equation}
and 
\begin{equation}
\label{eq:idea0000xxxxx010x1aaaaaaaa02}
{\delta^{\mathcal{O}}_2}(\alpha,\tau) := {\tau\,  {\langle\!\langle \, {\mathcal{G}^{\mathcal{O}}_{1 - \alpha}}(t) \rangle\!\rangle_{\tau}} } - {|1 - \alpha|}{\mathcal{O}_\alpha}({\rho_0}\|{\rho_{\tau}}) ~,
\end{equation}
while for the symmetrized relative entropies one gets
\begin{equation}
\label{eq:idea0000xxxxx010x1aaaaaaaa03}
{\delta^{\mathcal{O}}_3}(\alpha,\tau) := {\delta^{\mathcal{O}}_1}(\alpha,\tau) + {\delta^{\mathcal{O}}_2}(\alpha,\tau) ~,
\end{equation}
with $\mathcal{O} \equiv \{ \text{R} , \text{H} \}$. We note that Eqs.~\eqref{eq:idea0000xxxxx010x1aaaaaaaa01},~\eqref{eq:idea0000xxxxx010x1aaaaaaaa02}, and~\eqref{eq:idea0000xxxxx010x1aaaaaaaa03} will quantify how much the bounds deviate from the actual value of RRE and TRE entropies, for both nonsymmetric and symmetric cases. Hereafter, we will set the initial single-qubit state parameterized as $\{r,\theta,\phi\} = \{1/4,\pi/4,\pi/4\}$, and also fixing the ratio $\Delta/v = 0.5$. Figure~\ref{fig:figure000xxx00xxxx0002222} shows the plot of the normalized quantity ${\widetilde{\delta}^{\mathcal{O}}_l}(\alpha,\tau) = {{\delta}^{\mathcal{O}}_l}(\alpha,\tau)/\text{max}\left[{{\delta}^{\mathcal{O}}_l}(\alpha,\tau)\right]$, with $l = \{1,2,3\}$. On the one hand, Figs.~\ref{fig:figure000xxx00xxxx0002222}(a) and~\ref{fig:figure000xxx00xxxx0002222}(b) show that, for $0 \leq \alpha \leq 1$, both quantities ${\widetilde{\delta}^{\text{R}}_1}(\alpha,\tau)$ and ${\widetilde{\delta}^{\text{R}}_2}(\alpha,\tau)$ approaches zero for short times ($0 \lesssim \tau \lesssim 2$). On the other hand, Figs.~\ref{fig:figure000xxx00xxxx0002222}(d) and~\ref{fig:figure000xxx00xxxx0002222}(e) indicate that, as the time increases, the figure of merit ${\widetilde{\delta}^{\text{H}}_1}(\alpha,\tau)$ will remain close to zero as long as $\alpha \lesssim 0.4$, while ${\widetilde{\delta}^{\text{H}}_2}(\alpha,\tau)$ approaches a small value for $\alpha \gtrsim 0.4$ during the range $0\lesssim \tau \lesssim 4$. Finally, Figs.~\ref{fig:figure000xxx00xxxx0002222}(c) and~\ref{fig:figure000xxx00xxxx0002222}(f) suggest that, for the chosen initial parameters, both quantities ${\widetilde{\delta}^{\text{R}}_3}(\alpha,\tau)$ and ${\widetilde{\delta}^{\text{H}}_3}(\alpha,\tau)$ behave similarly, being slightly tight in the time window $0 \lesssim \tau \lesssim 2$, for $0 \leq \alpha \leq 1$. Nonetheless, we stress that a more ge\-ne\-ral analysis requires varying those parameters to include a larger class of initial states and Hamiltonians. Of course, this subject should deserve further investigation, also including a detailed study of Eqs.~\eqref{eq:idea0000xxxxx010x1aaaaaaaa01},~\eqref{eq:idea0000xxxxx010x1aaaaaaaa02} and~\eqref{eq:idea0000xxxxx010x1aaaaaaaa03}, as well as the family of QSLs bounds presented in Sec.~\ref{sec:sec0003}, for higher-dimensional systems.


\section{Discussion}
\label{sec:conclusions000xxx001}

The main contribution of this paper is to provide an upper bound on ge\-ne\-ra\-li\-zed entropies when the physical system undergoes a unitary transformation. As the first application of our bound, we derived a family of QSLs in Eqs.~\eqref{eq:QSL_asymmetric},~\eqref{eq:QSL_asymmetric2}, and~\eqref{eq:QSL_symmetric}. From this result, the minimum time required for the state transformation is inversely proportional to a quantity that involves the average energy of the system [see Eq.~\eqref{eq:idea0000xxxxx010x10002}]. In turn, this determines the evolution speed, while being directly proportional to $\mathcal{O}_{\alpha}$, and thus implying the entropies play the role of distances. Furthermore, our derivations of QSLs, based on RRE, is first step toward \emph{resource speed limits} quantifying consumption of resource in single shot scenarios~\cite{arXiv:2004.03078}.

Another interesting connection can be built based on asymmetry monotones (AMs), which in turn characterize conservation laws for general quantum systems, in the sense of Noether's theorem~\cite{Marvian2014}. In short, AM is a function $f\, : \mathcal{B}(\mathcal{H}) \rightarrow \mathbb{R}$ that quantifies how much the state of the system breaks a given symmetry. Mathematically, the action of a symmetry group $G$ on a quantum system is described by the operation $\mathcal{U}_{g}(\rho) = {U_g} \, \rho\,  {U_g^{\dagger}}$, where the variable $g$ labels the group elements. The key idea behind AMs is to recog\-nize the orbit of each quantum state as an en\-co\-ding process, while the map $\mathcal{U}_{g}(\rho_{0}) \rightarrow \mathcal{U}_{g}(\rho_{\tau})$ is viewed as data processing. This implies that we can employ any contractive information measure to characterize the orbit of each state, which leads to an AM satisfying $f(\mathcal{U}_{g}(\rho)) \leq f(\rho)$~\cite{Marvian2014}.

Considering the range of $\alpha$ in which RRE and TRE are contractive under the action of a completely-positive and trace-preserving map, we immediately see the symmetrized relative entropies in Eq.~\eqref{eq:idea0000xxxxx000x1} (as well as its nonsymmetric versions) define an entire family of AMs. Indeed, the standard relative entropy ($\alpha \rightarrow 1$) was previously considered as an asymmetry monotone~\cite{PhysRevA.94.052324}. 

Next, since Eqs.~\eqref{eq:idea0000xxxxx010x1},~\eqref{eq:nonsym000222020202}, and~\eqref{eq:sym000222020202} are valid for unitary transformations encoding any unknown parameter into the state of the system, i.e., it goes beyond the paradigmatic case of time evolutions, thus we can replace $t$ by the group variable $g$. Indeed, this is due to the fact the unitary representation of the symmetry group leads to the evolution equation $d{\rho_{g}}/{d g} = -i[K,\rho_{g}]$, with $K$ being the generator of the transformation. Therefore, our results provide upper bounds on how much the state breaks the symmetry generated by $K$. This sets upper bounds of how much the conservation of the associated physical quantity can be broken. In the specific case of QSLs, this implies the minimum evolution time is determined by the asymmetry measure respective to nonsymmetric and symmetric RRE and TRE, which in turn stands as a measure of how much the initial state breaks the time-translation symmetry [see Eqs.~\eqref{eq:QSL_asymmetric},~\eqref{eq:QSL_asymmetric2}, and~\eqref{eq:QSL_symmetric}].

Finally, we will discuss the relation between our results and the concept of nonequilibrium entropy production. We begin by setting the initial state of the system as a thermal one, $\rho_{0} = \rho_{\beta} = \exp{-\beta H_0}/\mathcal{Z}$, where $\mathcal{Z}$ is the partition function, and $H_0$ is the ``bare'' Hamiltonian of the system, i.e., $[{H_0},{H_t}] \neq 0$ for all $t \neq 0$. From Eq.~\eqref{eq:QSL_RE0000}, one may verify the lower bound on the time of evolution is proportional to $\text{S}(\rho_{\tau}\| \rho_{\beta})$, which in turn stands as the entropy production associated with the process under consideration. Therefore, a natural question arises about the extension of this connection to ge\-ne\-ral entropies and systems. Indeed, such a general picture could be addressed by exploiting the entire family of second laws of thermodynamics based on RRE~\cite{Brandao3275,PhysRevE.99.050101}. This may open an avenue for the comprehension of QSLs~\cite{1945_JPhysURSS_9_249,1992_PhysicaD_120_188,2013_PhysRevLett_110_050402,PhysRevLett.110.050403,2013_PhysRevLett_111_010402,PhysRevLett.124.110601,arXiv:2001.05418}, asymmetry monotones~\cite{PhysRevA.94.052324}, and quantum ther\-mo\-dy\-na\-mics~\cite{PhysRevE.98.032106,PhysRevLett.123.230603,PhysRevResearch.2.023377,arXiv:2007.01882}, based on a strictly geometric framework.

The results presented here raise another questions. First, we could consider the extension of our results to open quantum systems. Moreover, given the link between quantum co\-he\-ren\-ce and TRE~\cite{Zhao2018}, we could investigate the trade-off among entropy production, QSL and quantum coherence in this scenario. Furthermore, since our results also apply for the min-entropy, i.e., the limit $\alpha\rightarrow 0$ regarding to RRE, they can be employed in the single-shot information theo\-ry, where the relations among asymmetry, QSL, and thermodynamics can be further de\-ve\-lo\-ped into the nonasymptotic regime. 


\begin{acknowledgments}
D. P. P. and L. C. C. would like to acknowledge the financial support from the Brazilian ministries MEC and MCTIC, funding agencies CAPES and CNPq, and the Bra\-zi\-lian National Institute of Science and Technology of Quantum Information (INCT-IQ). This study was financed in part by the Coordena\c{c}\~{a}o de Aperfei\c{c}oamento de Pessoal de N\'{i}vel Superior, Brasil (CAPES), Finance Code 001. L. C. C. also acknowledges support from Spanish MCIU/AEI/FEDER (Grant No. PGC2018-095113-BI00), Basque Government IT986-16, Projects No. QMiCS (820505) and No. OpenSuperQ (820363) of the EU Flagship on Quantum Technologies, EU FET Open Grant Quromorphic (828826), the U.S. Department of Energy, Office of Science, Office of Advanced Scientific Computing Research (ASCR) quantum algorithm teams program, under field work proposal number ERKJ333, and the Shanghai STCSM (Grant No. 2019SHZDZX01-ZX04).
\end{acknowledgments}

\setcounter{equation}{0}
\setcounter{table}{0}
\setcounter{section}{0}
\numberwithin{equation}{section}
\makeatletter
\renewcommand{\thesection}{\Alph{section}} 
\renewcommand{\thesubsection}{\Alph{section}.\arabic{subsection}}
\def\@gobbleappendixname#1\csname thesubsection\endcsname{\Alph{section}.\arabic{subsection}}
\renewcommand{\theequation}{\Alph{section}\arabic{equation}}
\renewcommand{\thefigure}{\arabic{figure}}
\renewcommand{\bibnumfmt}[1]{[#1]}
\renewcommand{\citenumfont}[1]{#1}

\section*{Appendix}


\section{Bound on relative entropies}
\label{sec:B1}

In this Appendix we will present in details the derivation of the results discussed in Sec.~\ref{sec:sec0001}.

\subsection{Bounding $\alpha$-relative entropy}
\label{sec:B0}

Here we will derive a nontrivial lower bound on the $\alpha$-relative purity, which will be useful throughout this supplemental material. Let $\rho_1,\rho_2\in\Omega$ be two arbitrary density matrices, with $\Omega \subset \mathcal{H}$, where $\Omega = \{\rho \in \mathcal{H} \mid {\rho^{\dagger}} = \rho,~\rho\geq 0,~\text{Tr}(\rho) = 1\}$ sets the convex space of quantum states, while $\mathcal{H}$ is a $d$-dimensional Hilbert space, with $d = \text{dim}(\mathcal{H})$. The Tsallis relative entropy (TRE) is defined as ${\text{H}_{\alpha}}({\rho_1}\|{\rho_2}) = {(1 - \alpha)^{-1}}\left[ 1 - {g_{\alpha}}({\rho_1},{\rho_2}) \right] $ where ${g_{\alpha}}({\rho_1},{\rho_2}) = \text{Tr}({\rho_1^{\alpha}}{\rho_2^{1 - \alpha}})$ stands for the $\alpha$-relative purity~\cite{10.1063.1.4838856}. In particular, for $0 \leq \alpha \leq 1$, it has been proved that TRE is upper bounded as follows~\cite{Ruskai_1990}
\begin{equation}
\label{eq:00000002}
{\text{H}_{\alpha}}({\rho_1}\|{\rho_2}) \leq S({\rho_1}\|{\rho_2}) ~,
\end{equation}
where $S({\rho_1}\|{\rho_2}) = \text{Tr}[{\rho_1}(\ln{\rho_1} - \ln{\rho_2})] $ is the ``standard'' quantum relative entropy. Interestingly, from Eq.~\eqref{eq:00000002}, we readily derive the lower bound on $\alpha$-relative purity,
\begin{equation}
\label{eq:00000005}
{g_{\alpha}}({\rho_1},{\rho_2}) \geq 1 - (1 - \alpha)S({\rho_1}\|{\rho_2}) ~.
\end{equation}
Noteworthy, bound in Eq.~\eqref{eq:00000005} exhibits two important features. On the one hand, for $\alpha = 1$, it trivially saturates since ${g_1}({\rho_1},{\rho_2}) = 1$ for all states $\rho_1,\rho_2\in\Omega$. On the other hand, for $\alpha = 0$ one recovers the Klein's inequality, $S({\rho_1}\|{\rho_2}) \geq 0$, due to the fact that ${g_0}({\rho_1},{\rho_2}) = 1$ for all $\rho_1,\rho_2\in\Omega$, thus stating the quantum relative entropy is nonnegative. Next, quantum relative entropy satisfies the following lower bound~\cite{doi:10.1063/1.2044667,doi:10.1063/1.3657929}
\begin{equation}
\label{eq:00000006}
S({\rho_1}\|{\rho_2}) \leq - \ln({{\lambda_{\text{min}}}({\rho_2})}) ~,
\end{equation}
where ${\lambda_{\text{min}}}(\bullet)$ sets the minimum eigenvalue of the referred density matrix. Importantly, authors in Refs.~\cite{doi:10.1063/1.2044667,doi:10.1063/1.3657929} have derived a plethora of lower and upper bounds on $S({\rho_1}\|{\rho_2})$, which in turn depend on some distance measures as the operator norm, Schatten 1-norm, and Fr\"{o}benius norm. However, we shall stress the bound in Eq.~\eqref{eq:00000006} is more suitable for our purposes, mostly because it stands as one of the simplest nontrivial lower bounds on quantum relative entropy. Therefore, by combining Eqs.~\eqref{eq:00000005} and~\eqref{eq:00000006}, we thus have that
\begin{equation}
\label{eq:00000010}
{g_{\alpha}}({\rho_1},{\rho_2}) \geq 1 + (1 - \alpha) \ln({{\lambda_{\text{min}}}({\rho_2})})  ~,
\end{equation}
which implies the following upper bound on $\alpha$-relative purity:
\begin{equation}
\label{eq:00000011}
{[{{g_{\alpha}}({\rho_1},{\rho_2})}]^{-1}} \leq {[{1 + (1 - \alpha) \ln({{\lambda_{\text{min}}}({\rho_2})})}]^{-1}}  ~.
\end{equation}


\subsection{Bound on non-symmetric relative entropies}
\label{sec:B1A}

Here we will address the upper bound for nonsymmetric quantum relative entropies. From now on, we will focus on both R\'{e}nyi, ${\text{R}_{\alpha}}({\rho_t}\|{\rho_0})$, and Tsallis, ${\text{H}_{\alpha}}({\rho_t}\|{\rho_0})$, relative entropies, where $\rho_0$ is the initial state of the system, and ${\rho_t} = {U_t}{\rho_0}{U_t^{\dagger}}$ its respective evolved state, with ${U_t} = \mathcal{T} {e^{-i {\int_0^t} \, ds \, {H_s}}}$ being the unitary time-ordered evolution operator. The absolute value of the time derivative of R\'{e}nyi relative entropy (RRE) of states $\rho_0$ and $\rho_t$ is given by
\begin{equation}
\label{eq:nonsym0000xxxxx001}
 \left|\frac{d}{dt}{\text{R}_{\alpha}}({\rho_t}\|{\rho_0})\right| = \frac{1}{|1 - \alpha|\, {{g}_{\alpha}}({\rho_t},{\rho_0})} \left| \frac{d}{dt}{{g}_{\alpha}}({\rho_t},{\rho_0}) \right| ~.
\end{equation}
In Appendix~\ref{sec:B0} we have derived an upper bound on the inverse of the $\alpha$-relative purity, which in turn will exhbit a logarithmic dependence on the smallest eigenvalue of the initial state of the system. Indeed, by substituting Eq.~\eqref{eq:00000011} into~\eqref{eq:nonsym0000xxxxx001}, one obtains the following inequality
\begin{equation}
\label{eq:nonsym0000xxxxx002}
 \left|\frac{d}{dt}{\text{R}_{\alpha}}({\rho_t}\|{\rho_0})\right| \leq \frac{\left| \frac{d}{dt}{{g}_{\alpha}}({\rho_t},{\rho_0}) \right|}{|1 - \alpha| \left|1 + (1 - \alpha) \ln({{\lambda_{\text{min}}}({\rho_0})}) \right| }  ~.
\end{equation}
For completeness, the absolute value of the time derivative of Tsallis relative entropy (TRE) of states $\rho_0$ and $\rho_t$ is given by
\begin{equation}
\label{eq:nonsym0000xxxxx003}
\left|\frac{d}{dt}{\text{H}_{\alpha}}({\rho_t}\|{\rho_0})\right| = \frac{1}{|1 - \alpha|} \left| \frac{d}{dt}{{g}_{\alpha}}({\rho_t},{\rho_0}) \right| ~.
\end{equation}

Next, the time derivative of relative purity ${{g}_{\alpha}}({\rho_t},{\rho_0}) = \text{Tr}\left({\rho_t^{\alpha}}\,{\rho_0^{1 - \alpha}}\right)$ is evaluated as follows. Because ${{\rho}_t}$ evolves unitarily, it is possible to verify the operator ${{\rho}_t^{\alpha}} = {U_t}\,{{\rho}_0^{\alpha}}\,{U_t^{\dagger}}$ satisfies the von Neumann equation $d{\rho_t^{\alpha}}/dt = - i \left[{H_t},{{\rho}_t^{\alpha}}\right]$, where we used the identity ${U_t}({dU_t^{\dagger}}/{dt}) = - ({dU_t}/{dt}){U_t^{\dagger}} = - i {H_t}$. Hence, we thus have that
\begin{equation}
\label{eq:nonsym0000xxxxx005}
\frac{d}{dt}{{g}_{\alpha}}({{\rho}_t},{\rho_0}) = {-i}\,\text{Tr}\left({\rho_t^{\alpha}} \, [{\rho_0^{1 - \alpha}},{H_t}]\right) ~,
\end{equation}
where we have used the cyclic property of trace. By taking the absolute value of Eq.~\eqref{eq:nonsym0000xxxxx005}, and applying the Cauchy-Schwarz inequality, $|\text{Tr}({{\mathcal{A}}_1}{{\mathcal{A}}_2})| \leq {{\| {{\mathcal{A}}_1} \|}_2}{{\| {{\mathcal{A}}_2} \|}_2}$, with ${{\| {\mathcal{A}} \|}_2} = \sqrt{\text{Tr}({{\mathcal{A}}^{\dagger}} {\mathcal{A}})} $, one readily gets
\begin{equation}
\label{eq:nonsym0000xxxxx006}
\left| \frac{d}{dt}{{g}_{\alpha}}({{\rho}_t},{\rho_0}) \right| \leq {{\|{\rho_0^{\alpha}}\|}_2}\, {{\| [ {H_t}, {{\rho}_0^{1 - \alpha}} ] \|}_2} ~,
\end{equation}
where we have used that ${\|{\rho_t^{\alpha}}\|_2} = {\|{U_t}{\rho_0^{\alpha}}{U_t^{\dagger}}\|_2} = {\|{\rho_0^{\alpha}}\|_2}$, i.e., Schatten 2-norm is unitarily invariant. Hence, by substituting Eq.~\eqref{eq:nonsym0000xxxxx006} into Eqs.~\eqref{eq:nonsym0000xxxxx002} and~\eqref{eq:nonsym0000xxxxx003}, one finds the generalized upper bound 
\begin{equation}
\label{eq:nonsym0000xxxxx007}
 \left|\frac{d}{dt}{\mathcal{O}_{\alpha}}({\rho_t}\|{\rho_0})\right| \leq {|1 - \alpha|^{-1}} {\mathcal{G}^{\mathcal{O}}_{\alpha}}(t) ~,
\end{equation}
where we define the functional
\begin{equation}
\label{eq:nonsym0000xxxxx008}
{\mathcal{G}^{\mathcal{O}}_{\alpha}}(t) := {\Phi_{\alpha}^{\mathcal{O}}} \, {{\|{\rho_0^{\alpha}}\|}_2}\, {{\| [ {H_t}, {{\rho}_0^{1 - \alpha}} ] \|}_2}   ~,
\end{equation}
while the auxiliary function reads
\begin{equation}
\label{eq:nonsym0000xxxxx009}
{\Phi_{\alpha}^{\mathcal{O}}} = 
\begin{cases}
{\left|1 + (1 - \alpha) \ln({{\lambda_{\text{min}}}({\rho_0})}) \right|^{-1}} ~,& \mbox{for ${\mathcal{O}} \equiv {\text{R}}$} \\
1 ~,& \mbox{for ${\mathcal{O}} \equiv {\text{H}}$} ~.
\end{cases}
\end{equation}
Just to clarify, in the remainder of the paper the index $\mathcal{O} \equiv \{\text{R},\text{H}\}$ will label RRE and TRE, respectively. Finally, by integrating Eq.~\eqref{eq:nonsym0000xxxxx007} over the interval $0 \leq t \leq \tau$, we thus obtain the upper bound
\begin{equation}
\label{eq:nonsym0000xxxxx0010}
{\mathcal{O}_{\alpha}}({\rho_{\tau}} \| {\rho_0}) \leq {{|1 - \alpha|}^{-1}} \,  {\int_0^{\tau}} dt \,  {\mathcal{G}^{\mathcal{O}}_{\alpha}}(t) ~,
\end{equation}
where we have invoked the inequality $\left|{\int} dx f(x)\right| \leq {\int}dx |f(x)|$, and used the fact that both RRE and TRE are nonnegative, real-valued, information-theoretic measures. 

Importantly, an analogous bound can be derived for the case in which the states $\rho_0$ and $\rho_{\tau}$ are swapped. We will briefly sketch the proof since the calculations are similar to the previous discussion. 
The property ${g_{\alpha}}({\rho_0},{\rho_t}) = {g_{1 - \alpha}}({\rho_t},{\rho_0})$ for the relative purity implicates the skew symmetry 
\begin{equation}
\label{eq:nonsym0000xxxxx007bbbb}
{(1 - \alpha)}\, {\mathcal{O}_{\alpha}}({\rho_0}\|{\rho_t}) = \alpha\, {\mathcal{O}_{1 - \alpha}}({\rho_t}\|{\rho_0}) ~,
\end{equation} 
which holds for all ${\rho_0},{\rho_t}$ and $0 < \alpha < 1$. In turn, the identity in Eq.~\eqref{eq:nonsym0000xxxxx007bbbb} allow us to write down the time-derivative 
\begin{equation}
\label{eq:nonsym0000xxxxx007bbbb02}
\left|\frac{d}{dt}{\mathcal{O}_{\alpha}}({\rho_0}\|{\rho_t})\right| = \frac{\alpha}{|1 - \alpha|} \left| \frac{d}{dt} {\mathcal{O}_{1 - \alpha}}({\rho_t}\|{\rho_0}) \right| ~. 
\end{equation}
Invoking Eq.~\eqref{eq:nonsym0000xxxxx007} and mapping $\alpha \rightarrow 1 - \alpha$, one obtains $ \left|d{\mathcal{O}_{1 - \alpha}}({\rho_t}\|{\rho_0})/dt \right| \leq {\alpha^{-1}} {\mathcal{G}^{\mathcal{O}}_{1 - \alpha}}(t)$. Hence, by inserting this expression into the right-hand side of Eq.~\eqref{eq:nonsym0000xxxxx007bbbb02}, we thus have that
\begin{equation}
\label{eq:nonsym0000xxxxx007aaaa}
 \left|\frac{d}{dt}{\mathcal{O}_{\alpha}}({\rho_0}\|{\rho_t})\right| \leq {|1 - \alpha|^{-1}} {\mathcal{G}^{\mathcal{O}}_{1 - \alpha}}(t) ~,
\end{equation}
which implies the upper bound
\begin{equation}
\label{eq:nonsym0000xxxxx0010b}
{\mathcal{O}_{\alpha}}({\rho_0} \| {\rho_{\tau}}) \leq {{|1 - \alpha|}^{-1}} \,  {\int_0^{\tau}} dt \,  {\mathcal{G}^{\mathcal{O}}_{1 - \alpha}}(t) ~.
\end{equation}


\subsection{Bound on symmetric relative entropies}
\label{sec:B1B}

Here we will address the upper bound for sym\-me\-tric quantum relative entropies. We shall begin with the symmetric relative entropy
\begin{equation}
\label{eq:idea0000xxxxx002x1}
{\mathcal{O}_{\alpha}}({\rho_t}:{\rho_0}) := {\mathcal{O}_{\alpha}}({\rho_t}\|{\rho_0}) + {\mathcal{O}_{\alpha}}({\rho_0}\|{\rho_t}) ~,
\end{equation}
where $\rho_0$ is the initial state of the system, while ${\rho_t} = {U_t}{\rho_0}{U_t^{\dagger}}$ is the evolved state. Let $|d\, {\mathcal{O}_{\alpha}}({\rho_t}:{\rho_0})/dt |$ be the absolute value of the time-derivative of Eq.~\eqref{eq:idea0000xxxxx002x1}. Hence, by applying the triangle ine\-qua\-li\-ty $|{a_1} + {a_2}| \leq |{a_1}| + |{a_2}|$, one obtains
\begin{equation}
\label{eq:idea0000xxxxx003}
\left|\frac{d}{dt}{\mathcal{O}_{\alpha}}({\rho_t} : {\rho_0})\right| \leq \left|\frac{d}{dt}{\mathcal{O}_{\alpha}}({\rho_t}\|{\rho_0})\right| + \left|\frac{d}{dt}{\mathcal{O}_{\alpha}}({\rho_0}\|{\rho_t})\right| ~.
\end{equation}
Based on Eqs.~\eqref{eq:nonsym0000xxxxx007} and~\eqref{eq:nonsym0000xxxxx007aaaa}, it follows that
\begin{equation}
\label{eq:idea0000xxxxx007c}
\left|\frac{d}{dt}{\mathcal{O}_{\alpha}}({\rho_t} : {\rho_0})\right| \leq {|1 - \alpha|^{-1}} \left[ {\mathcal{G}^{\mathcal{O}}_{\alpha}}(t) + {\mathcal{G}^{\mathcal{O}}_{1 - \alpha}}(t)\right] ~.
\end{equation}
Finally, by integrating Eq.~\eqref{eq:idea0000xxxxx007c} over the interval $t \in [0,\tau]$, and using the fact that $\left|{\int} dx f(x)\right| \leq {\int}dx |f(x)|$, one finds the upper bound for symmetric relative entropies
\begin{equation}
\label{eq:idea0000xxxxx010}
\left|{\mathcal{O}_{\alpha}}({\rho_{\tau}} : {\rho_0})  \right| \leq {{|\alpha - 1|}^{-1}} {\int_0^{\tau}} dt \left[ {\mathcal{G}^{\mathcal{O}}_{\alpha}}(t) + {\mathcal{G}^{\mathcal{O}}_{1 - \alpha}}(t) \right] ~.
\end{equation}


\begin{widetext}
\section{Recovering relative entropy}
\label{app:otocAPD}

In this Appendix we present the details in the calculation of the limit $\alpha \rightarrow 1$. The idea here is going back few steps and pinpoint the main features needed to properly address such nontrivial limit. For the two states $\rho_0$ and ${\rho_t} = {U_t}{\rho_0}{U_t^{\dagger}}$, we shall begin evaluating the limiting case $\alpha \rightarrow 1$ of the time derivative of symmetric relative entropy defined in Eq.~\eqref{eq:idea0000xxxxx002x1}, i.e.,
\begin{equation}
\label{eq:appc0000000}
 {\lim_{\alpha \rightarrow 1}}\, \frac{d}{dt}{\mathcal{O}_{\alpha}}({\rho_t}:{\rho_0}) = {\lim_{\alpha \rightarrow 1}}\, \frac{d}{dt}{\mathcal{O}_{\alpha}}({\rho_t}\|{\rho_0}) + {\lim_{\alpha \rightarrow 1}}\, \frac{\alpha}{(1 - \alpha)}\frac{d}{dt}{\mathcal{O}_{1 - \alpha}}({\rho_t}\|{\rho_0}) ~,
\end{equation}
where we have used the fact that ${\mathcal{O}_{\alpha}}({\rho_0}\|{\rho_t})$ is skew symmetric [see Eq.~\eqref{eq:nonsym0000xxxxx007bbbb}]. Next, by taking the absolute value of Eq.~\eqref{eq:appc0000000}, then applying the triangle inequality $|{a_1} + {a_2}| \leq |{a_1}| + |{a_2}|$, and finally integrating the resulting expression over interval $t \in [0,\tau]$, we thus have that
\begin{align}
\label{eq:appc0000001}
{\int_0^{\tau}} dt  \left| {\lim_{\alpha \rightarrow 1}}\, \frac{d}{dt}{\mathcal{O}_{\alpha}}({\rho_t}:{\rho_0})\right| \leq {\int_0^{\tau}} dt  \left|{\lim_{\alpha \rightarrow 1}}\, \frac{d}{dt}{\mathcal{O}_{\alpha}}({\rho_t}\|{\rho_0})\right| + {\int_0^{\tau}} dt  \left| {\lim_{\alpha \rightarrow 1}}\, \frac{\alpha}{(1 - \alpha)}\frac{d}{dt}{\mathcal{O}_{1 - \alpha}}({\rho_t}\|{\rho_0})\right| ~.
\end{align}
Based on the definition of the relative entropy, $\text{S}(A\|B) = \text{Tr}[A(\ln{A} - \ln{B})]$, note that one may write down
\begin{equation}
\label{eq:appc0000002}
\left| \text{S}({\rho_{\tau}}\| {\rho_0} ) + \text{S}({\rho_0}\| {\rho_{\tau}} ) \right| 
= \left| {\lim_{\alpha \rightarrow 1}} ~ {\mathcal{O}_{\alpha}}({\rho_{\tau}} : {\rho_0})  \right| 
=  \left|  {\int_0^{\tau}} dt \, {\lim_{\alpha \rightarrow 1}}\,  \frac{d}{dt}{\mathcal{O}_{\alpha}}({\rho_t}:{\rho_0})\right| 
\leq {\int_0^{\tau}} dt  \left| {\lim_{\alpha \rightarrow 1}}\, \frac{d}{dt}{\mathcal{O}_{\alpha}}({\rho_t} : {\rho_0})\right| ~.
\end{equation}
Just to clarify, here we assume that RRE and TRE are continuous real-valued functions over the set $\alpha \in (0,1)\cup(1,+\infty)$ and $t \in [0, \tau]$. In this sense, we are formally able to switch the limit on parameter $\alpha$ with the integration sign over variable $t$. Thus, by combining Eqs.~\eqref{eq:appc0000001} and~\eqref{eq:appc0000002}, one readily obtains
\begin{align}
\label{eq:appc0000007}
\left| {\text{S}}({\rho_{\tau}}\| {\rho_0} ) + {\text{S}}({\rho_0}\| {\rho_{\tau}}) \right|  \leq {\int_0^{\tau}} dt \left|{\lim_{\alpha \rightarrow 1}}\, \frac{{\mathcal{E}_{\alpha}^{\mathcal{O}}}({\rho_t},{\rho_0})}{(1 - \alpha)} \frac{d}{dt}{{g}_{\alpha}}({{\rho}_t},{\rho_0}) \right| + {\int_0^{\tau}} dt  \left|{\lim_{\alpha \rightarrow 1}}\, \frac{{\mathcal{E}_{1 - \alpha}^{\mathcal{O}}}({\rho_t},{\rho_0})}{(1 - \alpha)} \frac{d}{dt}{{g}_{1 - \alpha}}({{\rho}_t},{\rho_0}) \right| ~,
\end{align}
where we have used that
\begin{equation}
\label{eq:appc0000004}
 \left|{\lim_{\alpha \rightarrow 1}}\, \frac{d}{dt}{\mathcal{O}_{\alpha}}({\rho_t}\|{\rho_0})\right| =  \left|{\lim_{\alpha \rightarrow 1}}\, \frac{{\mathcal{E}_{\alpha}^{\mathcal{O}}}({\rho_t},{\rho_0})}{(1 - \alpha)} \frac{d}{dt}{{g}_{\alpha}}({{\rho}_t},{\rho_0}) \right| ~,
\end{equation}
and
\begin{align}
\label{eq:appc0000005}
\left| {\lim_{\alpha \rightarrow 1}}\, \frac{\alpha}{(1 - \alpha)}\frac{d}{dt}{\mathcal{O}_{1 - \alpha}}({\rho_t}\|{\rho_0})\right| =  \left|{\lim_{\alpha \rightarrow 1}}\, \frac{{\mathcal{E}_{1 - \alpha}^{\mathcal{O}}}({\rho_t},{\rho_0})}{(1 - \alpha)} \frac{d}{dt}{{g}_{1 - \alpha}}({{\rho}_t},{\rho_0}) \right| ~,
\end{align}
with the auxiliary functional ${\mathcal{E}_s^{\mathcal{O}}}(A,B)$ defined as 
\begin{equation}
\label{eq:nonsym0000xxxxx009aaaaa}
{\mathcal{E}_{s}^{\mathcal{O}}}(A,B) = 
\begin{cases}
{\left[{g_s}(A,B)\right]^{-1}} ~,& \mbox{for ${\mathcal{O}} \equiv {\text{R}}$} \\
1 ~,& \mbox{for ${\mathcal{O}} \equiv {\text{H}}$} ~.
\end{cases}
\end{equation}
Interestingly, one may verify the right-hand side of Eq.~\eqref{eq:appc0000007} exhibits an indeterminacy in the limit $\alpha \rightarrow 1$. Indeed, since $(d{\rho_t^s}/dt) = - i \left[{H_t},{{\rho}_t^s}\right]$, it follows that 
\begin{equation}
{\lim_{\alpha \rightarrow 1}}\, [d{{g}_{\alpha}}({{\rho}_t},{\rho_0})/dt] = ({-i})\, {\lim_{\alpha \rightarrow 1}}\, \text{Tr}\left({\rho_0^{1 - \alpha}}\left[{H_t},{\rho_t^{\alpha}} \right]\right) = 0,
\end{equation}
and also 
\begin{equation}
{\lim_{\alpha \rightarrow 1}}\, [{d}{{g}_{1 - \alpha}}({{\rho}_t},{\rho_0})/dt] = ({-i})\, {\lim_{\alpha \rightarrow 1}}\, \text{Tr}\left({\rho_0^{\alpha}}[{H_t},{\rho_t^{1 - \alpha}}]\right) = 0 ~. 
\end{equation}
Similarly, one readily verifies that ${\lim_{\alpha \rightarrow 1}}\, (1 - \alpha){[{\mathcal{E}_{\alpha}^{\mathcal{O}}}({\rho_t},{\rho_0})]^{-1}} = 0$ and ${\lim_{\alpha \rightarrow 1}}\, (1 - \alpha){[{\mathcal{E}_{1 - \alpha}^{\mathcal{O}}}({\rho_t},{\rho_0})]^{-1}} = 0$, where we used the fact that ${\lim_{\alpha \rightarrow 1}}\, {[{\mathcal{E}_{\alpha}^{\mathcal{O}}}({\rho_t},{\rho_0})]^{-1}} = 1$, and ${\lim_{\alpha \rightarrow 1}}\, {[{\mathcal{E}_{1 - \alpha}^{\mathcal{O}}}({\rho_t},{\rho_0})]^{-1}} = 1$, for all $\mathcal{O} \equiv \{\text{R},\text{H} \}$. In this case, one obtains
\begin{equation}
\label{eq:appc0000010}
{\lim_{\alpha \rightarrow 1}}\, \frac{{\mathcal{E}_{\alpha}^{\mathcal{O}}}({\rho_t},{\rho_0})}{(1 - \alpha)} \frac{d}{dt}{{g}_{\alpha}}({{\rho}_t},{\rho_0}) \longrightarrow \frac{0}{0} ~,
\end{equation}
and 
\begin{equation}
\label{eq:appc0000011}
{\lim_{\alpha \rightarrow 1}}\, \frac{{\mathcal{E}_{1 - \alpha}^{\mathcal{O}}}({\rho_t},{\rho_0})}{(1 - \alpha)} \frac{d}{dt}{{g}_{1 - \alpha}}({{\rho}_t},{\rho_0}) \longrightarrow \frac{0}{0} ~,
\end{equation}
which in turn suggests the right-hand side of Eq.~\eqref{eq:appc0000007} is not well behaved. We note, however, this such issue is readily circumvented by applying the L'Hospital rule, leading us to
\begin{align}
\label{eq:appc0000012}
{\lim_{\alpha \rightarrow 1}}\, \frac{{\mathcal{E}_{\alpha}^{\mathcal{O}}}({\rho_t},{\rho_0})}{(1 - \alpha)} \frac{d}{dt}{{g}_{\alpha}}({{\rho}_t},{\rho_0})  =   {\lim_{\alpha \rightarrow 1}} ~ \frac{ d\left[ d\, {{g}_{\alpha}}({{\rho}_t},{\rho_0})/dt \right] /d\alpha }{ d \left( (1 - \alpha) \, {[{\mathcal{E}_{\alpha}^{\mathcal{O}}}({\rho_t},{\rho_0})]^{-1}} \right)/d\alpha } ~,
\end{align}
and 
\begin{align}
\label{eq:appc0000013}
{\lim_{\alpha \rightarrow 1}}\, \frac{{\mathcal{E}_{1 - \alpha}^{\mathcal{O}}}({\rho_t},{\rho_0})}{(1 - \alpha)} \frac{d}{dt}{{g}_{1 - \alpha}}({{\rho}_t},{\rho_0})  =   {\lim_{\alpha \rightarrow 1}} ~ \frac{ d\left[ d\, {{g}_{1 - \alpha}}({{\rho}_t},{\rho_0})/dt \right] /d\alpha }{ d \left( (1 - \alpha) \, {[{\mathcal{E}_{1 - \alpha}^{\mathcal{O}}}({\rho_t},{\rho_0})]^{-1}} \right)/d\alpha } ~.
\end{align}

In the following, we will discuss in details each contribution in the right-hand side of Eqs.~\eqref{eq:appc0000012} and~\eqref{eq:appc0000013}. Let us start by evaluating the following derivatives:
\begin{align}
\label{eq:appc0000027}
{\lim_{\alpha\rightarrow 1}} ~ \frac{d}{d\alpha}\left(\frac{d}{dt}{{g}_{\alpha}}({{\rho}_t},{\rho_0})\right) &= {-i}\, {\lim_{\alpha\rightarrow 1}} ~ \frac{d}{d\alpha}\left(\text{Tr}\left({\rho_0^{1 - \alpha}}\left[{H_t},{\rho_t^{\alpha}} \right]\right) \right) \nonumber\\
&= -i\, {\lim_{\alpha\rightarrow 1}} ~ {\sum_{j,\ell}}\, \frac{d}{d\alpha}\left( {p_j^{1 - \alpha}} {p_{\ell}^{\alpha}}\right) \langle{\psi_j}| \, [{H_t},{U_t}|{\psi_{\ell}}\rangle\langle{\psi_{\ell}}|{U_t^{\dagger}}]\,  |{\psi_j}\rangle  \nonumber\\
&=  -i\, {\sum_{j,\ell}}\, {p_{\ell}} (\ln{p_{\ell}} - \ln{p_j}) \, \langle{\psi_j}| \, [{H_t},{U_t}|{\psi_{\ell}}\rangle\langle{\psi_{\ell}}|{U_t^{\dagger}}]\,  |{\psi_j}\rangle \nonumber\\
&=  i \, \text{Tr}\left(\ln{\rho_0} \, [{H_t},{\rho_t} ] \right) ~,
\end{align}
and 
\begin{align}
\label{eq:appc0000028}
{\lim_{\alpha\rightarrow 1}} ~ \frac{d}{d\alpha}\left(\frac{d}{dt}{{g}_{1 - \alpha}}({{\rho}_t},{\rho_0}) \right) &= {-i}\, {\lim_{\alpha\rightarrow 1}} ~ \frac{d}{d\alpha}\left(\text{Tr}\left({\rho_0^{\alpha}}\left[{H_t},{\rho_t^{1 - \alpha}} \right]\right)\right) \nonumber\\
&= -i\, {\lim_{\alpha\rightarrow 1}} ~  {\sum_{j,\ell}} \, \frac{d}{d\alpha}\left( {p_j^{\alpha}} {p_{\ell}^{1 - \alpha}}\right) \langle{\psi_j}| \, [{H_t},{U_t}|{\psi_{\ell}}\rangle\langle{\psi_{\ell}}|{U_t^{\dagger}}]\,  |{\psi_j}\rangle \nonumber\\
&= i\, {\sum_{j,\ell}}\, {p_j} (\ln{p_{\ell}} - \ln{p_j}) \, \langle{\psi_j}| \, [{H_t},{U_t}|{\psi_{\ell}}\rangle\langle{\psi_{\ell}}|{U_t^{\dagger}}]\,  |{\psi_j}\rangle \nonumber\\
&=  - i\, \text{Tr}\left( {U_t}\ln{\rho_0} {U_t^{\dagger}}  [{H_t}, {\rho_0}]  \right) ~,
\end{align}
where we have used that $d\, ({p_j^{1 - \alpha}} {p_{\ell}^{\alpha}})/d\alpha = (\ln{p_{\ell}} - \ln{p_j}){p_j^{1 - \alpha}} {p_{\ell}^{\alpha}}$. Moving forward, note that
\begin{equation}
\label{eq:appc0000014}
\frac{d}{d\alpha} \left( (1 - \alpha) \, {[{\mathcal{E}_{\alpha}^{\mathcal{O}}}({\rho_t},{\rho_0})]^{-1}} \right) = - {[{\mathcal{E}_{\alpha}^{\mathcal{O}}}({\rho_t},{\rho_0})]^{-1}} + (1 - \alpha) \, \frac{d}{d\alpha}{[{\mathcal{E}_{\alpha}^{\mathcal{O}}}({\rho_t},{\rho_0})]^{-1}} ~,
\end{equation}
and
\begin{equation}
\label{eq:appc0000015}
\frac{d}{d\alpha} \left( (1 - \alpha) \, {[{\mathcal{E}_{1 - \alpha}^{\mathcal{O}}}({\rho_t},{\rho_0})]^{-1}} \right) = - {[{\mathcal{E}_{1 - \alpha}^{\mathcal{O}}}({\rho_t},{\rho_0})]^{-1}} + (1 - \alpha) \, \frac{d}{d\alpha}{[{\mathcal{E}_{1 - \alpha}^{\mathcal{O}}}({\rho_t},{\rho_0})]^{-1}} ~.
\end{equation}
From Eqs.~\eqref{eq:appc0000014} and~\eqref{eq:appc0000015}, we point out that ${\lim_{\alpha \rightarrow 1}}\, {[{\mathcal{E}_{\alpha}^{\mathcal{O}}}({\rho_t},{\rho_0})]^{-1}} = 1$, and ${\lim_{\alpha \rightarrow 1}}\, {[{\mathcal{E}_{1 - \alpha}^{\mathcal{O}}}({\rho_t},{\rho_0})]^{-1}} = 1$, for all $\mathcal{O} \equiv \{\text{R},\text{H} \}$. Hence, from now on it suffices to proceed by showing the derivatives $d\, {[{\mathcal{E}_{\alpha}^{\mathcal{O}}}({\rho_t},{\rho_0})]^{-1}}/d\alpha$, and $d\, {[{\mathcal{E}_{1 - \alpha}^{\mathcal{O}}}({\rho_t},{\rho_0})]^{-1}}/d\alpha$, are indeed well-behaved for $\alpha \rightarrow 1$. On the one hand, for Tsallis relative entropy the au\-xi\-li\-ary functional becomes ${\mathcal{E}_{\alpha}^{\text{H}}}({\rho_t},{\rho_0}) = {\mathcal{E}_{1 - \alpha}^{\text{H}}}({\rho_t},{\rho_0}) = 1$, for all $\alpha$, and the aforementioned derivatives are identically zero. In this case, from Eqs.~\eqref{eq:appc0000014} and~\eqref{eq:appc0000015}, one obtains
\begin{align}
\label{eq:appc0000016}
\frac{d}{d\alpha} \left( (1 - \alpha) \, {[{\mathcal{E}_{\alpha}^{\text{H}}}({\rho_t},{\rho_0})]^{-1}} \right) =\frac{d}{d\alpha} \left( (1 - \alpha) \, {[{\mathcal{E}_{1 - \alpha}^{\text{H}}}({\rho_t},{\rho_0})]^{-1}} \right) = -1 ~.
\end{align}
On the other hand, for R\'{e}nyi relative entropy the auxiliary functional behave as $ {[{\mathcal{E}_{\alpha}^{\text{R}}}({\rho_t},{\rho_0})]^{-1}} = {g_{\alpha}}({\rho_t},{\rho_0})$ and $ {[{\mathcal{E}_{1 - \alpha}^{\text{R}}}({\rho_t},{\rho_0})]^{-1}} = {g_{1 - \alpha}}({\rho_t},{\rho_0})$, and the calculation is more involved. To see this, let ${\rho_0} = {\sum_{\ell}}\, {p_{\ell}}|{\psi_{\ell}}\rangle\langle{\psi_{\ell}}|$ be the spectral decomposition of the initial state, with $0 \leq {p_{\ell}} \leq 1$ and ${\sum_{\ell}}\, {p_{\ell}} = 1$. In this case, given the evolved state ${\rho_t} = {U_t}\, {\rho_0}{U_t^{\dagger}}$, we thus have that ${\rho_t^{\alpha}} = {\sum_{\ell}}\, {p_{\ell}^{\alpha}} \, {U_t}|{\psi_{\ell}}\rangle\langle{\psi_{\ell}}|{U_t^{\dagger}}$, and the relative purity becomes ${g_{\alpha}}({\rho_t},{\rho_0}) = {\sum_{j,\ell}}\, {p_j^{1 - \alpha}} {p_{\ell}^{\alpha}} \, {| \langle{\psi_j}|{U_t}|{\psi_{\ell}}\rangle |^2}$. Hence, the derivative with respect to the parameter $\alpha$ is simply given by
\begin{align}
\label{eq:appc0000020}
{\lim_{\alpha \rightarrow 1}} ~ \frac{d}{d\alpha}{[{\mathcal{E}_{\alpha}^{\text{R}}}({\rho_t},{\rho_0})]^{-1}} &= {\lim_{\alpha \rightarrow 1}} ~ \frac{d}{d\alpha}\text{Tr}({\rho_t^{\alpha}}{\rho_0^{1 - \alpha}}) \nonumber\\ 
&= {\lim_{\alpha \rightarrow 1}} ~ {\sum_{j,\ell}}\, {p_{\ell}^{\alpha}} \, {p_j^{1 - \alpha}}(\ln{p_{\ell}} - \ln{p_j}) \, {| \langle{\psi_j}|{U_t}|{\psi_{\ell}}\rangle |^2} \nonumber\\
&= \text{S}({\rho_t} \| {\rho_0}) ~,
\end{align}
and
\begin{align}
\label{eq:appc0000021}
{\lim_{\alpha \rightarrow 1}} ~ \frac{d}{d\alpha}{[{\mathcal{E}_{1 - \alpha}^{\text{R}}}({\rho_t},{\rho_0})]^{-1}} &= {\lim_{\alpha \rightarrow 1}} ~ \frac{d}{d\alpha}\text{Tr}({\rho_t^{1 - \alpha}}{\rho_0^{\alpha}}) \nonumber\\
&= - {\lim_{\alpha \rightarrow 1}} ~ {\sum_{j,\ell}}\, {p_{\ell}^{1 - \alpha}} \, {p_j^{\alpha}}(\ln{p_{\ell}} - \ln{p_j}) \, {| \langle{\psi_j}|{U_t}|{\psi_{\ell}}\rangle |^2} \nonumber\\
&= \text{S}({\rho_0}\|{\rho_t}) ~.
\end{align}
Hence, by combining Eqs.~\eqref{eq:appc0000014},~\eqref{eq:appc0000015},~\eqref{eq:appc0000016},~\eqref{eq:appc0000020}, and~\eqref{eq:appc0000021}, we get the result
\begin{equation}
\label{eq:appc0000026}
{\lim_{\alpha \rightarrow 1}} ~\frac{d}{d\alpha} \left( (1 - \alpha) \, {[{\mathcal{E}_{\alpha}^{\mathcal{O}}}({\rho_t},{\rho_0})]^{-1}} \right) = {\lim_{\alpha \rightarrow 1}} ~\frac{d}{d\alpha} \left( (1 - \alpha) \, {[{\mathcal{E}_{1 - \alpha}^{\mathcal{O}}}({\rho_t},{\rho_0})]^{-1}} \right) = -1 ~.
\end{equation}
Therefore, by inserting Eqs.~\eqref{eq:appc0000027},~\eqref{eq:appc0000028},~\eqref{eq:appc0000016} and~\eqref{eq:appc0000026}, into Eqs.~\eqref{eq:appc0000012} and~\eqref{eq:appc0000013}, we conclude
\begin{equation}
\label{eq:appc0000031}
{\lim_{\alpha \rightarrow 1}}\, \frac{{\mathcal{E}_{\alpha}^{\mathcal{O}}}({\rho_t},{\rho_0})}{(1 - \alpha)} \frac{d}{dt}{{g}_{\alpha}}({{\rho}_t},{\rho_0}) = - i \,\text{Tr}\left(\ln{\rho_0} \, [{H_t},{\rho_t} ] \right)  ~,
\end{equation}
and 
\begin{equation}
\label{eq:appc0000032}
{\lim_{\alpha \rightarrow 1}}\, \frac{{\mathcal{E}_{1 - \alpha}^{\mathcal{O}}}({\rho_t},{\rho_0})}{(1 - \alpha)} \frac{d}{dt}{{g}_{1 - \alpha}}({{\rho}_t},{\rho_0}) =  i\, \text{Tr}\left( {U_t}\ln{\rho_0} {U_t^{\dagger}}  [{H_t}, {\rho_0}]  \right)  ~.
\end{equation}
Finally, by substituting Eqs.~\eqref{eq:appc0000031} and~\eqref{eq:appc0000032} into Eq.~\eqref{eq:appc0000007}, and then applying the Cauchy-Schwarz inequality, $|\text{Tr}({{\mathcal{A}}_1}{{\mathcal{A}}_2})| \leq {{\| {{\mathcal{A}}_1} \|}_2}{{\| {{\mathcal{A}}_2} \|}_2}$, with ${{\| {\mathcal{A}} \|}_2} = \sqrt{\text{Tr}({{\mathcal{A}}^{\dagger}} {\mathcal{A}})} $, it yields the result
\begin{align}
\label{eq:appc0000033}
\left| {\text{S}}({\rho_{\tau}}\| {\rho_0} ) + {\text{S}}({\rho_0}\| {\rho_{\tau}}) \right| \leq {\| \ln{\rho_0} \|_2}\, {\int_0^{\tau}} dt \, ( {\| [ {H_t} , {\rho_t} ] \|_2} + {\|[ {H_t} , {\rho_0} ] \|_2}) ~.
\end{align}


\section{Recovering min-relative entropy}
\label{app:otocminEntropy}

In this Appendix we will discuss the case $\alpha = 0$ for symmetric R\'{e}nyi relative entropy, which is related to the min-relative entropy 
\begin{equation}
\label{eq:min000xxx000}
{\text{R}_{0}}(\rho : \omega) := {\text{R}_0}(\rho\|\omega) + {\text{R}_0}(\omega\|\rho) ~,
\end{equation}
where $\text{R}_0(\rho\|\omega) = -\ln \text{Tr}(\Pi_{\rho}\, \omega)$, with ${\Pi_{\rho}}$ being the projector onto the support of the state $\rho$. Here we will focus on the time-independent initial state $\rho_0$, and its evolved state ${\rho_t} = {U_t}{\rho_0}{U_t^{\dagger}}$. By using the triangle inequality $|{a_1} + {a_2}| \leq |{a_1}| + |{a_2}|$, the absolute value of the time-derivative of Eq.~\eqref{eq:min000xxx000} is written as
\begin{equation}
\label{eq:min000xxx000x1}
\left| \frac{d}{dt}\, {\text{R}_{0}}({\rho_t} : {\rho_0}) \right| \leq \left| \frac{d}{dt}{\text{R}_0}({\rho_t}\|{\rho_0})\right| + \left|\frac{d}{dt}{\text{R}_0}({\rho_0}\|{\rho_t}) \right| ~.
\end{equation}

From now on we will discuss the evaluation of each contribution in right-hand side of Eq.~\eqref{eq:min000xxx000x1}. In order to do so, let ${\rho_0} = {\sum_j}\, {p_{\ell}} |{\psi_{\ell}}\rangle\langle{\psi_{\ell}}|$ be the spectral decomposition of the initial state $\rho_0$ into the basis $\{ |{\psi_{\ell}}\rangle \}_{\ell = 1,\ldots, d}$, with $0 \leq {p_{\ell}} \leq 1$ and ${\sum_{\ell}}\, {p_{\ell}} = 1$. By hypothesis, the support of $\rho_0$ has dimension ${d_{\rho_0}} := \text{dim}[\text{supp}({\rho_0})]$, and is given by $\text{supp}({\rho_0}) = \text{span}\{ |{\psi_{\ell}}\rangle : {p_{\ell}} \neq 0\}$. Thus, the projector onto the support of state $\rho_0$ is defined as ${\Pi_{\rho_0}} := {\sum_{ \ell : \, {p_{\ell}} \neq 0}} \, |{\psi_{\ell}}\rangle\langle{\psi_{\ell}}| $. The evolved state is given by ${\rho_t} = {U_t}{\rho_0}{U_t^{\dagger}} = {\sum_j}\, {p_{\ell}} |{\psi^t_{\ell}}\rangle\langle{\psi^t_{\ell}}|$, with $|{\psi^t_{\ell}}\rangle := {U_t}|{\psi_{\ell}}\rangle$, and its support is defined as $\text{supp}({\rho_t}) = \text{span}\{ |{\psi^t_{\ell}}\rangle : {p_{\ell}} \neq 0\}$. Noteworthy, since the unitary evolution does not change the purity of the initial state, i.e., both states $\rho_0$ and $\rho_t$ share the same set of eigenvalues, we thus have $\text{dim}[\text{supp}({\rho_t})] = \text{dim}[\text{supp}({\rho_0})]$. The projector ${\Pi_{\rho_t}}$ onto the support of $\rho_t$ read as 
\begin{align}
\label{eq:min000xxx005}
{\Pi_{\rho_t}} &= {\sum_{ \ell : \, {p_{\ell}} \neq 0}} \, |{\psi^t_{\ell}}\rangle\langle{\psi^t_{\ell}}| \nonumber\\
&=  {\sum_{ \ell : \, {p_{\ell}} \neq 0}} \, {U_t}|{\psi_{\ell}}\rangle\langle{\psi_{\ell}}| {U_t^{\dagger}} \nonumber\\
&= {U_t}\,{\Pi_{\rho_0}}{U_t^{\dagger}} ~.
\end{align}
Interestingly, starting from Eq.~\eqref{eq:min000xxx005}, the projector ${\Pi_{\rho_t}}$ fulfills the von Neumann-like equation $(d\, {\Pi_{\rho_t}}/dt) = - i\, \left[{H_t},{\Pi_{\rho_t}}\right]$, where we applied the identity ${U_t}({dU_t^{\dagger}}/{dt}) = - ({dU_t}/{dt}){U_t^{\dagger}}  = - i {H_t}$. Thus, the time derivative of min-relative entropy ${\text{R}_0}({\rho_t}\|{\rho_0}) = -\ln \text{Tr}({\Pi_{\rho_t}}{\rho_0})$ read as
\begin{align}
\label{eq:min000xxx006}
\frac{d}{dt}\,{\text{R}_0}({\rho_t}\|{\rho_0}) &= \frac{i\, \text{Tr}({\rho_0}\left[{H_t},{\Pi_{\rho_t}}\right]) }{\text{Tr}({\Pi_{\rho_t}}{\rho_0})} \nonumber\\
&= \frac{i\, \text{Tr}({U_t^{\dagger}}{\rho_0}\, {U_t}[{U_t^{\dagger}}{H_t}{U_t},{\Pi_{\rho_0}}]) }{\text{Tr}({\Pi_{\rho_0}}{U_t^{\dagger}}{\rho_0}\, {U_t})} ~,
\end{align}
where we have explicitly used the property obtained in Eq.~\eqref{eq:min000xxx005}. By taking the absolute value of Eq.~\eqref{eq:min000xxx006}, and thus applying the Cauchy-Schwarz inequality, $|\text{Tr}({{\mathcal{A}}_1}{{\mathcal{A}}_2})| \leq {{\| {{\mathcal{A}}_1} \|}_2}{{\| {{\mathcal{A}}_2} \|}_2}$, with ${{\| {\mathcal{A}} \|}_2} = \sqrt{\text{Tr}({{\mathcal{A}}^{\dagger}} {\mathcal{A}})} $, one obtains
\begin{align}
\label{eq:min000xxx007}
\left|\frac{d}{dt}\,{\text{R}_0}({\rho_t}\|{\rho_0})\right| \leq \frac{ {\|{\rho_0}\|_2} \, {\| [{U_t^{\dagger}}{H_t}{U_t},{\Pi_{\rho_0}}] \|_2} }{ |\text{Tr}({\Pi_{\rho_0}}{U_t^{\dagger}}{\rho_0}\, {U_t}) |} ~.
\end{align}

Let us now move to the the second term in the right-hand side of Eq.~\eqref{eq:min000xxx000x1}, which is related to the time-derivative of ${\text{R}_0}({\rho_0}\|{\rho_t}) = -\ln \text{Tr}({\Pi_{\rho_0}}\, {\rho_t})$. In this case, one readily obtains
\begin{align}
\label{eq:min000xxx001}
\frac{d}{dt}\,{\text{R}_0}({\rho_0}\|{\rho_t}) &= \frac{i\, \text{Tr}({\Pi_{\rho_0}} [{H_t},{\rho_t}]) }{\text{Tr}({\Pi_{\rho_0}}{\rho_t})} \nonumber\\
&= \frac{i \,\text{Tr}({U_t^{\dagger}}{\Pi_{\rho_0}}{U_t}[{U_t^{\dagger}}{H_t}{U_t},{\rho_0}]) }{ \text{Tr}({\Pi_{\rho_0}}{U_t}{\rho_0}{U_t^{\dagger}}) } ~,
\end{align}
where we used the fact that ${\rho_t}$ fulfills the von Neumann equation $(d{\rho_t}/dt) = - i\, \left[{H_t},{\rho_t}\right]$. By taking the absolute value of Eq.~\eqref{eq:min000xxx001}, and thus applying the aforementioned Cauchy-Schwarz inequality, one obtains
\begin{equation}
\label{eq:min000xxx002}
\left| \frac{d}{dt}\,{\text{R}_0}({\rho_0}\|{\rho_t})\right| \leq \frac{ {\|{\Pi_{\rho_0}}\|_2} {\| [{U_t^{\dagger}}{H_t}{U_t},{\rho_0}] \|_2} }{| \text{Tr}({\Pi_{\rho_0}} {U_t}{\rho_0}{U_t^{\dagger}}) |}  ~.
\end{equation}

Hence, by substituting Eqs.~\eqref{eq:min000xxx007} and~\eqref{eq:min000xxx002} into Eq.~\eqref{eq:min000xxx000x1}, we thus have that 
\begin{equation}
\label{eq:min000xxx000x10}
\left| \frac{d}{dt} \, {\text{R}_{0}}({\rho_t} : {\rho_0}) \right| \leq {\mathcal{Q}_0^t}({\rho_0},{\Pi_{\rho_0}}) + {\mathcal{Q}_0^t}({\Pi_{\rho_0}},{\rho_0})  ~,
\end{equation}
where the functional ${\mathcal{Q}_0}({\rho_t},\rho)$ is defined as follows
\begin{equation}
{\mathcal{Q}_0^t}(A,B) := \frac{ {\|A\|_2} \, {\| [{U_t^{\dagger}}{H_t}{U_t},B] \|_2} }{ |\text{Tr}(A {U_t}B{U_t^{\dagger}}) |} ~.
\end{equation}
Finally, by integrating Eq.~\eqref{eq:min000xxx000x10} over the interval $t \in [0,\tau]$, and thus applying the inequality $\left|{\int} dx f(x)\right| \leq {\int}dx |f(x)|$, one gets the inequality
\begin{equation}
\label{eq:min000xxx000x12}
\left| {\text{R}_{0}}({\rho_{\tau}} : {\rho_0}) \right| \leq {\int_0^{\tau}} \, dt \, \left[{\mathcal{Q}_0^t}({\rho_0},{\Pi_{\rho_0}}) + {\mathcal{Q}_0^t}({\Pi_{\rho_0}},{\rho_0}) \right] ~.
\end{equation}


\section{Example: single-qubit state}
\label{app:examplesAPD}


\subsection{Tsallis relative entropy (TRE)}
\label{sec:appF:sec000xxx002}

We shall begin showing that, for an initial single-qubit state evolving unitarily according the physical setting presented in Sec.~\ref{eq:singlequbit000000001}, the Tsallis relative entropy is given by
\begin{align}
\label{eq:app000xxx000xxx00120222}
{\text{H}_{\alpha}}({\rho_{\tau}}\|{\rho_0}) &= {\text{H}_{\alpha}}({\rho_0}\|{\rho_{\tau}})  \nonumber\\
&= \frac{1 - {g_{\alpha}}({\rho_{\tau}},{\rho_0})}{1 - \alpha} \nonumber\\
&=  \frac{{\xi_{\alpha}^-}{\xi_{1 - \alpha}^-} \, |1 - {({\hat{u}_{\tau}}\cdot\hat{r})^2}| \,  {\sin^2}(|{\vec{u}_{\tau}}| ) }{1 - \alpha} ~,
\end{align}
where the first equality follows from the property ${g_{\alpha}}({\rho_0},{\rho_{\tau}}) = {g_{1 - \alpha}}({\rho_{\tau}},{\rho_0}) = {g_{\alpha}}({\rho_{\tau}},{\rho_0})$ which holds for single-qubit states $\rho_0$ and $\rho_{\tau}$ [see Eq.~\eqref{eq:exampleQSL_xxx_005cc}]. From Eqs.~\eqref{eq:idea0000xxxxx010x10002} and~\eqref{eq:idea0000xxxxx010x100033333}, we thus have that [see details in Table~\ref{tab:table001}]
\begin{align}
\label{eq:app000xxx000xxx0013xxx00001}
{\langle\!\langle {\mathcal{G}^{\text{H}}_{\alpha}}(t) \rangle\!\rangle_{\tau}} &= {\Phi_{\alpha}^{\text{H}}}\, {\|{\rho_0^{1 - \alpha}}\|_2} \, {\langle\!\langle {\| [ {H_t}, {\rho_0^{\alpha}} ] \|_2} \rangle\!\rangle_{\tau}}  \nonumber\\
&= \sqrt{2\, {\xi_{2 - 2\alpha}^+}}  \,  {\xi_{\alpha}^-} \frac{1}{\tau} \, {\int_0^{\tau}} \, dt\,  \sqrt{1 - {({\hat{n}_t}\cdot\hat{r})^2}} ~,
\end{align}
and
\begin{align}
\label{eq:app000xxx000xxx0013xxx00002}
 {\langle\!\langle {\mathcal{G}^{\text{H}}_{1 - \alpha}}(t)\rangle\!\rangle_{\tau}} &=  {\Phi_{1 - \alpha}^{\text{H}}}\, {\|{\rho_0^{\alpha}}\|_2} \, {\langle\!\langle {\| [ {H_t}, {\rho_0^{1 - \alpha}} ] \|_2} \rangle\!\rangle_{\tau}} \nonumber\\
&= \sqrt{2{\xi_{2\alpha}^+}}  \, {\xi_{1 - \alpha}^-} \frac{1}{\tau} \, {\int_0^{\tau}} \, dt\,  \sqrt{1 - {({\hat{n}_t}\cdot\hat{r})^2}} ~.
\end{align}
Based on Eqs.~\eqref{eq:app000xxx000xxx00120222},~\eqref{eq:app000xxx000xxx0013xxx00001} and~\eqref{eq:app000xxx000xxx0013xxx00002}, the QSL time for TRE in Eqs.~\eqref{eq:QSL_asymmetric} and~\eqref{eq:QSL_asymmetric2} can be written, respectively, as
\begin{equation}
{\tau_{\alpha}^{\text{H}}}({\rho_{\tau}}\|{\rho_0}) = \frac{{\xi_{1 - \alpha}^-} \, |1 - {({\hat{u}_{\tau}}\cdot\hat{r})^2}| \,  {\sin^2}(|{\vec{u}_{\tau}}| ) }{ \sqrt{2\, {\xi_{2 - 2\alpha}^+}} \left( \frac{1}{\tau} \, {\int_0^{\tau}} \, dt\,  \sqrt{1 - {({\hat{n}_t}\cdot\hat{r})^2}} \, \right) } ~,
\end{equation}
and also
\begin{equation}
{\tau_{\alpha}^{\text{H}}}({\rho_0}\|{\rho_{\tau}}) = \frac{{\xi_{\alpha}^-}\, |1 - {({\hat{u}_{\tau}}\cdot\hat{r})^2}| \,  {\sin^2}(|{\vec{u}_{\tau}}| )}{ \sqrt{2\, {\xi_{2\alpha}^+}}  \left(\frac{1}{\tau} \, {\int_0^{\tau}} \, dt\,  \sqrt{1 - {({\hat{n}_t}\cdot\hat{r})^2}}\, \right) } ~,
\end{equation}
while the QSL time for symmetrized TRE in Eq.~\eqref{eq:QSL_symmetric} reads
\begin{equation}
\label{eq:app000xxx000xxx0014}
{\tau^{\text{H}}_{\alpha}}({\rho_0} : {\rho_{\tau}}) = \frac{\sqrt{2} \, {\xi_{\alpha}^-}{\xi_{1 - \alpha}^-} \, |1 - {({\hat{u}_{\tau}}\cdot\hat{r})^2}| \,  {\sin^2}(|{\vec{u}_{\tau}}| )}{ \left(\sqrt{\xi_{2 - 2\alpha}^+}  \,  {\xi_{\alpha}^-} + \sqrt{\xi_{2\alpha}^+}  \, {\xi_{1 - \alpha}^-} \right)  \left(\frac{1}{\tau} \, {\int_0^{\tau}} \, dt\,  \sqrt{1 - {({\hat{n}_t}\cdot\hat{r})^2}}\, \right)}  ~.
\end{equation}
In Fig.~\ref{fig:figure000xxx00xxxx001app} we plot the QSL time ${\tau^{\text{H}}_{\alpha}} = \max\{\tau^{\text{H}}_{\alpha}(\rho_\tau \| \rho_0), \tau^{\text{H}}_{\alpha} (\rho_0 \| \rho_\tau), {\tau^{\text{H}}_{\alpha}} (\rho_0:\rho_\tau)\}$, as a function of time $\tau$ and $\alpha$, for the initial single qubit state ${\rho_0} = (1/2)(\mathbb{I} + \vec{r}\cdot\vec{\sigma})$ with $\{r,\theta,\phi\} = \{1/4,\pi/4,\pi/4\}$, and also varying the ratio $(\text{a})$ $\Delta/v = 0.5$, $(\text{b})$ $\Delta/v = 1$, $(\text{c})$ $\Delta/v = 5$, and $(\text{d})$ $\Delta/v = 10$. 
\begin{figure*}[!t]
\begin{center}
\includegraphics[scale=0.8]{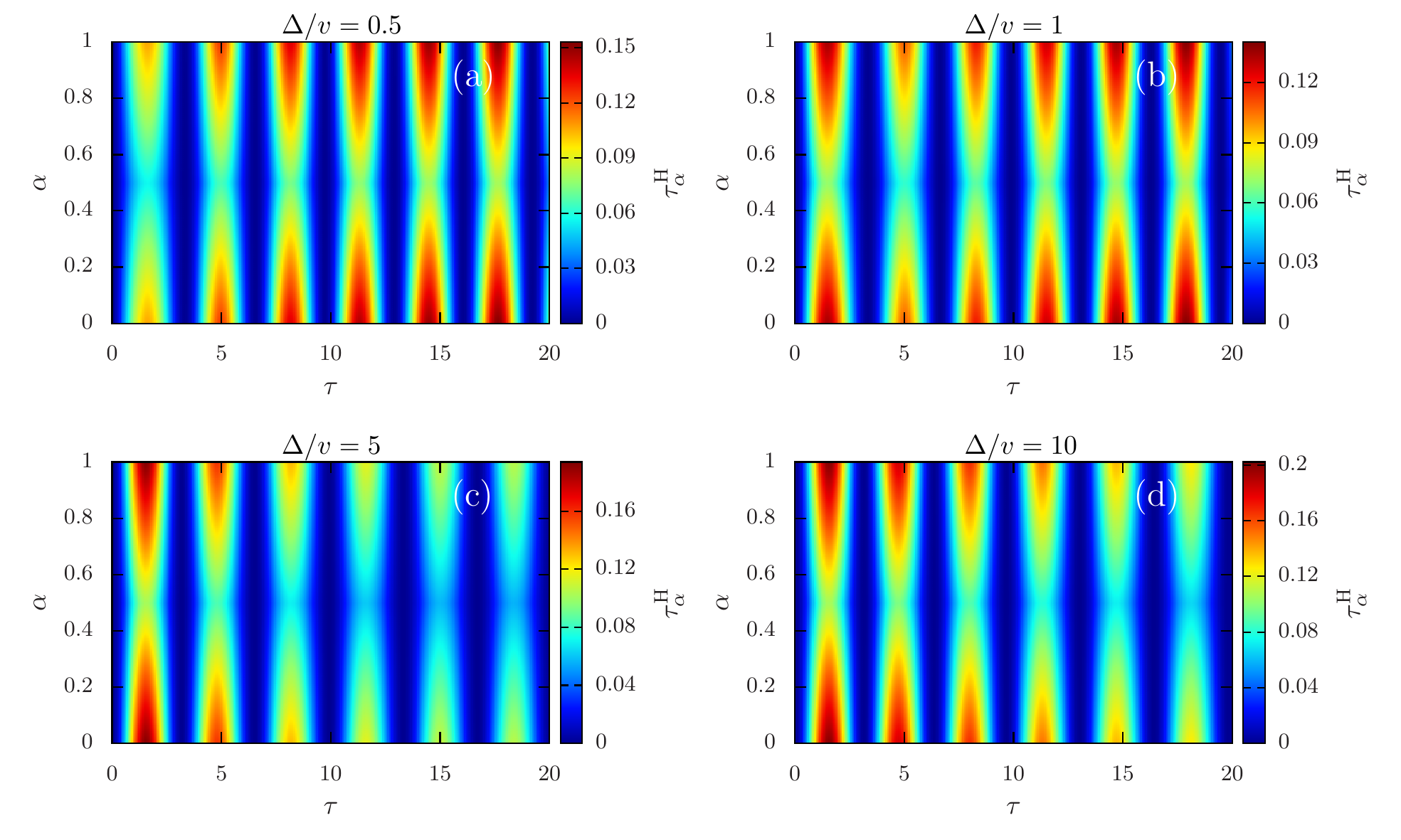}
\caption{(Color online) Density plot of QSL time ${\tau^{\text{H}}_{\alpha}}$, as a function of time $\tau$ and $\alpha$, respective to the unitary evolution generated by the time-dependent Hamiltonian ${H_t} = {\hat{n}_t}\cdot\vec{\sigma}$, with ${\hat{n}_t} = {N^{-1}}\{ \Delta,0,vt \}$ and $N := \sqrt{{\Delta^2} + {(vt)^2}}$. Here we choose the initial single qubit state ${\rho_0} = (1/2)(\mathbb{I} + \vec{r}\cdot\vec{\sigma})$ with $\{r,\theta,\phi\} = \{1/4,\pi/4,\pi/4\}$, and also setting the ratio [Fig.~$(\text{a})$] $\Delta/v = 0.5$, [Fig.~$(\text{b})$] $\Delta/v = 1$, [Fig.~$(\text{c})$] $\Delta/v = 5$, and [Fig.~$(\text{d})$] $\Delta/v = 10$.}
\label{fig:figure000xxx00xxxx001app}
\end{center}
\end{figure*}


\subsection{R\'{e}nyi relative entropy (RRE)}
\label{sec:appF:sec000xxx004a}

We shall begin showing that, for an initial single-qubit state evolving unitarily according the physical setting presented in Sec.~\ref{eq:singlequbit000000001}, the R\'{e}nyi relative entropy is given by
\begin{align}
\label{eq:app000xxx000xxx00000054}
{\text{R}_{\alpha}}({\rho_{\tau}}\|{\rho_0}) &= {\text{R}_{\alpha}}({\rho_0}\|{\rho_{\tau}})  \nonumber\\
&= \frac{ \ln[{g_{\alpha}}({\rho_{\tau}},{\rho_0})]}{ \alpha - 1} \nonumber\\
&= \frac{ \ln \left(1 - {\xi_{\alpha}^-}{\xi_{1 - \alpha}^-} \left(1 - {({\hat{u}_{\tau}}\cdot\hat{r})^2}\right) {\sin^2}(|{\vec{u}_{\tau}}| )\right) }{ \alpha - 1} ~,
\end{align}
where the first equality follows from the identity ${g_{\alpha}}({\rho_0},{\rho_{\tau}}) = {g_{1 - \alpha}}({\rho_{\tau}},{\rho_0}) = {g_{\alpha}}({\rho_{\tau}},{\rho_0})$, which holds for single-qubit states $\rho_0$ and $\rho_{\tau}$ [see Eq.~\eqref{eq:exampleQSL_xxx_005cc}]. From Eqs.~\eqref{eq:idea0000xxxxx010x10002} and~\eqref{eq:idea0000xxxxx010x100033333}, we thus have that [see details in Table~\ref{tab:table001}]
\begin{align}
\label{eq:app000xxx000xxx0000005401}
{\langle\!\langle {\mathcal{G}^{\text{R}}_{\alpha}}(t) \rangle\!\rangle_{\tau}} &= \frac{{\|{\rho_0^{1 - \alpha}}\|_2} \, {\langle\!\langle {\| [ {H_t}, {\rho_0^{\alpha}} ] \|_2} \rangle\!\rangle_{\tau}}}{\left|1 + (1 - \alpha) \ln({{\lambda_{\text{min}}}({\rho_0})}) \right|}  \nonumber\\
&= \frac{{\tau^{-1}}\, \sqrt{2\, {\xi_{2 - 2\alpha}^+}}  \,  {\xi_{\alpha}^-}  \, {\int_0^{\tau}} \, dt\,  \sqrt{1 - {({\hat{n}_t}\cdot\hat{r})^2}}}{\left| 1 + (1 - \alpha)\ln\left(\frac{1 - r}{2}\right) \right|} ~,
\end{align}
and
\begin{align}
\label{eq:app000xxx000xxx0000005402}
 {\langle\!\langle {\mathcal{G}^{\text{R}}_{1 - \alpha}}(t)\rangle\!\rangle_{\tau}} &=  \frac{{\|{\rho_0^{\alpha}}\|_2} \, {\langle\!\langle {\| [ {H_t}, {\rho_0^{1 - \alpha}} ] \|_2} \rangle\!\rangle_{\tau}}}{\left| 1 + \alpha \ln({{\lambda_{\text{min}}}({\rho_0})}) \right|} \nonumber\\
&= \frac{{\tau^{-1}}\, \sqrt{2{\xi_{2\alpha}^+}}  \, {\xi_{1 - \alpha}^-} \, {\int_0^{\tau}} \, dt\,  \sqrt{1 - {({\hat{n}_t}\cdot\hat{r})^2}}}{\left| 1 + \alpha\ln\left(\frac{1 - r}{2}\right) \right|} ~.
\end{align}
Based on Eqs.~\eqref{eq:app000xxx000xxx00000054},~\eqref{eq:app000xxx000xxx0000005401} and~\eqref{eq:app000xxx000xxx0000005402}, the QSL time for RRE in Eqs.~\eqref{eq:QSL_asymmetric} and~\eqref{eq:QSL_asymmetric2} can be written, respectively, as
\begin{equation}
\label{eq:app000xxx000xxx0000005403}
{\tau_{\alpha}^{\text{R}}}({\rho_{\tau}}\|{\rho_0}) = \frac{\left| \ln \left(1 - {\xi_{\alpha}^-}{\xi_{1 - \alpha}^-} \left(1 - {({\hat{u}_{\tau}}\cdot\hat{r})^2}\right) {\sin^2}(|{\vec{u}_{\tau}}| )\right) \right| }{\frac{ \sqrt{2\, {\xi_{2 - 2\alpha}^+}}  \,  {\xi_{\alpha}^-}  \left( \frac{1}{\tau} {\int_0^{\tau}} \, dt\,  \sqrt{1 - {({\hat{n}_t}\cdot\hat{r})^2}} \, \right) }{\left|1 + (1 - \alpha)\ln\left(\frac{1 - r}{2}\right) \right|} } ~,
\end{equation}
and
\begin{equation}
\label{eq:app000xxx000xxx0000005404}
{\tau_{\alpha}^{\text{R}}}({\rho_0}\|{\rho_{\tau}}) = \frac{\left| \ln \left(1 - {\xi_{\alpha}^-}{\xi_{1 - \alpha}^-} \left(1 - {({\hat{u}_{\tau}}\cdot\hat{r})^2}\right) {\sin^2}(|{\vec{u}_{\tau}}| )\right) \right| }{ \frac{ \sqrt{2{\xi_{2\alpha}^+}}  \, {\xi_{1 - \alpha}^-} \left(\frac{1}{\tau} {\int_0^{\tau}} \, dt\,  \sqrt{1 - {({\hat{n}_t}\cdot\hat{r})^2}} \, \right) }{\left| 1 + \alpha\ln\left(\frac{1 - r}{2}\right) \right|} } ~,
\end{equation}
while the QSL time for symmetrized RRE in Eq.~\eqref{eq:QSL_symmetric} reads
\begin{equation}
\label{eq:app000xxx000xxx0000005405}
{\tau^{\text{R}}_{\alpha}}({\rho_0} : {\rho_{\tau}}) = \frac{ \sqrt{2}\,  | \ln \left(1 - {\xi_{\alpha}^-}{\xi_{1 - \alpha}^-} \left(1 - {({\hat{u}_{\tau}}\cdot\hat{r})^2}\right) {\sin^2}(|{\vec{u}_{\tau}}| )\right) | }{ \left( \frac{\sqrt{\xi_{2 - 2\alpha}^+}  \,  {\xi_{\alpha}^-}}{\left| 1 + (1 - \alpha)\ln\left(\frac{1 - r}{2}\right)\right| } + \frac{\sqrt{\xi_{2\alpha}^+}  \, {\xi_{1 - \alpha}^-}}{\left| 1 + \alpha\ln\left(\frac{1 - r}{2}\right)\right| } \right) \left(  \frac{1}{\tau} \, {\int_0^{\tau}} \, dt\,  { \sqrt{1 - {({\hat{n}_t}\cdot\hat{r})^2}} } \, \right)  } ~.
\end{equation}
In Fig.~\ref{fig:figure000xxx00xxxx000app} we plot the QSL time ${\tau^{\text{R}}_{\alpha}} = \max\{\tau^{\text{R}}_{\alpha}(\rho_\tau \| \rho_0), \tau^{\text{R}}_{\alpha} (\rho_0 \| \rho_\tau), {\tau^{\text{R}}_{\alpha}} (\rho_0:\rho_\tau)\}$, as a function of time $\tau$ and $\alpha$, for the initial single qubit state ${\rho_0} = (1/2)(\mathbb{I} + \vec{r}\cdot\vec{\sigma})$ with $\{r,\theta,\phi\} = \{1/4,\pi/4,\pi/4\}$, and also varying the ratio $(\text{a})$ $\Delta/v = 0.5$, $(\text{b})$ $\Delta/v = 1$, $(\text{c})$ $\Delta/v = 5$, and $(\text{d})$ $\Delta/v = 10$.
\begin{figure*}[!t]
\begin{center}
\includegraphics[scale=0.8]{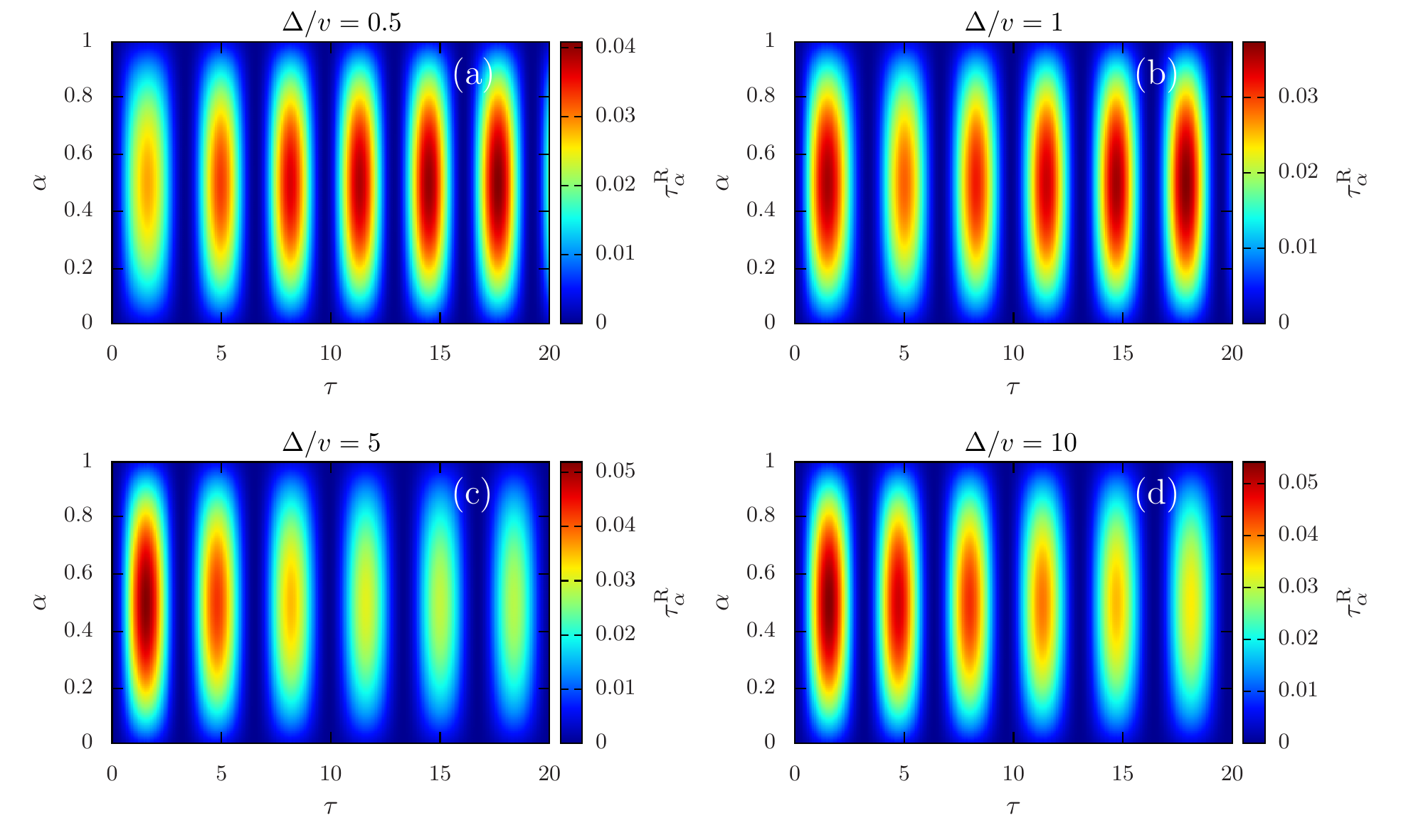}
\caption{(Color online) Density plot of QSL time ${\tau^{\text{R}}_{\alpha}}$, as a function of time $\tau$ and $\alpha$, respective to the unitary evolution generated by the time-dependent Hamiltonian ${H_t} = {\hat{n}_t}\cdot\vec{\sigma}$, with ${\hat{n}_t} = {N^{-1}}\{ \Delta,0,vt \}$ and $N := \sqrt{{\Delta^2} + {(vt)^2}}$. Here we choose the initial single qubit state ${\rho_0} = (1/2)(\mathbb{I} + \vec{r}\cdot\vec{\sigma})$ with $\{r,\theta,\phi\} = \{1/4,\pi/4,\pi/4\}$, and also setting the ratio [Fig.~$(\text{a})$] $\Delta/v = 0.5$, [Fig.~$(\text{b})$] $\Delta/v = 1$, [Fig.~$(\text{c})$] $\Delta/v = 5$, and [Fig.~$(\text{d})$] $\Delta/v = 10$.}
\label{fig:figure000xxx00xxxx000app}
\end{center}
\end{figure*}


\subsection{Relative entropy (RE)}
\label{sec:appF:sec000xxx003}

\begin{figure*}[!t]
\begin{center}
\includegraphics[scale=0.8]{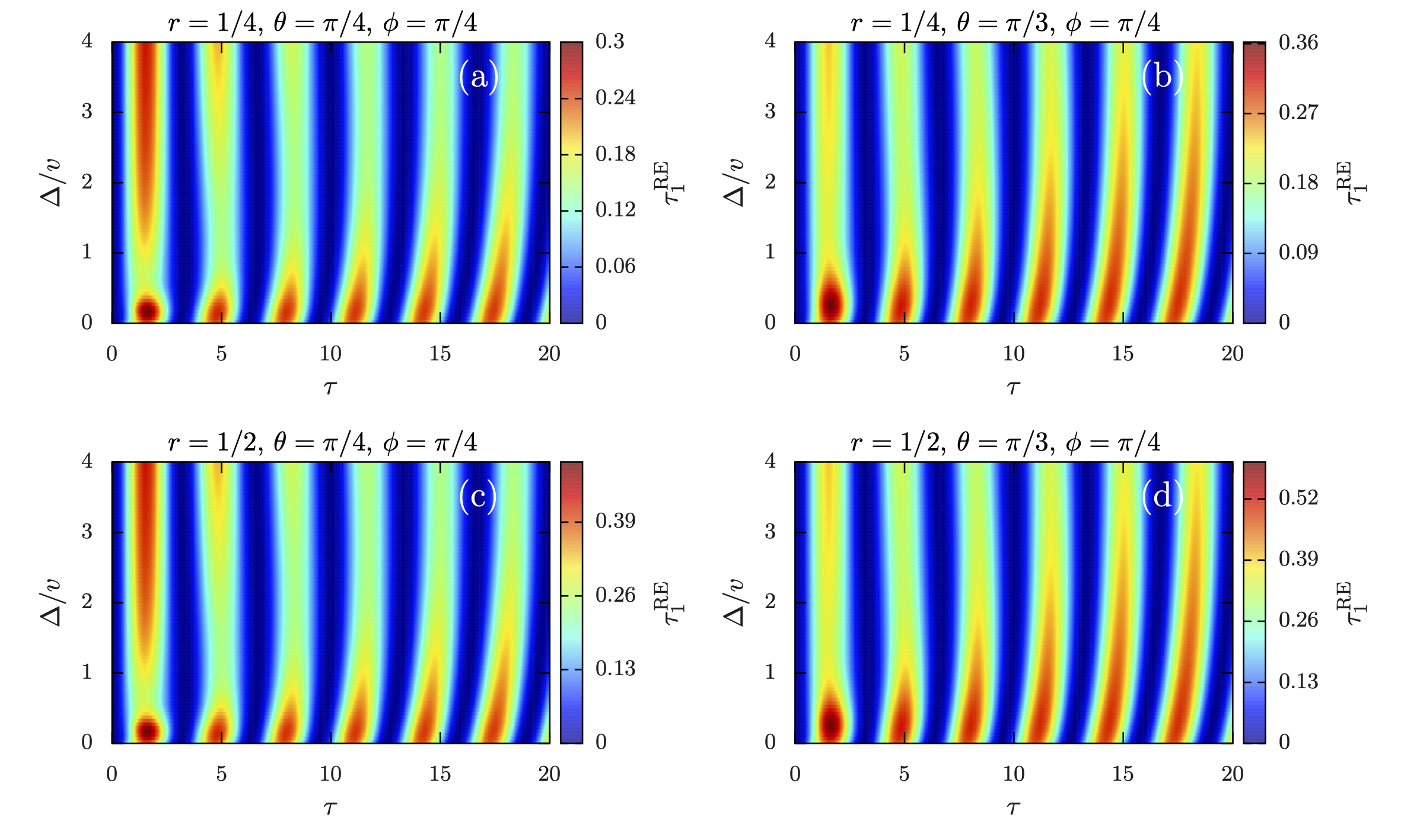}
\caption{(Color online) Density plot of QSL time ${\tau^{\text{RE}}_{1}}$, as a function of time $\tau$ and the ratio $\Delta/v$, respective to the unitary evolution generated by the time-dependent Hamiltonian ${H_t} = {\hat{n}_t}\cdot\vec{\sigma}$, with ${\hat{n}_t} = {N^{-1}}\{ \Delta,0,vt \}$ and $N := \sqrt{{\Delta^2} + {(vt)^2}}$. Here we choose the initial single qubit state ${\rho_0} = (1/2)(\mathbb{I} + \vec{r}\cdot\vec{\sigma})$ with [Fig.~$(\text{a})$] $\{r,\theta,\phi\} = \{1/4,\pi/4,\pi/4\}$, [Fig.~$(\text{b})$] $\{r,\theta,\phi\} = \{1/4,\pi/3,\pi/4\}$, [Fig.~$(\text{c})$] $\{r,\theta,\phi\} = \{1/2,\pi/4,\pi/4\}$, and [Fig.~$(\text{d})$] $\{r,\theta,\phi\} = \{1/2,\pi/3,\pi/4\}$.}
\label{fig:figure000xxx001app}
\end{center}
\end{figure*}
For the case of single-qubit states $\rho_0$ and $\rho_{\tau}$ discussed in Sec.~\ref{eq:singlequbit000000001}, it is possible to show that relative entropy fulfills the identity ${\text{S}}({\rho_{\tau}}\| {\rho_0}) = {\text{S}}({\rho_0}\| {\rho_{\tau}})$, with [see Table~\ref{tab:table001}]
\begin{equation}
\label{eq:app000xxx000xxx0020}
{\text{S}}({\rho_{\tau}}\| {\rho_0}) =  r \ln\left(\frac{1 + r}{1 - r}\right)\left(1 - {({\hat{u}_{\tau}}\cdot\hat{r})^2}\right){\sin^2}( |{\vec{u}_{\tau}}| )  ~.
\end{equation}
Next, by using the results in Table~\ref{tab:table001}, one readily obtains the time averages
\begin{equation}
\label{eq:app000xxx000xxx0021}
 {\langle\!\langle \, {\| [ {H_t} , {\rho_t} ] \|_2} \, \rangle\!\rangle_{\tau}} = \sqrt{2} \, r \, \frac{1}{\tau}\, {\int_0^{\tau}} {dt} \, \sqrt{1 - {({\hat{\mu}_t}\cdot\hat{r})^2} } ~,
\end{equation}
and
\begin{equation}
\label{eq:app000xxx000xxx002102}
 {\langle\!\langle \, {\|[ {H_t} , {\rho_0} ] \|_2} \, \rangle\!\rangle_{\tau}} = \sqrt{2} \, r \, \frac{1}{\tau}\, {\int_0^{\tau}} {dt} \, \sqrt{1 - {({\hat{n}_t}\cdot\hat{r})^2} }  ~.
\end{equation}
Based on Eqs.~\eqref{eq:app000xxx000xxx0020},~\eqref{eq:app000xxx000xxx0021} and~\eqref{eq:app000xxx000xxx002102}, the QSL time related to the relative entropy in Eq.~\eqref{eq:QSL_RE0000} implies that
\begin{equation}
\label{eq:app000xxx000xxx0019}
{\tau^{\text{RE}}_{1}}({\rho_{\tau}} \| {\rho_0}) = \frac{\ln\left(\frac{1 + r}{1 - r}\right)\left(1 - {({\hat{u}_{\tau}}\cdot\hat{r})^2}\right){\sin^2}( |{\vec{u}_{\tau}}| )}{ \sqrt{2} \, \sqrt{{\ln^2}\left(\frac{1 - r}{2}\right) + {\ln^2}\left(\frac{1 + r}{2}\right)} \left( \frac{1}{\tau}\, {\int_0^{\tau}} {dt} \, \sqrt{1 - {({\hat{\mu}_t}\cdot\hat{r})^2} } \right)  } ~,
\end{equation}
and by swapping the arrangement of states $\rho_0$ and $\rho_{\tau}$, it follows
\begin{equation}
\label{eq:app000xxx000xxx001902}
{\tau^{\text{RE}}_{1}}({\rho_0} \| {\rho_{\tau}}) = \frac{\ln\left(\frac{1 + r}{1 - r}\right)\left(1 - {({\hat{u}_{\tau}}\cdot\hat{r})^2}\right){\sin^2}( |{\vec{u}_{\tau}}| )}{\sqrt{2}\,  \sqrt{{\ln^2}\left(\frac{1 - r}{2}\right) + {\ln^2}\left(\frac{1 + r}{2}\right)} \left( \frac{1}{\tau}\, {\int_0^{\tau}} {dt} \, \sqrt{1 - {({\hat{n}_t}\cdot\hat{r})^2} } \right) } ~,
\end{equation}
while the QSL time for the symmetrized relative entropy in Eq.~\eqref{eq:QSL_RE0000v00002} reads
\begin{equation}
\label{eq:app000xxx000xxx0022}
{\tau^{\text{RE}}_{1}}({\rho_0} : {\rho_{\tau}}) = \frac{\sqrt{2} \, \ln\left(\frac{1 + r}{1 - r}\right)\left(1 - {({\hat{u}_{\tau}}\cdot\hat{r})^2}\right){\sin^2}( |{\vec{u}_{\tau}}| )}{\sqrt{{\ln^2}\left(\frac{1 - r}{2}\right) + {\ln^2}\left(\frac{1 + r}{2}\right)}  \left[  \frac{1}{\tau}\, {\int_0^{\tau}} {dt} \, \left( \sqrt{1 - {({\hat{\mu}_t}\cdot\hat{r})^2} } +  \sqrt{1 - {({\hat{n}_t}\cdot\hat{r})^2} } \,  \right) \right]  }   ~.
\end{equation}
In Fig.~\ref{fig:figure000xxx001app} we plot the QSL time ${\tau^{\text{RE}}_1} = \max\{{\tau^{\text{RE}}_1}(\rho_\tau \| \rho_0), {\tau^{\text{RE}}_1} (\rho_0 \| \rho_\tau), {\tau^{\text{RE}}_1} (\rho_0:\rho_\tau)\}$, as a function of time $\tau$ and $\Delta/v$, for the initial single qubit state ${\rho_0} = (1/2)(\mathbb{I} + \vec{r}\cdot\vec{\sigma})$ with $(\text{a})$ $\{r,\theta,\phi\} = \{1/4,\pi/4,\pi/4\}$, $(\text{b})$ $\{r,\theta,\phi\} = \{1/4,\pi/3,\pi/4\}$, $(\text{c})$ $\{r,\theta,\phi\} = \{1/2,\pi/4,\pi/4\}$, and $(\text{d})$ $\{r,\theta,\phi\} = \{1/2,\pi/3,\pi/4\}$.


\subsection{Min-relative entropy}
\label{sec:appF:sec000xxx004}

Here we will present the details of the QSL time related to the min-relative entropy, with the latter defined as $\text{R}_0(\rho\|\omega) = -\ln \text{Tr}(\Pi_{\rho}\, \omega)$, with ${\Pi_{\rho}}$ being the projector onto the support of the state $\rho$. From now on, we will choose the initial state $\rho_0$ being a pure one, i.e., a non -full-rank density matrix. In particular, for a single-qubit state such a condition is equivalent to imposing the purity value $r = 1$, i.e., $\rho_0 = (1/2)(\mathbb{I} + \hat{r}\cdot\vec{\sigma})$, which in turn implies the spectral decomposition ${\rho_0} = {\sum_{\ell = \pm}} \, {p_{\ell}} |{\psi_{\ell}}\rangle\langle{\psi_{\ell}}|$, with eigenvalues ${p_+} = 1$ and ${p_-} = 0$, and eigenstates $|{\psi_+}\rangle = |\theta,\phi\rangle$ and $|{\psi_-}\rangle = |\theta - \pi,\phi\rangle$, with $|\theta,\phi\rangle := \cos(\theta/2)|0\rangle + {e^{-i\phi}}\sin(\theta/2)|1\rangle$. Just to clarify, here $|0\rangle = {\left( 1 \quad 0 \right)}^{\textsf{T}}$ and $|1\rangle = {\left( 0 \quad 1 \right)}^{\textsf{T}}$ define the standard states of the computational basis. In this case, the projector onto the support of $\rho_0$ read as ${\Pi_{\rho_0}} = |{\psi_+}\rangle\langle{\psi_+}|$. Moving forward, one may proceed the calculation as follows
\begin{align}
\label{eq:app000xxx000xxx0025}
\text{Tr}({\Pi_{\rho_0}}{U_{\tau}}{\rho_0}{U_{\tau}^{\dagger}}) &= \text{Tr}({\rho_0}{U_{\tau}}{\Pi_{\rho_0}}{U_{\tau}^{\dagger}}) \nonumber\\
& = {|\langle{\psi_+}|{U_{\tau}} |{\psi_+}\rangle|^2} \nonumber\\
&= 1 - (1 - {({\hat{u}_t}\cdot\hat{r})^2})\, {\sin^2}(|{\vec{u}_t}|) ~.
\end{align}
Next, by using the result of Eq.~\eqref{eq:app000xxx000xxx0025}, the min-relative entropy becomes
\begin{align}
\label{eq:app000xxx000xxx0026}
{\text{R}_0}({\rho_{\tau}}\|{\rho_0}) &= {\text{R}_0}({\rho_0}\|{\rho_{\tau}}) \nonumber\\
&= -  \ln ( 1 - (1 - {({\hat{u}_t}\cdot\hat{r})^2})\, {\sin^2}(|{\vec{u}_t}|) ) ~.
\end{align}
Furthermore, given that $\langle{\psi_+}|{({U_t^{\dagger}}{H_t}{U_t})^2}|{\psi_+}\rangle = 1 + {\varpi^2} + 2\varpi ({\hat{\mu}_t}\cdot\hat{r})$ and also ${\langle{\psi_+}|{U_t^{\dagger}}{H_t}{U_t}|{\psi_+}\rangle} = \varpi + {\hat{\mu}_t}\cdot\hat{r} $, one obtains
\begin{align}
\label{eq:app000xxx000xxx0027}
{\| [{U_t^{\dagger}}{H_t}{U_t},{\rho_0}] \|_2^2} &= {\| [{U_t^{\dagger}}{H_t}{U_t},{\Pi_{\rho_0}}] \|_2^2} \nonumber\\
&= 2 \left( \langle{\psi_+}|{({U_t^{\dagger}}{H_t}{U_t})^2}|{\psi_+}\rangle - {\langle{\psi_+}|{U_t^{\dagger}}{H_t}{U_t}|{\psi_+}\rangle^2} \right) \nonumber\\
&= 2\left(1 - {({\hat{\mu}_t}\cdot\hat{r})^2}\right) ~.
\end{align}
\begin{figure*}[!t]
\begin{center}
\includegraphics[scale=0.8]{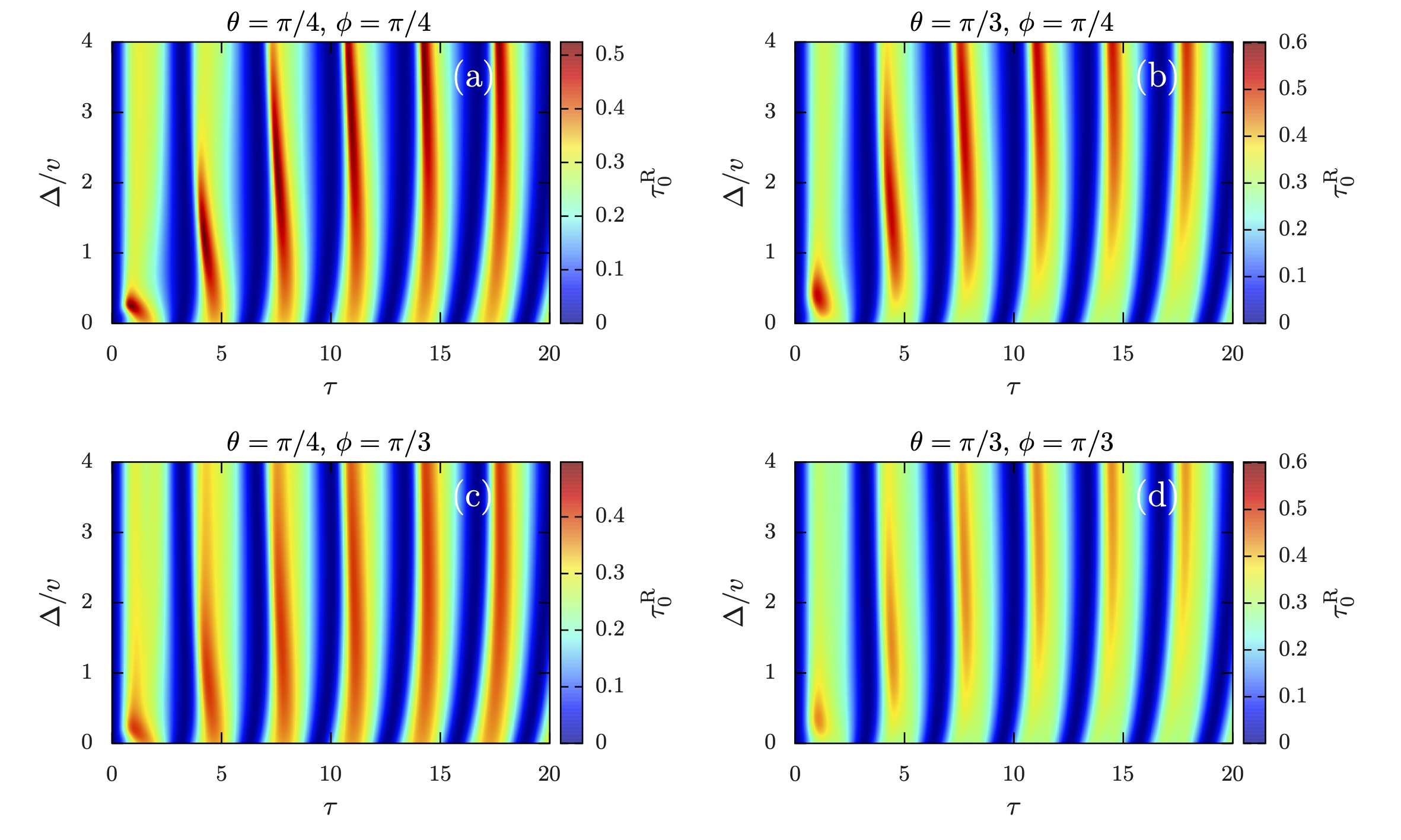}
\caption{(Color online) Density plot of QSL time ${\tau^{\text{R}}_{0}}$, as a function of time $\tau$ and the ratio $\Delta/v$, respective to the unitary evolution generated by the time-dependent Hamiltonian ${H_t} = {\hat{n}_t}\cdot\vec{\sigma}$, with ${\hat{n}_t} = {N^{-1}}\{ \Delta,0,vt \}$ and $N := \sqrt{{\Delta^2} + {(vt)^2}}$. Here we choose the initial pure single qubit state ${\rho_0} = (1/2)(\mathbb{I} + \hat{r}\cdot\vec{\sigma})$ with $r = 1$ and [Fig.~$(\text{a})$] $\{\theta,\phi\} = \{\pi/4,\pi/4\}$; [Fig.~$(\text{b})$] $\{\theta,\phi\} = \{\pi/3,\pi/4\}$; [Fig.~$(\text{c})$] $\{\theta,\phi\} = \{\pi/4,\pi/3\}$; and [Fig.~$(\text{d})$] $\{\theta,\phi\} = \{\pi/3,\pi/3\}$.}
\label{fig:figure000xxx004app}
\end{center}
\end{figure*}
From Eqs.~\eqref{eq:app000xxx000xxx0025}, and~\eqref{eq:app000xxx000xxx0027}, and also using the Schatten $2$-norms ${\|{\rho_0}\|_2} = {\| \Pi_{\rho_0} \|_2} = 1$, we thus conclude that ${\mathcal{Q}_0^t}({\rho_0},{\Pi_{\rho_0}}) = {\mathcal{Q}_0^t}({\Pi_{\rho_0}},{\rho_0})$ [see Eq.~\eqref{eq:min000xxx000x1201aaaaa}], which implies the following
\begin{align}
\label{eq:app000xxx000xxx0028}
{\langle\!\langle {\mathcal{Q}_0^t}({\rho_0},{\Pi_{\rho_0}}) \rangle\!\rangle_{\tau}} &= {\langle\!\langle  {\mathcal{Q}_0^t}({\Pi_{\rho_0}},{\rho_0})  \rangle\!\rangle_{\tau}} \nonumber\\
&= \frac{\sqrt{2}}{\tau} \, {\int_0^{\tau}} dt\, \frac{\sqrt{1 - {({\hat{\mu}_t}\cdot\hat{r})^2}} }{| 1 - (1 - {({\hat{u}_t}\cdot\hat{r})^2})\, {\sin^2}(|{\vec{u}_t}|) |} ~.
\end{align}
Based on Eqs.~\eqref{eq:app000xxx000xxx0026} and~\eqref{eq:app000xxx000xxx0028}, the QSL time for min-entropy defined in Eqs.~\eqref{eq:min000xxx000x1200002a} and~\eqref{eq:min000xxx000x1200002b} satisfy the following constraint
\begin{align}
{\tau_{0}^{\text{R}}}({\rho_{\tau}} \| {\rho_0})  &= {\tau_{0}^{\text{R}}}({\rho_0} \| {\rho_{\tau}}) \nonumber\\
&= \frac{ | \ln ( 1 - (1 - {({\hat{u}_t}\cdot\hat{r})^2})\, {\sin^2}(|{\vec{u}_t}|) ) |}{\sqrt{2} \left( \frac{1}{\tau} \, {\int_0^{\tau}} dt\, \frac{\sqrt{1 - {({\hat{\mu}_t}\cdot\hat{r})^2}} }{| 1 - (1 - {({\hat{u}_t}\cdot\hat{r})^2})\, {\sin^2}(|{\vec{u}_t}|) |} \, \right)} ~,
\end{align}
while the QSL time for the symmetrized min-relative entropy reads [see Eq.~\eqref{eq:min000xxx000x1200002c}]
\begin{equation}
\label{eq:app000xxx000xxx0029}
{\tau^{\text{R}}_{0}}({\rho_0} : {\rho_{\tau}}) = \frac{ | \ln ( 1 - (1 - {({\hat{u}_t}\cdot\hat{r})^2})\, {\sin^2}(|{\vec{u}_t}|) ) |}{\sqrt{2} \left( \frac{1}{\tau} \, {\int_0^{\tau}} dt\, \frac{\sqrt{1 - {({\hat{\mu}_t}\cdot\hat{r})^2}} }{| 1 - (1 - {({\hat{u}_t}\cdot\hat{r})^2})\, {\sin^2}(|{\vec{u}_t}|) |} \, \right)}  ~.
\end{equation}
In Fig.~\ref{fig:figure000xxx004app} we plot the QSL time ${\tau^{\text{R}}_{0}} = \max\{\tau_{0}^{\text{R}}(\rho_\tau \| \rho_0), {\tau_{0}^{\text{R}}}(\rho_0 \| \rho_\tau), \tau_{0}^{\text{R}}({\rho_0} : {\rho_{\tau}}) \}$, as a function of time $\tau$ and $\Delta/v$, for the initial pure single qubit state ${\rho_0} = (1/2)(\mathbb{I} + \hat{r}\cdot\vec{\sigma})$ with $r = 1$ and $(\text{a})$ $\{\theta,\phi\} = \{\pi/4,\pi/4\}$; $(\text{b})$ $\{\theta,\phi\} = \{\pi/3,\pi/4\}$; $(\text{c})$ $\{\theta,\phi\} = \{\pi/4,\pi/3\}$; and $(\text{d})$ $\{\theta,\phi\} = \{\pi/3,\pi/3\}$.
\end{widetext}



\begin{thebibliography}{88}%
\makeatletter
\providecommand \@ifxundefined [1]{%
 \@ifx{#1\undefined}
}%
\providecommand \@ifnum [1]{%
 \ifnum #1\expandafter \@firstoftwo
 \else \expandafter \@secondoftwo
 \fi
}%
\providecommand \@ifx [1]{%
 \ifx #1\expandafter \@firstoftwo
 \else \expandafter \@secondoftwo
 \fi
}%
\providecommand \natexlab [1]{#1}%
\providecommand \enquote  [1]{``#1''}%
\providecommand \bibnamefont  [1]{#1}%
\providecommand \bibfnamefont [1]{#1}%
\providecommand \citenamefont [1]{#1}%
\providecommand \href@noop [0]{\@secondoftwo}%
\providecommand \href [0]{\begingroup \@sanitize@url \@href}%
\providecommand \@href[1]{\@@startlink{#1}\@@href}%
\providecommand \@@href[1]{\endgroup#1\@@endlink}%
\providecommand \@sanitize@url [0]{\catcode `\\12\catcode `\$12\catcode
  `\&12\catcode `\#12\catcode `\^12\catcode `\_12\catcode `\%12\relax}%
\providecommand \@@startlink[1]{}%
\providecommand \@@endlink[0]{}%
\providecommand \url  [0]{\begingroup\@sanitize@url \@url }%
\providecommand \@url [1]{\endgroup\@href {#1}{\urlprefix }}%
\providecommand \urlprefix  [0]{URL }%
\providecommand \Eprint [0]{\href }%
\providecommand \doibase [0]{http://dx.doi.org/}%
\providecommand \selectlanguage [0]{\@gobble}%
\providecommand \bibinfo  [0]{\@secondoftwo}%
\providecommand \bibfield  [0]{\@secondoftwo}%
\providecommand \translation [1]{[#1]}%
\providecommand \BibitemOpen [0]{}%
\providecommand \bibitemStop [0]{}%
\providecommand \bibitemNoStop [0]{.\EOS\space}%
\providecommand \EOS [0]{\spacefactor3000\relax}%
\providecommand \BibitemShut  [1]{\csname bibitem#1\endcsname}%
\let\auto@bib@innerbib\@empty
\bibitem [{\citenamefont {Shannon}(1948)}]{Shannon1948}%
  \BibitemOpen
  \bibfield  {author} {\bibinfo {author} {\bibfnamefont {C.~E.}\ \bibnamefont
  {Shannon}},\ }\bibfield  {title} {\enquote {\bibinfo {title} {A
  {M}athematical {T}heory of {C}ommunication},}\ }\href
  {https://ieeexplore.ieee.org/document/6773024} {\bibfield  {journal}
  {\bibinfo  {journal} {The Bell System Technical Journal}\ }\textbf {\bibinfo
  {volume} {27}},\ \bibinfo {pages} {379} (\bibinfo {year} {1948})}\BibitemShut
  {NoStop}%
\bibitem [{\citenamefont {Dimitrov}\ \emph {et~al.}(2011)\citenamefont
  {Dimitrov}, \citenamefont {Lazar},\ and\ \citenamefont
  {Victor}}]{Dimitrov2011}%
  \BibitemOpen
  \bibfield  {author} {\bibinfo {author} {\bibfnamefont {A.~G.}\ \bibnamefont
  {Dimitrov}}, \bibinfo {author} {\bibfnamefont {A.~A.}\ \bibnamefont {Lazar}},
  \ and\ \bibinfo {author} {\bibfnamefont {J.~D.}\ \bibnamefont {Victor}},\
  }\bibfield  {title} {\enquote {\bibinfo {title} {Information theory in
  neuroscience},}\ }\href {\doibase 10.1007/s10827-011-0314-3} {\bibfield
  {journal} {\bibinfo  {journal} {J. Comput. Neurosci.}\ }\textbf {\bibinfo
  {volume} {30}},\ \bibinfo {pages} {1} (\bibinfo {year} {2011})}\BibitemShut
  {NoStop}%
\bibitem [{\citenamefont {Kempf}(2018)}]{Kempf2018}%
  \BibitemOpen
  \bibfield  {author} {\bibinfo {author} {\bibfnamefont {A.}~\bibnamefont
  {Kempf}},\ }\bibfield  {title} {\enquote {\bibinfo {title} {Quantum
  {G}ravity, {I}nformation {T}heory and the {CMB}},}\ }\href {\doibase
  10.1007/s10701-018-0163-2} {\bibfield  {journal} {\bibinfo  {journal} {Found.
  Phys.}\ }\textbf {\bibinfo {volume} {48}},\ \bibinfo {pages} {1191} (\bibinfo
  {year} {2018})}\BibitemShut {NoStop}%
\bibitem [{\citenamefont {Konishi}(2020)}]{Konishi_2020}%
  \BibitemOpen
  \bibfield  {author} {\bibinfo {author} {\bibfnamefont {E.}~\bibnamefont
  {Konishi}},\ }\bibfield  {title} {\enquote {\bibinfo {title} {Holographic
  interpretation of {S}hannon entropy of coherence of quantum pure states},}\
  }\href {\doibase 10.1209/0295-5075/129/11006} {\bibfield  {journal} {\bibinfo
   {journal} {{EPL}}\ }\textbf {\bibinfo {volume} {129}},\ \bibinfo {pages}
  {11006} (\bibinfo {year} {2020})}\BibitemShut {NoStop}%
\bibitem [{\citenamefont {Goold}\ \emph {et~al.}(2016)\citenamefont {Goold},
  \citenamefont {Huber}, \citenamefont {Riera}, \citenamefont {del Rio},\ and\
  \citenamefont {Skrzypczyk}}]{Goold2016}%
  \BibitemOpen
  \bibfield  {author} {\bibinfo {author} {\bibfnamefont {J.}~\bibnamefont
  {Goold}}, \bibinfo {author} {\bibfnamefont {M.}~\bibnamefont {Huber}},
  \bibinfo {author} {\bibfnamefont {A.}~\bibnamefont {Riera}}, \bibinfo
  {author} {\bibfnamefont {L.}~\bibnamefont {del Rio}}, \ and\ \bibinfo
  {author} {\bibfnamefont {P.}~\bibnamefont {Skrzypczyk}},\ }\bibfield  {title}
  {\enquote {\bibinfo {title} {The role of quantum information in
  thermodynamics{\textemdash}a topical review},}\ }\href {\doibase
  10.1088/1751-8113/49/14/143001} {\bibfield  {journal} {\bibinfo  {journal}
  {J. Phys. A: Math. Theor.}\ }\textbf {\bibinfo {volume} {49}},\ \bibinfo
  {pages} {143001} (\bibinfo {year} {2016})}\BibitemShut {NoStop}%
\bibitem [{\citenamefont {Zhou}\ \emph {et~al.}(2013)\citenamefont {Zhou},
  \citenamefont {Cai},\ and\ \citenamefont {Tong}}]{Zhou2013}%
  \BibitemOpen
  \bibfield  {author} {\bibinfo {author} {\bibfnamefont {R.}~\bibnamefont
  {Zhou}}, \bibinfo {author} {\bibfnamefont {R.}~\bibnamefont {Cai}}, \ and\
  \bibinfo {author} {\bibfnamefont {G.}~\bibnamefont {Tong}},\ }\bibfield
  {title} {\enquote {\bibinfo {title} {Applications of entropy in finance: {A}
  review},}\ }\href {\doibase 10.3390/e15114909} {\bibfield  {journal}
  {\bibinfo  {journal} {Entropy}\ }\textbf {\bibinfo {volume} {15}},\ \bibinfo
  {pages} {4909} (\bibinfo {year} {2013})}\BibitemShut {NoStop}%
\bibitem [{\citenamefont {Seoane}\ and\ \citenamefont
  {Sol\'{e}}(2018)}]{Seoane2018}%
  \BibitemOpen
  \bibfield  {author} {\bibinfo {author} {\bibfnamefont {L.~F.}\ \bibnamefont
  {Seoane}}\ and\ \bibinfo {author} {\bibfnamefont {R.~V.}\ \bibnamefont
  {Sol\'{e}}},\ }\bibfield  {title} {\enquote {\bibinfo {title} {Information
  theory, predictability and the emergence of complex life},}\ }\href {\doibase
  10.1098/rsos.172221} {\bibfield  {journal} {\bibinfo  {journal} {R. Soc. Open
  Sci.}\ }\textbf {\bibinfo {volume} {5}},\ \bibinfo {pages} {172221} (\bibinfo
  {year} {2018})}\BibitemShut {NoStop}%
\bibitem [{\citenamefont {Shiraishi}\ \emph {et~al.}(2018)\citenamefont
  {Shiraishi}, \citenamefont {Funo},\ and\ \citenamefont
  {Saito}}]{PhysRevLett.121.070601}%
  \BibitemOpen
  \bibfield  {author} {\bibinfo {author} {\bibfnamefont {N.}~\bibnamefont
  {Shiraishi}}, \bibinfo {author} {\bibfnamefont {K.}~\bibnamefont {Funo}}, \
  and\ \bibinfo {author} {\bibfnamefont {K.}~\bibnamefont {Saito}},\ }\bibfield
   {title} {\enquote {\bibinfo {title} {Speed {L}imit for {C}lassical
  {S}tochastic {P}rocesses},}\ }\href {\doibase 10.1103/PhysRevLett.121.070601}
  {\bibfield  {journal} {\bibinfo  {journal} {Phys. Rev. Lett.}\ }\textbf
  {\bibinfo {volume} {121}},\ \bibinfo {pages} {070601} (\bibinfo {year}
  {2018})}\BibitemShut {NoStop}%
\bibitem [{\citenamefont {Tsallis}(1988)}]{Tsallis1988}%
  \BibitemOpen
  \bibfield  {author} {\bibinfo {author} {\bibfnamefont {C.}~\bibnamefont
  {Tsallis}},\ }\bibfield  {title} {\enquote {\bibinfo {title} {Possible
  generalization of {B}oltzmann-{G}ibbs statistics},}\ }\href {\doibase
  10.1007/BF01016429} {\bibfield  {journal} {\bibinfo  {journal} {J. Stat.
  Phys.}\ }\textbf {\bibinfo {volume} {52}},\ \bibinfo {pages} {479} (\bibinfo
  {year} {1988})}\BibitemShut {NoStop}%
\bibitem [{\citenamefont {R\'{e}nyi}(1961)}]{Renyi1961}%
  \BibitemOpen
  \bibfield  {author} {\bibinfo {author} {\bibfnamefont {A.}~\bibnamefont
  {R\'{e}nyi}},\ }\bibfield  {title} {\enquote {\bibinfo {title} {{O}n
  {M}easures of {E}ntropy and {I}nformation},}\ }in\ \href
  {https://projecteuclid.org/euclid.bsmsp/1200512181} {\emph {\bibinfo
  {booktitle} {Proc. Fourth Berkeley Symp. on Math. Statist. and Prob., {V}ol.
  1: {C}ontributions to the {T}heory of {S}tatistics}}}\ (\bibinfo  {publisher}
  {University of {C}a\-li\-for\-nia {P}ress},\ \bibinfo {address} {Berkeley},\
  \bibinfo {year} {1961})\ pp.\ \bibinfo {pages} {547--561}\BibitemShut
  {NoStop}%
\bibitem [{\citenamefont {Umegaki}(1962)}]{umegaki1962}%
  \BibitemOpen
  \bibfield  {author} {\bibinfo {author} {\bibfnamefont {H.}~\bibnamefont
  {Umegaki}},\ }\bibfield  {title} {\enquote {\bibinfo {title} {Conditional
  expectation in an operator algebra. {I}{V}. {E}ntropy and information},}\
  }\href {\doibase 10.2996/kmj/1138844604} {\bibfield  {journal} {\bibinfo
  {journal} {Kodai Math. Sem. Rep.}\ }\textbf {\bibinfo {volume} {14}},\
  \bibinfo {pages} {59} (\bibinfo {year} {1962})}\BibitemShut {NoStop}%
\bibitem [{\citenamefont {van Erven}\ and\ \citenamefont
  {Harremos}(2014)}]{6832827ErvenHarremos}%
  \BibitemOpen
  \bibfield  {author} {\bibinfo {author} {\bibfnamefont {T.}~\bibnamefont {van
  Erven}}\ and\ \bibinfo {author} {\bibfnamefont {P.}~\bibnamefont
  {Harremos}},\ }\bibfield  {title} {\enquote {\bibinfo {title} {R\'{e}nyi
  {D}ivergence and {K}ullback-{L}eibler {D}ivergence},}\ }\href {\doibase
  10.1109/TIT.2014.2320500} {\bibfield  {journal} {\bibinfo  {journal} {IEEE
  Trans. Inf. Theory}\ }\textbf {\bibinfo {volume} {60}},\ \bibinfo {pages}
  {3797} (\bibinfo {year} {2014})}\BibitemShut {NoStop}%
\bibitem [{\citenamefont {Kawai}\ \emph {et~al.}(2007)\citenamefont {Kawai},
  \citenamefont {Parrondo},\ and\ \citenamefont {Van~den Broeck}}]{Kawai2007}%
  \BibitemOpen
  \bibfield  {author} {\bibinfo {author} {\bibfnamefont {R.}~\bibnamefont
  {Kawai}}, \bibinfo {author} {\bibfnamefont {J.~M.~R.}\ \bibnamefont
  {Parrondo}}, \ and\ \bibinfo {author} {\bibfnamefont {C.}~\bibnamefont
  {Van~den Broeck}},\ }\bibfield  {title} {\enquote {\bibinfo {title}
  {Dissipation: {T}he {P}hase-{S}pace {P}erspective},}\ }\href {\doibase
  10.1103/PhysRevLett.98.080602} {\bibfield  {journal} {\bibinfo  {journal}
  {Phys. Rev. Lett.}\ }\textbf {\bibinfo {volume} {98}},\ \bibinfo {pages}
  {080602} (\bibinfo {year} {2007})}\BibitemShut {NoStop}%
\bibitem [{\citenamefont {Vedral}\ \emph {et~al.}(1997)\citenamefont {Vedral},
  \citenamefont {Plenio}, \citenamefont {Rippin},\ and\ \citenamefont
  {Knight}}]{Vedral1997}%
  \BibitemOpen
  \bibfield  {author} {\bibinfo {author} {\bibfnamefont {V.}~\bibnamefont
  {Vedral}}, \bibinfo {author} {\bibfnamefont {M.~B.}\ \bibnamefont {Plenio}},
  \bibinfo {author} {\bibfnamefont {M.~A.}\ \bibnamefont {Rippin}}, \ and\
  \bibinfo {author} {\bibfnamefont {P.~L.}\ \bibnamefont {Knight}},\ }\bibfield
   {title} {\enquote {\bibinfo {title} {Quantifying {E}ntanglement},}\ }\href
  {\doibase 10.1103/PhysRevLett.78.2275} {\bibfield  {journal} {\bibinfo
  {journal} {Phys. Rev. Lett.}\ }\textbf {\bibinfo {volume} {78}},\ \bibinfo
  {pages} {2275} (\bibinfo {year} {1997})}\BibitemShut {NoStop}%
\bibitem [{\citenamefont {Bu}\ \emph {et~al.}(2017)\citenamefont {Bu},
  \citenamefont {Singh}, \citenamefont {Fei}, \citenamefont {Pati},\ and\
  \citenamefont {Wu}}]{Bu2017}%
  \BibitemOpen
  \bibfield  {author} {\bibinfo {author} {\bibfnamefont {K.}~\bibnamefont
  {Bu}}, \bibinfo {author} {\bibfnamefont {U.}~\bibnamefont {Singh}}, \bibinfo
  {author} {\bibfnamefont {S.-M.}\ \bibnamefont {Fei}}, \bibinfo {author}
  {\bibfnamefont {A.~K.}\ \bibnamefont {Pati}}, \ and\ \bibinfo {author}
  {\bibfnamefont {J.}~\bibnamefont {Wu}},\ }\bibfield  {title} {\enquote
  {\bibinfo {title} {{M}aximum {R}elative {E}ntropy of {C}oherence: {A}n
  {O}perational {C}oherence {M}easure},}\ }\href {\doibase
  10.1103/PhysRevLett.119.150405} {\bibfield  {journal} {\bibinfo  {journal}
  {Phys. Rev. Lett.}\ }\textbf {\bibinfo {volume} {119}},\ \bibinfo {pages}
  {150405} (\bibinfo {year} {2017})}\BibitemShut {NoStop}%
\bibitem [{\citenamefont {Shao}\ \emph {et~al.}(2017)\citenamefont {Shao},
  \citenamefont {Li}, \citenamefont {Luo},\ and\ \citenamefont
  {Xi}}]{0253-6102-67-6-631}%
  \BibitemOpen
  \bibfield  {author} {\bibinfo {author} {\bibfnamefont {L-H}\ \bibnamefont
  {Shao}}, \bibinfo {author} {\bibfnamefont {Y.-M.}\ \bibnamefont {Li}},
  \bibinfo {author} {\bibfnamefont {Y.}~\bibnamefont {Luo}}, \ and\ \bibinfo
  {author} {\bibfnamefont {Z.-J.}\ \bibnamefont {Xi}},\ }\bibfield  {title}
  {\enquote {\bibinfo {title} {Quantum {C}oherence {Q}uantifiers {B}ased on
  {R}\'{e}nyi $\alpha$-{R}elative {E}ntropy},}\ }\href {\doibase
  10.1088/0253-6102/67/6/631} {\bibfield  {journal} {\bibinfo  {journal}
  {Commun. Theor. Phys.}\ }\textbf {\bibinfo {volume} {67}},\ \bibinfo {pages}
  {631} (\bibinfo {year} {2017})}\BibitemShut {NoStop}%
\bibitem [{\citenamefont {Shiraishi}\ and\ \citenamefont
  {Saito}(2019)}]{PhysRevLett.123.110603}%
  \BibitemOpen
  \bibfield  {author} {\bibinfo {author} {\bibfnamefont {N.}~\bibnamefont
  {Shiraishi}}\ and\ \bibinfo {author} {\bibfnamefont {K.}~\bibnamefont
  {Saito}},\ }\bibfield  {title} {\enquote {\bibinfo {title}
  {Information-{T}heoretical {B}ound of the {I}rreversibility in {T}hermal
  {R}elaxation {P}rocesses},}\ }\href {\doibase 10.1103/PhysRevLett.123.110603}
  {\bibfield  {journal} {\bibinfo  {journal} {Phys. Rev. Lett.}\ }\textbf
  {\bibinfo {volume} {123}},\ \bibinfo {pages} {110603} (\bibinfo {year}
  {2019})}\BibitemShut {NoStop}%
\bibitem [{\citenamefont {Kolchinsky}\ and\ \citenamefont
  {Wolpert}(2020)}]{arXiv:2008.10764}%
  \BibitemOpen
  \bibfield  {author} {\bibinfo {author} {\bibfnamefont {A.}~\bibnamefont
  {Kolchinsky}}\ and\ \bibinfo {author} {\bibfnamefont {D.~H.}\ \bibnamefont
  {Wolpert}},\ }\bibfield  {title} {\enquote {\bibinfo {title} {Entropy
  production and thermodynamics of information under protocol constraints},}\
  }\href {https://www.arxiv.org/abs/2008.10764} {\bibfield  {journal} {\bibinfo
   {journal} {arXiv:2008.10764}\ } (\bibinfo {year} {2020})}\BibitemShut
  {NoStop}%
\bibitem [{\citenamefont {Scandi}\ \emph {et~al.}(2020)\citenamefont {Scandi},
  \citenamefont {Miller}, \citenamefont {Anders},\ and\ \citenamefont
  {Perarnau-Llobet}}]{PhysRevResearch.2.023377}%
  \BibitemOpen
  \bibfield  {author} {\bibinfo {author} {\bibfnamefont {M.}~\bibnamefont
  {Scandi}}, \bibinfo {author} {\bibfnamefont {H.~J.~D.}\ \bibnamefont
  {Miller}}, \bibinfo {author} {\bibfnamefont {J.}~\bibnamefont {Anders}}, \
  and\ \bibinfo {author} {\bibfnamefont {M.}~\bibnamefont {Perarnau-Llobet}},\
  }\bibfield  {title} {\enquote {\bibinfo {title} {Quantum work statistics
  close to equilibrium},}\ }\href {\doibase 10.1103/PhysRevResearch.2.023377}
  {\bibfield  {journal} {\bibinfo  {journal} {Phys. Rev. Research}\ }\textbf
  {\bibinfo {volume} {2}},\ \bibinfo {pages} {023377} (\bibinfo {year}
  {2020})}\BibitemShut {NoStop}%
\bibitem [{\citenamefont {Marvian}\ and\ \citenamefont
  {Spekkens}(2014)}]{Marvian2014}%
  \BibitemOpen
  \bibfield  {author} {\bibinfo {author} {\bibfnamefont {I.}~\bibnamefont
  {Marvian}}\ and\ \bibinfo {author} {\bibfnamefont {R.~W.}\ \bibnamefont
  {Spekkens}},\ }\bibfield  {title} {\enquote {\bibinfo {title} {Extending
  {N}oether's theorem by quantifying the asymmetry of quantum states},}\ }\href
  {\doibase 10.1038/ncomms4821} {\bibfield  {journal} {\bibinfo  {journal}
  {Nat. Commun}\ }\textbf {\bibinfo {volume} {5}},\ \bibinfo {pages} {3821}
  (\bibinfo {year} {2014})}\BibitemShut {NoStop}%
\bibitem [{\citenamefont {Brand\~{a}o}\ \emph {et~al.}(2015)\citenamefont
  {Brand\~{a}o}, \citenamefont {Horodecki}, \citenamefont {Ng}, \citenamefont
  {Oppenheim},\ and\ \citenamefont {Wehner}}]{Brandao3275}%
  \BibitemOpen
  \bibfield  {author} {\bibinfo {author} {\bibfnamefont {F.}~\bibnamefont
  {Brand\~{a}o}}, \bibinfo {author} {\bibfnamefont {M.}~\bibnamefont
  {Horodecki}}, \bibinfo {author} {\bibfnamefont {N.}~\bibnamefont {Ng}},
  \bibinfo {author} {\bibfnamefont {J.}~\bibnamefont {Oppenheim}}, \ and\
  \bibinfo {author} {\bibfnamefont {S.}~\bibnamefont {Wehner}},\ }\bibfield
  {title} {\enquote {\bibinfo {title} {The second laws of quantum
  thermodynamics},}\ }\href {\doibase 10.1073/pnas.1411728112} {\bibfield
  {journal} {\bibinfo  {journal} {Proc. Natl. Acad. Sci.}\ }\textbf {\bibinfo
  {volume} {112}},\ \bibinfo {pages} {3275} (\bibinfo {year}
  {2015})}\BibitemShut {NoStop}%
\bibitem [{\citenamefont {Bernamonti}\ \emph {et~al.}(2018)\citenamefont
  {Bernamonti}, \citenamefont {Galli}, \citenamefont {Myers},\ and\
  \citenamefont {Oppenheim}}]{Bernamonti2018}%
  \BibitemOpen
  \bibfield  {author} {\bibinfo {author} {\bibfnamefont {A.}~\bibnamefont
  {Bernamonti}}, \bibinfo {author} {\bibfnamefont {F.}~\bibnamefont {Galli}},
  \bibinfo {author} {\bibfnamefont {R.~C.}\ \bibnamefont {Myers}}, \ and\
  \bibinfo {author} {\bibfnamefont {J.}~\bibnamefont {Oppenheim}},\ }\bibfield
  {title} {\enquote {\bibinfo {title} {Holographic second laws of black hole
  thermodynamics},}\ }\href {\doibase 10.1007/JHEP07(2018)111} {\bibfield
  {journal} {\bibinfo  {journal} {J. High Energ. Phys.}\ }\textbf {\bibinfo
  {volume} {2018}},\ \bibinfo {pages} {111} (\bibinfo {year}
  {2018})}\BibitemShut {NoStop}%
\bibitem [{\citenamefont {Csisz\'{a}r}(1995)}]{Csiszar1995}%
  \BibitemOpen
  \bibfield  {author} {\bibinfo {author} {\bibfnamefont {I.}~\bibnamefont
  {Csisz\'{a}r}},\ }\bibfield  {title} {\enquote {\bibinfo {title} {Generalized
  cutoff rates and {R}\'{e}nyi's information measures},}\ }\href {\doibase
  10.1109/18.370121} {\bibfield  {journal} {\bibinfo  {journal} {IEEE Trans.
  Inf. Theory}\ }\textbf {\bibinfo {volume} {41}},\ \bibinfo {pages} {26}
  (\bibinfo {year} {1995})}\BibitemShut {NoStop}%
\bibitem [{\citenamefont {Seshadreesan}\ \emph {et~al.}(2018)\citenamefont
  {Seshadreesan}, \citenamefont {Lami},\ and\ \citenamefont
  {Wilde}}]{doi:10.1063/1.5007167}%
  \BibitemOpen
  \bibfield  {author} {\bibinfo {author} {\bibfnamefont {K.~P.}\ \bibnamefont
  {Seshadreesan}}, \bibinfo {author} {\bibfnamefont {L.}~\bibnamefont {Lami}},
  \ and\ \bibinfo {author} {\bibfnamefont {M.~M.}\ \bibnamefont {Wilde}},\
  }\bibfield  {title} {\enquote {\bibinfo {title} {R\'{e}nyi relative entropies
  of quantum {G}aussian states},}\ }\href {\doibase 10.1063/1.5007167}
  {\bibfield  {journal} {\bibinfo  {journal} {J. Math. Phys.}\ }\textbf
  {\bibinfo {volume} {59}},\ \bibinfo {pages} {072204} (\bibinfo {year}
  {2018})}\BibitemShut {NoStop}%
\bibitem [{\citenamefont {Coles}\ \emph {et~al.}(2019)\citenamefont {Coles},
  \citenamefont {Katariya}, \citenamefont {Lloyd}, \citenamefont {Marvian},\
  and\ \citenamefont {Wilde}}]{PhysRevLett.122.100401}%
  \BibitemOpen
  \bibfield  {author} {\bibinfo {author} {\bibfnamefont {P.~J.}\ \bibnamefont
  {Coles}}, \bibinfo {author} {\bibfnamefont {V.}~\bibnamefont {Katariya}},
  \bibinfo {author} {\bibfnamefont {S.}~\bibnamefont {Lloyd}}, \bibinfo
  {author} {\bibfnamefont {I.}~\bibnamefont {Marvian}}, \ and\ \bibinfo
  {author} {\bibfnamefont {M.~M.}\ \bibnamefont {Wilde}},\ }\bibfield  {title}
  {\enquote {\bibinfo {title} {Entropic {E}nergy-{T}ime {U}ncertainty
  {R}elation},}\ }\href {\doibase 10.1103/PhysRevLett.122.100401} {\bibfield
  {journal} {\bibinfo  {journal} {Phys. Rev. Lett.}\ }\textbf {\bibinfo
  {volume} {122}},\ \bibinfo {pages} {100401} (\bibinfo {year}
  {2019})}\BibitemShut {NoStop}%
\bibitem [{\citenamefont {Pires}\ \emph {et~al.}(2020)\citenamefont {Pires},
  \citenamefont {Smerzi},\ and\ \citenamefont
  {Macr\`{\i}}}]{PhysRevA.102.012429}%
  \BibitemOpen
  \bibfield  {author} {\bibinfo {author} {\bibfnamefont {D.~P.}\ \bibnamefont
  {Pires}}, \bibinfo {author} {\bibfnamefont {A.}~\bibnamefont {Smerzi}}, \
  and\ \bibinfo {author} {\bibfnamefont {T.}~\bibnamefont {Macr\`{\i}}},\
  }\bibfield  {title} {\enquote {\bibinfo {title} {Relating relative r\'enyi
  entropies and {W}igner-{Y}anase-{D}yson skew information to generalized
  multiple quantum coherences},}\ }\href {\doibase 10.1103/PhysRevA.102.012429}
  {\bibfield  {journal} {\bibinfo  {journal} {Phys. Rev. A}\ }\textbf {\bibinfo
  {volume} {102}},\ \bibinfo {pages} {012429} (\bibinfo {year}
  {2020})}\BibitemShut {NoStop}%
\bibitem [{\citenamefont {Tsallis}\ and\ \citenamefont
  {Brigatti}(2004)}]{Tsallis2004}%
  \BibitemOpen
  \bibfield  {author} {\bibinfo {author} {\bibfnamefont {C.}~\bibnamefont
  {Tsallis}}\ and\ \bibinfo {author} {\bibfnamefont {E.}~\bibnamefont
  {Brigatti}},\ }\bibfield  {title} {\enquote {\bibinfo {title} {Nonextensive
  statistical mechanics: {A} brief introduction},}\ }\href {\doibase
  10.1007/s00161-004-0174-4} {\bibfield  {journal} {\bibinfo  {journal}
  {Continuum Mech. Thermodyn.}\ }\textbf {\bibinfo {volume} {16}},\ \bibinfo
  {pages} {223} (\bibinfo {year} {2004})}\BibitemShut {NoStop}%
\bibitem [{\citenamefont {Furuichi}\ \emph {et~al.}(2004)\citenamefont
  {Furuichi}, \citenamefont {Yanagi},\ and\ \citenamefont
  {Kuriyama}}]{doi:10.1063_1.1805729}%
  \BibitemOpen
  \bibfield  {author} {\bibinfo {author} {\bibfnamefont {S.}~\bibnamefont
  {Furuichi}}, \bibinfo {author} {\bibfnamefont {K.}~\bibnamefont {Yanagi}}, \
  and\ \bibinfo {author} {\bibfnamefont {K.}~\bibnamefont {Kuriyama}},\
  }\bibfield  {title} {\enquote {\bibinfo {title} {Fundamental properties of
  {T}sallis relative entropy},}\ }\href {\doibase 10.1063/1.1805729} {\bibfield
   {journal} {\bibinfo  {journal} {J. Math. Phys.}\ }\textbf {\bibinfo {volume}
  {45}},\ \bibinfo {pages} {4868} (\bibinfo {year} {2004})}\BibitemShut
  {NoStop}%
\bibitem [{\citenamefont {Rastegin}(2016)}]{PhysRevA.93.032136}%
  \BibitemOpen
  \bibfield  {author} {\bibinfo {author} {\bibfnamefont {A.~E.}\ \bibnamefont
  {Rastegin}},\ }\bibfield  {title} {\enquote {\bibinfo {title}
  {Quantum-coherence quantifiers based on the {T}sallis relative
  $\ensuremath{\alpha}$ entropies},}\ }\href {\doibase
  10.1103/PhysRevA.93.032136} {\bibfield  {journal} {\bibinfo  {journal} {Phys.
  Rev. A}\ }\textbf {\bibinfo {volume} {93}},\ \bibinfo {pages} {032136}
  (\bibinfo {year} {2016})}\BibitemShut {NoStop}%
\bibitem [{\citenamefont {Rastegin}(2013)}]{ARastegin_MathPhysAnalGeom_16_213}%
  \BibitemOpen
  \bibfield  {author} {\bibinfo {author} {\bibfnamefont {A.~E.}\ \bibnamefont
  {Rastegin}},\ }\bibfield  {title} {\enquote {\bibinfo {title} {Bounds of the
  {P}insker and {F}annes {T}ypes on the {T}sallis {R}elative {E}ntropy},}\
  }\href {\doibase 10.1007/s11040-013-9128-z} {\bibfield  {journal} {\bibinfo
  {journal} {Math. Phys. Anal. Geom.}\ }\textbf {\bibinfo {volume} {16}},\
  \bibinfo {pages} {213} (\bibinfo {year} {2013})}\BibitemShut {NoStop}%
\bibitem [{\citenamefont {Vedral}(2002)}]{RevModPhys.74.197}%
  \BibitemOpen
  \bibfield  {author} {\bibinfo {author} {\bibfnamefont {V.}~\bibnamefont
  {Vedral}},\ }\bibfield  {title} {\enquote {\bibinfo {title} {The role of
  relative entropy in quantum information theory},}\ }\href {\doibase
  10.1103/RevModPhys.74.197} {\bibfield  {journal} {\bibinfo  {journal} {Rev.
  Mod. Phys.}\ }\textbf {\bibinfo {volume} {74}},\ \bibinfo {pages} {197--234}
  (\bibinfo {year} {2002})}\BibitemShut {NoStop}%
\bibitem [{\citenamefont {Baez}\ and\ \citenamefont
  {Pollard}(2016)}]{Baez_2016}%
  \BibitemOpen
  \bibfield  {author} {\bibinfo {author} {\bibfnamefont {J.}~\bibnamefont
  {Baez}}\ and\ \bibinfo {author} {\bibfnamefont {B.}~\bibnamefont {Pollard}},\
  }\bibfield  {title} {\enquote {\bibinfo {title} {Relative {E}ntropy in
  {B}iological {S}ystems},}\ }\href {\doibase 10.3390/e18020046} {\bibfield
  {journal} {\bibinfo  {journal} {Entropy}\ }\textbf {\bibinfo {volume} {18}},\
  \bibinfo {pages} {46} (\bibinfo {year} {2016})}\BibitemShut {NoStop}%
\bibitem [{\citenamefont {Renner}(2006)}]{arXiv:0512258}%
  \BibitemOpen
  \bibfield  {author} {\bibinfo {author} {\bibfnamefont {R.}~\bibnamefont
  {Renner}},\ }\bibfield  {title} {\enquote {\bibinfo {title} {Security of
  {Q}uantum {K}ey {D}istribution},}\ }\href
  {https://arxiv.org/abs/quant-ph/0512258} {\bibfield  {journal} {\bibinfo
  {journal} {arXiv:quant-ph/0512258}\ } (\bibinfo {year} {2006})}\BibitemShut
  {NoStop}%
\bibitem [{\citenamefont {Guarnieri}\ \emph {et~al.}(2019)\citenamefont
  {Guarnieri}, \citenamefont {Ng}, \citenamefont {Modi}, \citenamefont
  {Eisert}, \citenamefont {Paternostro},\ and\ \citenamefont
  {Goold}}]{PhysRevE.99.050101}%
  \BibitemOpen
  \bibfield  {author} {\bibinfo {author} {\bibfnamefont {G.}~\bibnamefont
  {Guarnieri}}, \bibinfo {author} {\bibfnamefont {N.~H.~Y.}\ \bibnamefont
  {Ng}}, \bibinfo {author} {\bibfnamefont {K.}~\bibnamefont {Modi}}, \bibinfo
  {author} {\bibfnamefont {J.}~\bibnamefont {Eisert}}, \bibinfo {author}
  {\bibfnamefont {M.}~\bibnamefont {Paternostro}}, \ and\ \bibinfo {author}
  {\bibfnamefont {J.}~\bibnamefont {Goold}},\ }\bibfield  {title} {\enquote
  {\bibinfo {title} {Quantum work statistics and resource theories: {B}ridging
  the gap through {R}\'enyi divergences},}\ }\href {\doibase
  10.1103/PhysRevE.99.050101} {\bibfield  {journal} {\bibinfo  {journal} {Phys.
  Rev. E}\ }\textbf {\bibinfo {volume} {99}},\ \bibinfo {pages} {050101}
  (\bibinfo {year} {2019})}\BibitemShut {NoStop}%
\bibitem [{\citenamefont {Marvian}\ \emph {et~al.}(2016)\citenamefont
  {Marvian}, \citenamefont {Spekkens},\ and\ \citenamefont
  {Zanardi}}]{PhysRevA.93.052331}%
  \BibitemOpen
  \bibfield  {author} {\bibinfo {author} {\bibfnamefont {I.}~\bibnamefont
  {Marvian}}, \bibinfo {author} {\bibfnamefont {R.~W.}\ \bibnamefont
  {Spekkens}}, \ and\ \bibinfo {author} {\bibfnamefont {P.}~\bibnamefont
  {Zanardi}},\ }\bibfield  {title} {\enquote {\bibinfo {title} {Quantum speed
  limits, coherence, and asymmetry},}\ }\href {\doibase
  10.1103/PhysRevA.93.052331} {\bibfield  {journal} {\bibinfo  {journal} {Phys.
  Rev. A}\ }\textbf {\bibinfo {volume} {93}},\ \bibinfo {pages} {052331}
  (\bibinfo {year} {2016})}\BibitemShut {NoStop}%
\bibitem [{\citenamefont {Abe}\ and\ \citenamefont {Okamoto}(2001)}]{Abe2001}%
  \BibitemOpen
  \bibfield  {author} {\bibinfo {author} {\bibfnamefont {S.}~\bibnamefont
  {Abe}}\ and\ \bibinfo {author} {\bibfnamefont {Y.}~\bibnamefont {Okamoto}},\
  }\href {\doibase 10.1007/3-540-40919-X} {\emph {\bibinfo {title}
  {Nonextensive {S}tatistical {M}echanics and {I}ts {A}pplications}}}\
  (\bibinfo  {publisher} {Springer},\ \bibinfo {address} {Berlin},\ \bibinfo
  {year} {2001})\BibitemShut {NoStop}%
\bibitem [{\citenamefont {M\"{u}ller-Lennert}\ \emph
  {et~al.}(2013)\citenamefont {M\"{u}ller-Lennert}, \citenamefont {Dupuis},
  \citenamefont {Szehr}, \citenamefont {Fehr},\ and\ \citenamefont
  {Tomamichel}}]{10.1063.1.4838856}%
  \BibitemOpen
  \bibfield  {author} {\bibinfo {author} {\bibfnamefont {M.}~\bibnamefont
  {M\"{u}ller-Lennert}}, \bibinfo {author} {\bibfnamefont {F.}~\bibnamefont
  {Dupuis}}, \bibinfo {author} {\bibfnamefont {O.}~\bibnamefont {Szehr}},
  \bibinfo {author} {\bibfnamefont {S.}~\bibnamefont {Fehr}}, \ and\ \bibinfo
  {author} {\bibfnamefont {M.}~\bibnamefont {Tomamichel}},\ }\bibfield  {title}
  {\enquote {\bibinfo {title} {On quantum {R}\'{e}nyi entropies: {A} new
  generalization and some properties},}\ }\href {\doibase 10.1063/1.4838856}
  {\bibfield  {journal} {\bibinfo  {journal} {J. Math. Phys.}\ }\textbf
  {\bibinfo {volume} {54}},\ \bibinfo {pages} {122203} (\bibinfo {year}
  {2013})}\BibitemShut {NoStop}%
\bibitem [{\citenamefont {Audenaert}(2014)}]{DBLPjournals_qic_Audenaert01}%
  \BibitemOpen
  \bibfield  {author} {\bibinfo {author} {\bibfnamefont {K.~M.~R.}\
  \bibnamefont {Audenaert}},\ }\bibfield  {title} {\enquote {\bibinfo {title}
  {Comparisons between quantum state distinguishability measures},}\ }\href
  {\doibase 10.26421/QIC14.1-2} {\bibfield  {journal} {\bibinfo  {journal}
  {Quantum {I}nf. {C}omput.}\ }\textbf {\bibinfo {volume} {14}},\ \bibinfo
  {pages} {31} (\bibinfo {year} {2014})}\BibitemShut {NoStop}%
\bibitem [{\citenamefont {Petz}(1986)}]{PETZ198657}%
  \BibitemOpen
  \bibfield  {author} {\bibinfo {author} {\bibfnamefont {D.}~\bibnamefont
  {Petz}},\ }\bibfield  {title} {\enquote {\bibinfo {title} {Quasi-entropies
  for finite quantum systems},}\ }\href {\doibase 10.1016/0034-4877(86)90067-4}
  {\bibfield  {journal} {\bibinfo  {journal} {Rep. Math. Phys.}\ }\textbf
  {\bibinfo {volume} {23}},\ \bibinfo {pages} {57} (\bibinfo {year}
  {1986})}\BibitemShut {NoStop}%
\bibitem [{\citenamefont {Frank}\ and\ \citenamefont
  {Lieb}(2013)}]{doi:10.1063/1.4838835}%
  \BibitemOpen
  \bibfield  {author} {\bibinfo {author} {\bibfnamefont {R.~L.}\ \bibnamefont
  {Frank}}\ and\ \bibinfo {author} {\bibfnamefont {E.~H.}\ \bibnamefont
  {Lieb}},\ }\bibfield  {title} {\enquote {\bibinfo {title} {Monotonicity of a
  relative {R}\'{e}nyi entropy},}\ }\href {\doibase 10.1063/1.4838835}
  {\bibfield  {journal} {\bibinfo  {journal} {J. Math. Phys.}\ }\textbf
  {\bibinfo {volume} {54}},\ \bibinfo {pages} {122201} (\bibinfo {year}
  {2013})}\BibitemShut {NoStop}%
\bibitem [{\citenamefont {Beigi}(2013)}]{doi:10.1063/1.4838855}%
  \BibitemOpen
  \bibfield  {author} {\bibinfo {author} {\bibfnamefont {S.}~\bibnamefont
  {Beigi}},\ }\bibfield  {title} {\enquote {\bibinfo {title} {Sandwiched
  {R}\'{e}nyi divergence satisfies data processing inequality},}\ }\href
  {\doibase 10.1063/1.4838855} {\bibfield  {journal} {\bibinfo  {journal} {J.
  Math. Phys.}\ }\textbf {\bibinfo {volume} {54}},\ \bibinfo {pages} {122202}
  (\bibinfo {year} {2013})}\BibitemShut {NoStop}%
\bibitem [{\citenamefont {Datta}\ and\ \citenamefont
  {Leditzky}(2014)}]{Datta_2014}%
  \BibitemOpen
  \bibfield  {author} {\bibinfo {author} {\bibfnamefont {N.}~\bibnamefont
  {Datta}}\ and\ \bibinfo {author} {\bibfnamefont {F.}~\bibnamefont
  {Leditzky}},\ }\bibfield  {title} {\enquote {\bibinfo {title} {A limit of the
  quantum {R}{\'{e}}nyi divergence},}\ }\href {\doibase
  10.1088/1751-8113/47/4/045304} {\bibfield  {journal} {\bibinfo  {journal} {J.
  Phy. A: Math. Theor.}\ }\textbf {\bibinfo {volume} {47}},\ \bibinfo {pages}
  {045304} (\bibinfo {year} {2014})}\BibitemShut {NoStop}%
\bibitem [{\citenamefont {Audenaert}\ and\ \citenamefont
  {Datta}(2015)}]{10.1063.1.4906367}%
  \BibitemOpen
  \bibfield  {author} {\bibinfo {author} {\bibfnamefont {K.~M.~R.}\
  \bibnamefont {Audenaert}}\ and\ \bibinfo {author} {\bibfnamefont
  {N.}~\bibnamefont {Datta}},\ }\bibfield  {title} {\enquote {\bibinfo {title}
  {$\alpha$-$z$-{R}\'{e}nyi relative entropies},}\ }\href {\doibase
  10.1063/1.4906367} {\bibfield  {journal} {\bibinfo  {journal} {J. Math.
  Phys.}\ }\textbf {\bibinfo {volume} {56}},\ \bibinfo {pages} {022202}
  (\bibinfo {year} {2015})}\BibitemShut {NoStop}%
\bibitem [{\citenamefont {Luo}\ and\ \citenamefont
  {Zhang}(2004)}]{PhysRevA.69.032106}%
  \BibitemOpen
  \bibfield  {author} {\bibinfo {author} {\bibfnamefont {S.}~\bibnamefont
  {Luo}}\ and\ \bibinfo {author} {\bibfnamefont {Q.}~\bibnamefont {Zhang}},\
  }\bibfield  {title} {\enquote {\bibinfo {title} {Informational distance on
  quantum-state space},}\ }\href {\doibase 10.1103/PhysRevA.69.032106}
  {\bibfield  {journal} {\bibinfo  {journal} {Phys. Rev. A}\ }\textbf {\bibinfo
  {volume} {69}},\ \bibinfo {pages} {032106} (\bibinfo {year}
  {2004})}\BibitemShut {NoStop}%
\bibitem [{\citenamefont {Gibilisco}\ and\ \citenamefont
  {Isola}(2003)}]{Gibilisco_Isola_10.1063_1.1598279}%
  \BibitemOpen
  \bibfield  {author} {\bibinfo {author} {\bibfnamefont {P.}~\bibnamefont
  {Gibilisco}}\ and\ \bibinfo {author} {\bibfnamefont {T.}~\bibnamefont
  {Isola}},\ }\bibfield  {title} {\enquote {\bibinfo {title} {{W}igner-{Y}anase
  information on quantum state space: {T}he geometric approach},}\ }\href
  {\doibase 10.1063/1.1598279} {\bibfield  {journal} {\bibinfo  {journal} {J.
  Math. Phys.}\ }\textbf {\bibinfo {volume} {44}},\ \bibinfo {pages}
  {3752--3762} (\bibinfo {year} {2003})}\BibitemShut {NoStop}%
\bibitem [{\citenamefont {Jen\v{c}ov\'{a}}(2004)}]{Jencova_10.1063_1.1689000}%
  \BibitemOpen
  \bibfield  {author} {\bibinfo {author} {\bibfnamefont {A.}~\bibnamefont
  {Jen\v{c}ov\'{a}}},\ }\bibfield  {title} {\enquote {\bibinfo {title}
  {Geodesic distances on density matrices},}\ }\href {\doibase
  10.1063/1.1689000} {\bibfield  {journal} {\bibinfo  {journal} {J. Math.
  Phys.}\ }\textbf {\bibinfo {volume} {45}},\ \bibinfo {pages} {1787--1794}
  (\bibinfo {year} {2004})}\BibitemShut {NoStop}%
\bibitem [{\citenamefont {Amig\'{o}}\ \emph {et~al.}(2018)\citenamefont
  {Amig\'{o}}, \citenamefont {Balogh},\ and\ \citenamefont
  {Hern\'{a}ndez}}]{Amigo2018}%
  \BibitemOpen
  \bibfield  {author} {\bibinfo {author} {\bibfnamefont {J.~M.}\ \bibnamefont
  {Amig\'{o}}}, \bibinfo {author} {\bibfnamefont {S.~G.}\ \bibnamefont
  {Balogh}}, \ and\ \bibinfo {author} {\bibfnamefont {S.}~\bibnamefont
  {Hern\'{a}ndez}},\ }\bibfield  {title} {\enquote {\bibinfo {title} {A {B}rief
  {R}eview of {G}eneralized {E}ntropies},}\ }\href {\doibase 10.3390/e20110813}
  {\bibfield  {journal} {\bibinfo  {journal} {Entropy}\ }\textbf {\bibinfo
  {volume} {20}},\ \bibinfo {pages} {813} (\bibinfo {year} {2018})}\BibitemShut
  {NoStop}%
\bibitem [{\citenamefont {{Datta}}(2009)}]{4957651_Datta}%
  \BibitemOpen
  \bibfield  {author} {\bibinfo {author} {\bibfnamefont {N.}~\bibnamefont
  {{Datta}}},\ }\bibfield  {title} {\enquote {\bibinfo {title} {Min- and
  {M}ax-{R}elative {E}ntropies and a {N}ew {E}ntanglement {M}onotone},}\ }\href
  {\doibase 10.1109/TIT.2009.2018325} {\bibfield  {journal} {\bibinfo
  {journal} {IEEE Trans. Inf. Theory}\ }\textbf {\bibinfo {volume} {55}},\
  \bibinfo {pages} {2816} (\bibinfo {year} {2009})}\BibitemShut {NoStop}%
\bibitem [{\citenamefont {Mosonyi}\ and\ \citenamefont
  {F.~Hiai}(2011)}]{5730573}%
  \BibitemOpen
  \bibfield  {author} {\bibinfo {author} {\bibfnamefont {M.}~\bibnamefont
  {Mosonyi}}\ and\ \bibinfo {author} {\bibfnamefont {F.}~\bibnamefont
  {F.~Hiai}},\ }\bibfield  {title} {\enquote {\bibinfo {title} {On the
  {Q}uantum {R}\'{e}nyi {R}e\-la\-ti\-ve {E}ntropies and {R}elated {C}apacity
  {F}ormulas},}\ }\href {\doibase 10.1109/TIT.2011.2110050} {\bibfield
  {journal} {\bibinfo  {journal} {IEEE Trans. Inf. Theory}\ }\textbf {\bibinfo
  {volume} {57}},\ \bibinfo {pages} {2474} (\bibinfo {year}
  {2011})}\BibitemShut {NoStop}%
\bibitem [{\citenamefont {Sagawa}(2012)}]{Sagawa2012}%
  \BibitemOpen
  \bibfield  {author} {\bibinfo {author} {\bibfnamefont {T.}~\bibnamefont
  {Sagawa}},\ }\enquote {\bibinfo {title} {Second {L}aw-{L}ike {I}nequalities
  with {Q}uantum {R}elative {E}ntropy: {A}n {I}ntroduction},}\ in\ \href
  {\doibase 10.1142/9789814425193_0003} {\emph {\bibinfo {booktitle} {Lectures
  on {Q}uantum {C}omputing, {T}hermodynamics and {S}tatistical {P}hysics}}}\
  (\bibinfo  {publisher} {World Scientific},\ \bibinfo {address} {New York},\
  \bibinfo {year} {2012})\ pp.\ \bibinfo {pages} {125--190}\BibitemShut
  {NoStop}%
\bibitem [{\citenamefont {Abe}(2003{\natexlab{a}})}]{ABE2003336}%
  \BibitemOpen
  \bibfield  {author} {\bibinfo {author} {\bibfnamefont {S.}~\bibnamefont
  {Abe}},\ }\bibfield  {title} {\enquote {\bibinfo {title} {Monotonic decrease
  of the quantum nonadditive divergence by projective measurements},}\ }\href
  {\doibase 10.1016/S0375-9601(03)00682-0} {\bibfield  {journal} {\bibinfo
  {journal} {Phys. Lett. A}\ }\textbf {\bibinfo {volume} {312}},\ \bibinfo
  {pages} {336} (\bibinfo {year} {2003}{\natexlab{a}})}\BibitemShut {NoStop}%
\bibitem [{\citenamefont {Abe}(2003{\natexlab{b}})}]{PhysRevA.68.032302}%
  \BibitemOpen
  \bibfield  {author} {\bibinfo {author} {\bibfnamefont {S.}~\bibnamefont
  {Abe}},\ }\bibfield  {title} {\enquote {\bibinfo {title} {Nonadditive
  generalization of the quantum {K}ullback-{L}eibler divergence for measuring
  the degree of purification},}\ }\href {\doibase 10.1103/PhysRevA.68.032302}
  {\bibfield  {journal} {\bibinfo  {journal} {Phys. Rev. A}\ }\textbf {\bibinfo
  {volume} {68}},\ \bibinfo {pages} {032302} (\bibinfo {year}
  {2003}{\natexlab{b}})}\BibitemShut {NoStop}%
\bibitem [{\citenamefont {Virosztek}(2019)}]{arXiv:1910.10447}%
  \BibitemOpen
  \bibfield  {author} {\bibinfo {author} {\bibfnamefont {D.}~\bibnamefont
  {Virosztek}},\ }\bibfield  {title} {\enquote {\bibinfo {title} {The metric
  property of the quantum {J}ensen-{S}hannon divergence},}\ }\href
  {https://www.arxiv.org/abs/1910.10447} {\bibfield  {journal} {\bibinfo
  {journal} {arXiv:1910.10447}\ } (\bibinfo {year} {2019})}\BibitemShut
  {NoStop}%
\bibitem [{\citenamefont {B\"{o}ttcher}\ and\ \citenamefont
  {Wenzel}(2008)}]{BOTTCHER20081864}%
  \BibitemOpen
  \bibfield  {author} {\bibinfo {author} {\bibfnamefont {A.}~\bibnamefont
  {B\"{o}ttcher}}\ and\ \bibinfo {author} {\bibfnamefont {D.}~\bibnamefont
  {Wenzel}},\ }\bibfield  {title} {\enquote {\bibinfo {title} {The {F}robenius
  norm and the commutator},}\ }\href {\doibase 10.1016/j.laa.2008.05.020}
  {\bibfield  {journal} {\bibinfo  {journal} {Linear Algebra Appl.}\ }\textbf
  {\bibinfo {volume} {429}},\ \bibinfo {pages} {1864} (\bibinfo {year}
  {2008})}\BibitemShut {NoStop}%
\bibitem [{\citenamefont {Audenaert}(2010)}]{AUDENAERT20101126}%
  \BibitemOpen
  \bibfield  {author} {\bibinfo {author} {\bibfnamefont {K.~M.~R.}\
  \bibnamefont {Audenaert}},\ }\bibfield  {title} {\enquote {\bibinfo {title}
  {Variance bounds, with an application to norm bounds for commutators},}\
  }\href {\doibase 10.1016/j.laa.2009.10.022} {\bibfield  {journal} {\bibinfo
  {journal} {Linear Algebra Appl.}\ }\textbf {\bibinfo {volume} {432}},\
  \bibinfo {pages} {1126} (\bibinfo {year} {2010})}\BibitemShut {NoStop}%
\bibitem [{\citenamefont {Fong}\ \emph {et~al.}(2011)\citenamefont {Fong},
  \citenamefont {Lok},\ and\ \citenamefont {Cheng}}]{FONG20111193}%
  \BibitemOpen
  \bibfield  {author} {\bibinfo {author} {\bibfnamefont {K.-S.}\ \bibnamefont
  {Fong}}, \bibinfo {author} {\bibfnamefont {I.-K.}\ \bibnamefont {Lok}}, \
  and\ \bibinfo {author} {\bibfnamefont {C.-M.}\ \bibnamefont {Cheng}},\
  }\bibfield  {title} {\enquote {\bibinfo {title} {A note on the norm of the
  commutator and the norm of ${X}{Y} - {Y}{X^T}$},}\ }\href {\doibase
  10.1016/j.laa.2011.02.028} {\bibfield  {journal} {\bibinfo  {journal} {Linear
  Algebra Appl.}\ }\textbf {\bibinfo {volume} {435}},\ \bibinfo {pages} {1193}
  (\bibinfo {year} {2011})}\BibitemShut {NoStop}%
\bibitem [{\citenamefont {Mandelstam}\ and\ \citenamefont
  {Tamm}(1945)}]{1945_JPhysURSS_9_249}%
  \BibitemOpen
  \bibfield  {author} {\bibinfo {author} {\bibfnamefont {L.}~\bibnamefont
  {Mandelstam}}\ and\ \bibinfo {author} {\bibfnamefont {I.}~\bibnamefont
  {Tamm}},\ }\bibfield  {title} {\enquote {\bibinfo {title} {The uncertainty
  relation between energy and time in non-relativistic quantum mechanics},}\
  }\href@noop {} {\bibfield  {journal} {\bibinfo  {journal} {J. Phys. USSR}\
  }\textbf {\bibinfo {volume} {9}},\ \bibinfo {pages} {249} (\bibinfo {year}
  {1945})}\BibitemShut {NoStop}%
\bibitem [{\citenamefont {Margolus}\ and\ \citenamefont
  {Levitin}(1998)}]{1992_PhysicaD_120_188}%
  \BibitemOpen
  \bibfield  {author} {\bibinfo {author} {\bibfnamefont {N.}~\bibnamefont
  {Margolus}}\ and\ \bibinfo {author} {\bibfnamefont {L.~B.}\ \bibnamefont
  {Levitin}},\ }\bibfield  {title} {\enquote {\bibinfo {title} {The maximum
  speed of dynamical evolution},}\ }\href {\doibase
  10.1016/S0167-2789(98)00054-2} {\bibfield  {journal} {\bibinfo  {journal}
  {Physica D}\ }\textbf {\bibinfo {volume} {120}},\ \bibinfo {pages} {188}
  (\bibinfo {year} {1998})}\BibitemShut {NoStop}%
\bibitem [{\citenamefont {Levitin}\ and\ \citenamefont
  {Toffoli}(2009)}]{PhysRevLett.103.160502}%
  \BibitemOpen
  \bibfield  {author} {\bibinfo {author} {\bibfnamefont {L.~B.}\ \bibnamefont
  {Levitin}}\ and\ \bibinfo {author} {\bibfnamefont {T.}~\bibnamefont
  {Toffoli}},\ }\bibfield  {title} {\enquote {\bibinfo {title} {Fundamental
  {L}imit on the {R}ate of {Q}uantum {D}ynamics: {T}he {U}nified {B}ound {I}s
  {T}ight},}\ }\href {\doibase 10.1103/PhysRevLett.103.160502} {\bibfield
  {journal} {\bibinfo  {journal} {Phys. Rev. Lett.}\ }\textbf {\bibinfo
  {volume} {103}},\ \bibinfo {pages} {160502} (\bibinfo {year}
  {2009})}\BibitemShut {NoStop}%
\bibitem [{\citenamefont {Taddei}\ \emph {et~al.}(2013)\citenamefont {Taddei},
  \citenamefont {Escher}, \citenamefont {Davidovich},\ and\ \citenamefont
  {de~Matos~Filho}}]{2013_PhysRevLett_110_050402}%
  \BibitemOpen
  \bibfield  {author} {\bibinfo {author} {\bibfnamefont {M.~M.}\ \bibnamefont
  {Taddei}}, \bibinfo {author} {\bibfnamefont {B.~M.}\ \bibnamefont {Escher}},
  \bibinfo {author} {\bibfnamefont {L.}~\bibnamefont {Davidovich}}, \ and\
  \bibinfo {author} {\bibfnamefont {R.~L.}\ \bibnamefont {de~Matos~Filho}},\
  }\bibfield  {title} {\enquote {\bibinfo {title} {Quantum {S}peed {L}imit for
  {P}hysical {P}rocesses},}\ }\href {\doibase 10.1103/PhysRevLett.110.050402}
  {\bibfield  {journal} {\bibinfo  {journal} {Phys. Rev. Lett.}\ }\textbf
  {\bibinfo {volume} {110}},\ \bibinfo {pages} {050402} (\bibinfo {year}
  {2013})}\BibitemShut {NoStop}%
\bibitem [{\citenamefont {del Campo}\ \emph {et~al.}(2013)\citenamefont {del
  Campo}, \citenamefont {Egusquiza}, \citenamefont {Plenio},\ and\
  \citenamefont {Huelga}}]{PhysRevLett.110.050403}%
  \BibitemOpen
  \bibfield  {author} {\bibinfo {author} {\bibfnamefont {A.}~\bibnamefont {del
  Campo}}, \bibinfo {author} {\bibfnamefont {I.~L.}\ \bibnamefont {Egusquiza}},
  \bibinfo {author} {\bibfnamefont {M.~B.}\ \bibnamefont {Plenio}}, \ and\
  \bibinfo {author} {\bibfnamefont {S.~F.}\ \bibnamefont {Huelga}},\ }\bibfield
   {title} {\enquote {\bibinfo {title} {Quantum {S}peed {L}imits in {O}pen
  {S}ystem {D}y\-na\-mics},}\ }\href {\doibase 10.1103/PhysRevLett.110.050403}
  {\bibfield  {journal} {\bibinfo  {journal} {Phys. Rev. Lett.}\ }\textbf
  {\bibinfo {volume} {110}},\ \bibinfo {pages} {050403} (\bibinfo {year}
  {2013})}\BibitemShut {NoStop}%
\bibitem [{\citenamefont {Deffner}\ and\ \citenamefont
  {Lutz}(2013)}]{2013_PhysRevLett_111_010402}%
  \BibitemOpen
  \bibfield  {author} {\bibinfo {author} {\bibfnamefont {S.}~\bibnamefont
  {Deffner}}\ and\ \bibinfo {author} {\bibfnamefont {E.}~\bibnamefont {Lutz}},\
  }\bibfield  {title} {\enquote {\bibinfo {title} {Quantum {S}peed {L}imit for
  {N}on-{M}arkovian {D}ynamics},}\ }\href {\doibase
  10.1103/PhysRevLett.111.010402} {\bibfield  {journal} {\bibinfo  {journal}
  {Phys. Rev. Lett.}\ }\textbf {\bibinfo {volume} {111}},\ \bibinfo {pages}
  {010402} (\bibinfo {year} {2013})}\BibitemShut {NoStop}%
\bibitem [{\citenamefont {Deffner}\ and\ \citenamefont
  {Campbell}(2017)}]{Deffner_2017}%
  \BibitemOpen
  \bibfield  {author} {\bibinfo {author} {\bibfnamefont {S.}~\bibnamefont
  {Deffner}}\ and\ \bibinfo {author} {\bibfnamefont {S.}~\bibnamefont
  {Campbell}},\ }\bibfield  {title} {\enquote {\bibinfo {title} {Quantum speed
  limits: from {H}eisenberg's uncertainty principle to optimal quantum
  control},}\ }\href {\doibase 10.1088/1751-8121/aa86c6} {\bibfield  {journal}
  {\bibinfo  {journal} {J. Phys. A: Math. Theor.}\ }\textbf {\bibinfo {volume}
  {50}},\ \bibinfo {pages} {453001} (\bibinfo {year} {2017})}\BibitemShut
  {NoStop}%
\bibitem [{\citenamefont {Shanahan}\ \emph {et~al.}(2018)\citenamefont
  {Shanahan}, \citenamefont {Chenu}, \citenamefont {Margolus},\ and\
  \citenamefont {del Campo}}]{PhysRevLett.120.070401}%
  \BibitemOpen
  \bibfield  {author} {\bibinfo {author} {\bibfnamefont {B.}~\bibnamefont
  {Shanahan}}, \bibinfo {author} {\bibfnamefont {A.}~\bibnamefont {Chenu}},
  \bibinfo {author} {\bibfnamefont {N.}~\bibnamefont {Margolus}}, \ and\
  \bibinfo {author} {\bibfnamefont {A.}~\bibnamefont {del Campo}},\ }\bibfield
  {title} {\enquote {\bibinfo {title} {Quantum {S}peed {L}imits across the
  {Q}uantum-to-{C}lassical {T}ransition},}\ }\href {\doibase
  10.1103/PhysRevLett.120.070401} {\bibfield  {journal} {\bibinfo  {journal}
  {Phys. Rev. Lett.}\ }\textbf {\bibinfo {volume} {120}},\ \bibinfo {pages}
  {070401} (\bibinfo {year} {2018})}\BibitemShut {NoStop}%
\bibitem [{\citenamefont {Okuyama}\ and\ \citenamefont
  {Ohzeki}(2018)}]{PhysRevLett.120.070402}%
  \BibitemOpen
  \bibfield  {author} {\bibinfo {author} {\bibfnamefont {M.}~\bibnamefont
  {Okuyama}}\ and\ \bibinfo {author} {\bibfnamefont {M.}~\bibnamefont
  {Ohzeki}},\ }\bibfield  {title} {\enquote {\bibinfo {title} {Quantum {S}peed
  {L}imit is {N}ot {Q}uantum},}\ }\href {\doibase
  10.1103/PhysRevLett.120.070402} {\bibfield  {journal} {\bibinfo  {journal}
  {Phys. Rev. Lett.}\ }\textbf {\bibinfo {volume} {120}},\ \bibinfo {pages}
  {070402} (\bibinfo {year} {2018})}\BibitemShut {NoStop}%
\bibitem [{\citenamefont {Shao}\ \emph {et~al.}(2020)\citenamefont {Shao},
  \citenamefont {Liu}, \citenamefont {Zhang}, \citenamefont {Yuan},\ and\
  \citenamefont {Liu}}]{PhysRevResearch.2.023299}%
  \BibitemOpen
  \bibfield  {author} {\bibinfo {author} {\bibfnamefont {Y.}~\bibnamefont
  {Shao}}, \bibinfo {author} {\bibfnamefont {B.}~\bibnamefont {Liu}}, \bibinfo
  {author} {\bibfnamefont {M.}~\bibnamefont {Zhang}}, \bibinfo {author}
  {\bibfnamefont {H.}~\bibnamefont {Yuan}}, \ and\ \bibinfo {author}
  {\bibfnamefont {J.}~\bibnamefont {Liu}},\ }\bibfield  {title} {\enquote
  {\bibinfo {title} {Operational definition of a quantum speed limit},}\ }\href
  {\doibase 10.1103/PhysRevResearch.2.023299} {\bibfield  {journal} {\bibinfo
  {journal} {Phys. Rev. Research}\ }\textbf {\bibinfo {volume} {2}},\ \bibinfo
  {pages} {023299} (\bibinfo {year} {2020})}\BibitemShut {NoStop}%
\bibitem [{\citenamefont {Kobayashi}\ and\ \citenamefont
  {Yamamoto}(2020)}]{PhysRevA.102.042606}%
  \BibitemOpen
  \bibfield  {author} {\bibinfo {author} {\bibfnamefont {K.}~\bibnamefont
  {Kobayashi}}\ and\ \bibinfo {author} {\bibfnamefont {N.}~\bibnamefont
  {Yamamoto}},\ }\bibfield  {title} {\enquote {\bibinfo {title} {Quantum speed
  limit for robust state characterization and engineering},}\ }\href {\doibase
  10.1103/PhysRevA.102.042606} {\bibfield  {journal} {\bibinfo  {journal}
  {Phys. Rev. A}\ }\textbf {\bibinfo {volume} {102}},\ \bibinfo {pages}
  {042606} (\bibinfo {year} {2020})}\BibitemShut {NoStop}%
\bibitem [{\citenamefont {Puebla}\ \emph {et~al.}(2020)\citenamefont {Puebla},
  \citenamefont {Deffner},\ and\ \citenamefont
  {Campbell}}]{PhysRevResearch.2.032020}%
  \BibitemOpen
  \bibfield  {author} {\bibinfo {author} {\bibfnamefont {R.}~\bibnamefont
  {Puebla}}, \bibinfo {author} {\bibfnamefont {S.}~\bibnamefont {Deffner}}, \
  and\ \bibinfo {author} {\bibfnamefont {S.}~\bibnamefont {Campbell}},\
  }\bibfield  {title} {\enquote {\bibinfo {title} {Kibble-{Z}urek scaling in
  quantum speed limits for shortcuts to adiabaticity},}\ }\href {\doibase
  10.1103/PhysRevResearch.2.032020} {\bibfield  {journal} {\bibinfo  {journal}
  {Phys. Rev. Research}\ }\textbf {\bibinfo {volume} {2}},\ \bibinfo {pages}
  {032020} (\bibinfo {year} {2020})}\BibitemShut {NoStop}%
\bibitem [{\citenamefont {Mohan}\ and\ \citenamefont
  {Pati}(2020)}]{arXiv:2006.14523}%
  \BibitemOpen
  \bibfield  {author} {\bibinfo {author} {\bibfnamefont {B.}~\bibnamefont
  {Mohan}}\ and\ \bibinfo {author} {\bibfnamefont {A.~K.}\ \bibnamefont
  {Pati}},\ }\bibfield  {title} {\enquote {\bibinfo {title} {Reverse {Q}uantum
  {S}peed {L}imit: {H}ow {S}low {Q}uantum {B}attery can {D}ischarge?}}\ }\href
  {https://arxiv.org/abs/2006.14523} {\bibfield  {journal} {\bibinfo  {journal}
  {arXiv:2006.14523}\ } (\bibinfo {year} {2020})}\BibitemShut {NoStop}%
\bibitem [{\citenamefont {del Campo}(2020)}]{arXiv:2007.15019}%
  \BibitemOpen
  \bibfield  {author} {\bibinfo {author} {\bibfnamefont {A.}~\bibnamefont {del
  Campo}},\ }\bibfield  {title} {\enquote {\bibinfo {title} {Probing {Q}uantum
  {S}peed {L}imits with {U}ltracold {G}ases},}\ }\href
  {https://arxiv.org/abs/2007.15019} {\bibfield  {journal} {\bibinfo  {journal}
  {arXiv:2007.15019}\ } (\bibinfo {year} {2020})}\BibitemShut {NoStop}%
\bibitem [{\citenamefont {Lam}\ \emph {et~al.}(2020)\citenamefont {Lam},
  \citenamefont {Peter}, \citenamefont {Groh}, \citenamefont {Alt},
  \citenamefont {Robens}, \citenamefont {Meschede}, \citenamefont {Negretti},
  \citenamefont {Montangero}, \citenamefont {Calarco},\ and\ \citenamefont
  {Alberti}}]{arXiv:2009.02231}%
  \BibitemOpen
  \bibfield  {author} {\bibinfo {author} {\bibfnamefont {M.~R.}\ \bibnamefont
  {Lam}}, \bibinfo {author} {\bibfnamefont {N.}~\bibnamefont {Peter}}, \bibinfo
  {author} {\bibfnamefont {T.}~\bibnamefont {Groh}}, \bibinfo {author}
  {\bibfnamefont {W.}~\bibnamefont {Alt}}, \bibinfo {author} {\bibfnamefont
  {C.}~\bibnamefont {Robens}}, \bibinfo {author} {\bibfnamefont
  {D.}~\bibnamefont {Meschede}}, \bibinfo {author} {\bibfnamefont
  {A.}~\bibnamefont {Negretti}}, \bibinfo {author} {\bibfnamefont
  {S.}~\bibnamefont {Montangero}}, \bibinfo {author} {\bibfnamefont
  {T.}~\bibnamefont {Calarco}}, \ and\ \bibinfo {author} {\bibfnamefont
  {A.}~\bibnamefont {Alberti}},\ }\bibfield  {title} {\enquote {\bibinfo
  {title} {Demonstration of quantum brachistochrones between distant states of
  an atom},}\ }\href {https://arxiv.org/abs/2009.02231} {\bibfield  {journal}
  {\bibinfo  {journal} {arXiv:2009.02231}\ } (\bibinfo {year}
  {2020})}\BibitemShut {NoStop}%
\bibitem [{\citenamefont {O'Connor}\ \emph {et~al.}(2020)\citenamefont
  {O'Connor}, \citenamefont {Guarnieri},\ and\ \citenamefont
  {Campbell}}]{arXiv:2011.05232}%
  \BibitemOpen
  \bibfield  {author} {\bibinfo {author} {\bibfnamefont {E.}~\bibnamefont
  {O'Connor}}, \bibinfo {author} {\bibfnamefont {G.}~\bibnamefont {Guarnieri}},
  \ and\ \bibinfo {author} {\bibfnamefont {S.}~\bibnamefont {Campbell}},\
  }\bibfield  {title} {\enquote {\bibinfo {title} {Action quantum speed
  limits},}\ }\href {https://arxiv.org/abs/2011.05232} {\bibfield  {journal}
  {\bibinfo  {journal} {arXiv:2011.05232}\ } (\bibinfo {year}
  {2020})}\BibitemShut {NoStop}%
\bibitem [{\citenamefont {Campaioli}\ \emph {et~al.}(2020)\citenamefont
  {Campaioli}, \citenamefont {Yu}, \citenamefont {Pollock},\ and\ \citenamefont
  {Modi}}]{arXiv:2004.03078}%
  \BibitemOpen
  \bibfield  {author} {\bibinfo {author} {\bibfnamefont {F.}~\bibnamefont
  {Campaioli}}, \bibinfo {author} {\bibfnamefont {C.-S.}\ \bibnamefont {Yu}},
  \bibinfo {author} {\bibfnamefont {F.~A.}\ \bibnamefont {Pollock}}, \ and\
  \bibinfo {author} {\bibfnamefont {K.}~\bibnamefont {Modi}},\ }\bibfield
  {title} {\enquote {\bibinfo {title} {Resource speed limits: {M}aximal rate of
  resource variation},}\ }\href {https://www.arxiv.org/abs/2004.03078}
  {\bibfield  {journal} {\bibinfo  {journal} {arXiv:2004.03078}\ } (\bibinfo
  {year} {2020})}\BibitemShut {NoStop}%
\bibitem [{\citenamefont {Girolami}(2014)}]{PhysRevLett.113.170401}%
  \BibitemOpen
  \bibfield  {author} {\bibinfo {author} {\bibfnamefont {D.}~\bibnamefont
  {Girolami}},\ }\bibfield  {title} {\enquote {\bibinfo {title} {Observable
  {M}easure of {Q}uantum {C}oherence in {F}inite {D}imensional {S}ystems},}\
  }\href {\doibase 10.1103/PhysRevLett.113.170401} {\bibfield  {journal}
  {\bibinfo  {journal} {Phys. Rev. Lett.}\ }\textbf {\bibinfo {volume} {113}},\
  \bibinfo {pages} {170401} (\bibinfo {year} {2014})}\BibitemShut {NoStop}%
\bibitem [{\citenamefont {Wigner}\ and\ \citenamefont
  {Yanase}(1963)}]{Wigner1963}%
  \BibitemOpen
  \bibfield  {author} {\bibinfo {author} {\bibfnamefont {E.~P.}\ \bibnamefont
  {Wigner}}\ and\ \bibinfo {author} {\bibfnamefont {M.~M.}\ \bibnamefont
  {Yanase}},\ }\bibfield  {title} {\enquote {\bibinfo {title} {Information
  contents of distributions},}\ }\href {\doibase 10.1073/pnas.49.6.910}
  {\bibfield  {journal} {\bibinfo  {journal} {Proc. Natl. Acad. Sci.}\ }\textbf
  {\bibinfo {volume} {49}},\ \bibinfo {pages} {910} (\bibinfo {year}
  {1963})}\BibitemShut {NoStop}%
\bibitem [{\citenamefont {Campaioli}\ \emph {et~al.}(2018)\citenamefont
  {Campaioli}, \citenamefont {Pollock}, \citenamefont {Binder},\ and\
  \citenamefont {Modi}}]{PhysRevLett.120.060409}%
  \BibitemOpen
  \bibfield  {author} {\bibinfo {author} {\bibfnamefont {F.}~\bibnamefont
  {Campaioli}}, \bibinfo {author} {\bibfnamefont {F.~A.}\ \bibnamefont
  {Pollock}}, \bibinfo {author} {\bibfnamefont {Felix~C.}\ \bibnamefont
  {Binder}}, \ and\ \bibinfo {author} {\bibfnamefont {K.}~\bibnamefont
  {Modi}},\ }\bibfield  {title} {\enquote {\bibinfo {title} {Tightening
  {Q}uantum {S}peed {L}imits for {A}lmost {A}ll {S}tates},}\ }\href {\doibase
  10.1103/PhysRevLett.120.060409} {\bibfield  {journal} {\bibinfo  {journal}
  {Phys. Rev. Lett.}\ }\textbf {\bibinfo {volume} {120}},\ \bibinfo {pages}
  {060409} (\bibinfo {year} {2018})}\BibitemShut {NoStop}%
\bibitem [{\citenamefont {Campaioli}\ \emph {et~al.}(2019)\citenamefont
  {Campaioli}, \citenamefont {Pollock},\ and\ \citenamefont
  {Modi}}]{Campaioli2019tightrobust}%
  \BibitemOpen
  \bibfield  {author} {\bibinfo {author} {\bibfnamefont {F.}~\bibnamefont
  {Campaioli}}, \bibinfo {author} {\bibfnamefont {F.~A.}\ \bibnamefont
  {Pollock}}, \ and\ \bibinfo {author} {\bibfnamefont {K.}~\bibnamefont
  {Modi}},\ }\bibfield  {title} {\enquote {\bibinfo {title} {Tight, robust, and
  feasible quantum speed limits for open dynamics},}\ }\href {\doibase
  10.22331/q-2019-08-05-168} {\bibfield  {journal} {\bibinfo  {journal}
  {{Quantum}}\ }\textbf {\bibinfo {volume} {3}},\ \bibinfo {pages} {168}
  (\bibinfo {year} {2019})}\BibitemShut {NoStop}%
\bibitem [{\citenamefont {Yang}\ \emph {et~al.}(2017)\citenamefont {Yang},
  \citenamefont {Pang},\ and\ \citenamefont {Jordan}}]{PhysRevA.96.020301}%
  \BibitemOpen
  \bibfield  {author} {\bibinfo {author} {\bibfnamefont {J.}~\bibnamefont
  {Yang}}, \bibinfo {author} {\bibfnamefont {S.}~\bibnamefont {Pang}}, \ and\
  \bibinfo {author} {\bibfnamefont {A.~N.}\ \bibnamefont {Jordan}},\ }\bibfield
   {title} {\enquote {\bibinfo {title} {Quantum parameter estimation with the
  {L}andau-{Z}ener transition},}\ }\href {\doibase 10.1103/PhysRevA.96.020301}
  {\bibfield  {journal} {\bibinfo  {journal} {Phys. Rev. A}\ }\textbf {\bibinfo
  {volume} {96}},\ \bibinfo {pages} {020301} (\bibinfo {year}
  {2017})}\BibitemShut {NoStop}%
\bibitem [{\citenamefont {Marvian}\ and\ \citenamefont
  {Spekkens}(2016)}]{PhysRevA.94.052324}%
  \BibitemOpen
  \bibfield  {author} {\bibinfo {author} {\bibfnamefont {I.}~\bibnamefont
  {Marvian}}\ and\ \bibinfo {author} {\bibfnamefont {R.~W.}\ \bibnamefont
  {Spekkens}},\ }\bibfield  {title} {\enquote {\bibinfo {title} {How to
  quantify coherence: {D}istinguishing speakable and unspeakable notions},}\
  }\href {\doibase 10.1103/PhysRevA.94.052324} {\bibfield  {journal} {\bibinfo
  {journal} {Phys. Rev. A}\ }\textbf {\bibinfo {volume} {94}},\ \bibinfo
  {pages} {052324} (\bibinfo {year} {2016})}\BibitemShut {NoStop}%
\bibitem [{\citenamefont {Fogarty}\ \emph {et~al.}(2020)\citenamefont
  {Fogarty}, \citenamefont {Deffner}, \citenamefont {Busch},\ and\
  \citenamefont {Campbell}}]{PhysRevLett.124.110601}%
  \BibitemOpen
  \bibfield  {author} {\bibinfo {author} {\bibfnamefont {T.}~\bibnamefont
  {Fogarty}}, \bibinfo {author} {\bibfnamefont {S.}~\bibnamefont {Deffner}},
  \bibinfo {author} {\bibfnamefont {T.}~\bibnamefont {Busch}}, \ and\ \bibinfo
  {author} {\bibfnamefont {S.}~\bibnamefont {Campbell}},\ }\bibfield  {title}
  {\enquote {\bibinfo {title} {Orthogonality {C}atastrophe as a {C}onsequence
  of the {Q}uantum {S}peed {L}imit},}\ }\href {\doibase
  10.1103/PhysRevLett.124.110601} {\bibfield  {journal} {\bibinfo  {journal}
  {Phys. Rev. Lett.}\ }\textbf {\bibinfo {volume} {124}},\ \bibinfo {pages}
  {110601} (\bibinfo {year} {2020})}\BibitemShut {NoStop}%
\bibitem [{\citenamefont {Nicholson}\ \emph {et~al.}(2020)\citenamefont
  {Nicholson}, \citenamefont {Garcia-Pintos}, \citenamefont {del Campo},\ and\
  \citenamefont {Green}}]{arXiv:2001.05418}%
  \BibitemOpen
  \bibfield  {author} {\bibinfo {author} {\bibfnamefont {S.~B.}\ \bibnamefont
  {Nicholson}}, \bibinfo {author} {\bibfnamefont {L.~P.}\ \bibnamefont
  {Garcia-Pintos}}, \bibinfo {author} {\bibfnamefont {A.}~\bibnamefont {del
  Campo}}, \ and\ \bibinfo {author} {\bibfnamefont {J.~R.}\ \bibnamefont
  {Green}},\ }\bibfield  {title} {\enquote {\bibinfo {title} {Time-information
  uncertainty relations in thermodynamics},}\ }\href {\doibase
  10.1038/s41567-020-0981-y} {\bibfield  {journal} {\bibinfo  {journal} {Nat.
  Phys.}\ }\textbf {\bibinfo {volume} {16}},\ \bibinfo {pages} {1211} (\bibinfo
  {year} {2020})}\BibitemShut {NoStop}%
\bibitem [{\citenamefont {Nicholson}\ \emph {et~al.}(2018)\citenamefont
  {Nicholson}, \citenamefont {del Campo},\ and\ \citenamefont
  {Green}}]{PhysRevE.98.032106}%
  \BibitemOpen
  \bibfield  {author} {\bibinfo {author} {\bibfnamefont {S.~B.}\ \bibnamefont
  {Nicholson}}, \bibinfo {author} {\bibfnamefont {A.}~\bibnamefont {del
  Campo}}, \ and\ \bibinfo {author} {\bibfnamefont {J.~R.}\ \bibnamefont
  {Green}},\ }\bibfield  {title} {\enquote {\bibinfo {title} {Nonequilibrium
  uncertainty principle from information geometry},}\ }\href {\doibase
  10.1103/PhysRevE.98.032106} {\bibfield  {journal} {\bibinfo  {journal} {Phys.
  Rev. E}\ }\textbf {\bibinfo {volume} {98}},\ \bibinfo {pages} {032106}
  (\bibinfo {year} {2018})}\BibitemShut {NoStop}%
\bibitem [{\citenamefont {Miller}\ \emph {et~al.}(2019)\citenamefont {Miller},
  \citenamefont {Scandi}, \citenamefont {Anders},\ and\ \citenamefont
  {Perarnau-Llobet}}]{PhysRevLett.123.230603}%
  \BibitemOpen
  \bibfield  {author} {\bibinfo {author} {\bibfnamefont {H.~J.~D.}\
  \bibnamefont {Miller}}, \bibinfo {author} {\bibfnamefont {M.}~\bibnamefont
  {Scandi}}, \bibinfo {author} {\bibfnamefont {J.}~\bibnamefont {Anders}}, \
  and\ \bibinfo {author} {\bibfnamefont {M.}~\bibnamefont {Perarnau-Llobet}},\
  }\bibfield  {title} {\enquote {\bibinfo {title} {Work {F}luctuations in
  {S}low {P}rocesses: {Q}uantum {S}ignatures and {O}ptimal {C}ontrol},}\ }\href
  {\doibase 10.1103/PhysRevLett.123.230603} {\bibfield  {journal} {\bibinfo
  {journal} {Phys. Rev. Lett.}\ }\textbf {\bibinfo {volume} {123}},\ \bibinfo
  {pages} {230603} (\bibinfo {year} {2019})}\BibitemShut {NoStop}%
\bibitem [{\citenamefont {Miller}\ \emph {et~al.}(2020)\citenamefont {Miller},
  \citenamefont {Guarnieri}, \citenamefont {Mitchison},\ and\ \citenamefont
  {Goold}}]{arXiv:2007.01882}%
  \BibitemOpen
  \bibfield  {author} {\bibinfo {author} {\bibfnamefont {H.~J.~D.}\
  \bibnamefont {Miller}}, \bibinfo {author} {\bibfnamefont {G.}~\bibnamefont
  {Guarnieri}}, \bibinfo {author} {\bibfnamefont {M.~T.}\ \bibnamefont
  {Mitchison}}, \ and\ \bibinfo {author} {\bibfnamefont {J.}~\bibnamefont
  {Goold}},\ }\bibfield  {title} {\enquote {\bibinfo {title} {Quantum
  {F}luctuations {H}inder {F}inite-{T}ime {I}nformation {E}rasure near the
  {L}andauer {L}imit},}\ }\href {\doibase 10.1103/PhysRevLett.125.160602}
  {\bibfield  {journal} {\bibinfo  {journal} {Phys. Rev. Lett.}\ }\textbf
  {\bibinfo {volume} {125}},\ \bibinfo {pages} {160602} (\bibinfo {year}
  {2020})}\BibitemShut {NoStop}%
\bibitem [{\citenamefont {Zhao}\ and\ \citenamefont {Yu}(2018)}]{Zhao2018}%
  \BibitemOpen
  \bibfield  {author} {\bibinfo {author} {\bibfnamefont {H.}~\bibnamefont
  {Zhao}}\ and\ \bibinfo {author} {\bibfnamefont {C.-S.}\ \bibnamefont {Yu}},\
  }\bibfield  {title} {\enquote {\bibinfo {title} {Coherence measure in terms
  of the {T}sallis relative $\alpha$ entropy},}\ }\href {\doibase
  10.1038/s41598-017-18692-1} {\bibfield  {journal} {\bibinfo  {journal} {Sci.
  Rep.}\ }\textbf {\bibinfo {volume} {8}},\ \bibinfo {pages} {299} (\bibinfo
  {year} {2018})}\BibitemShut {NoStop}%
\bibitem [{\citenamefont {Ruskai}\ and\ \citenamefont
  {Stillinger}(1990)}]{Ruskai_1990}%
  \BibitemOpen
  \bibfield  {author} {\bibinfo {author} {\bibfnamefont {M.~B.}\ \bibnamefont
  {Ruskai}}\ and\ \bibinfo {author} {\bibfnamefont {F.~H.}\ \bibnamefont
  {Stillinger}},\ }\bibfield  {title} {\enquote {\bibinfo {title} {Convexity
  inequalities for estimating free energy and relative entropy},}\ }\href
  {\doibase 10.1088/0305-4470/23/12/023} {\bibfield  {journal} {\bibinfo
  {journal} {J. Phys. A: Math. Gen.}\ }\textbf {\bibinfo {volume} {23}},\
  \bibinfo {pages} {2421} (\bibinfo {year} {1990})}\BibitemShut {NoStop}%
\bibitem [{\citenamefont {Audenaert}\ and\ \citenamefont
  {Eisert}(2005)}]{doi:10.1063/1.2044667}%
  \BibitemOpen
  \bibfield  {author} {\bibinfo {author} {\bibfnamefont {K.~M.~R.}\
  \bibnamefont {Audenaert}}\ and\ \bibinfo {author} {\bibfnamefont
  {J.}~\bibnamefont {Eisert}},\ }\bibfield  {title} {\enquote {\bibinfo {title}
  {Continuity bounds on the quantum relative entropy},}\ }\href {\doibase
  10.1063/1.2044667} {\bibfield  {journal} {\bibinfo  {journal} {J. Math.
  Phys.}\ }\textbf {\bibinfo {volume} {46}},\ \bibinfo {pages} {102104}
  (\bibinfo {year} {2005})}\BibitemShut {NoStop}%
\bibitem [{\citenamefont {Audenaert}\ and\ \citenamefont
  {Eisert}(2011)}]{doi:10.1063/1.3657929}%
  \BibitemOpen
  \bibfield  {author} {\bibinfo {author} {\bibfnamefont {K.~M.~R.}\
  \bibnamefont {Audenaert}}\ and\ \bibinfo {author} {\bibfnamefont
  {J.}~\bibnamefont {Eisert}},\ }\bibfield  {title} {\enquote {\bibinfo {title}
  {Continuity bounds on the quantum relative entropy -- {II}},}\ }\href
  {\doibase 10.1063/1.3657929} {\bibfield  {journal} {\bibinfo  {journal} {J.
  Math. Phys.}\ }\textbf {\bibinfo {volume} {52}},\ \bibinfo {pages} {112201}
  (\bibinfo {year} {2011})}\BibitemShut {NoStop}%
\end{thebibliography}

%

\end{document}